\documentclass[11pt]{article}
\usepackage{graphicx}
\usepackage[utf8]{inputenc} 
\pdfoutput=1

\usepackage{amsmath,amsfonts,amssymb,epsfig,empheq}
\usepackage{slashed,xcolor}
\usepackage[symbol]{footmisc}

\numberwithin{equation}{section}

\usepackage{graphicx}
\usepackage{cancel}
\usepackage{cite}
\usepackage{color,tikz}

\usepackage{xspace}

\def\gev{\ \mathrm{GeV}}

\newcommand{\exclude}[1]{}
\def\nn{\nonumber}

\long\def\symbolfootnote[#1]#2{\begingroup%
\def\thefootnote{\fnsymbol{footnote}}\footnote[#1]{#2}\endgroup}

\def\bra{\langle}
\def\ket{\rangle}

\def\ov{\overline}

\def\msbar{{\ov {\rm MS}}}
\def\drbar{{\ensuremath{ \overline{\rm DR}'}}}

\def\lag{\mathcal{L}}
\def\lagr{\mathcal{L}}

\def\lt{\tilde{\lambda}}
\def\blog{\overline{\log}}
\def\llog{\overline{\log}}

\def\db{\ov{\delta}}

\def\twomat[#1,#2][#3,#4]{\left( \begin{array}{cc} #1 & #2 \\ #3 & #4 \end{array} \right)}
\def\threemat[#1,#2,#3][#4,#5,#6][#7,#8,#9]{\left( \begin{array}{ccc} #1 & #2 & #3\\ #4 & #5 & #6 \\ #7 & #8 & #9 \end{array} \right)}
\def\twovec[#1,#2]{\left( \begin{array}{c} #1  \\ #2 \end{array} \right)}
\def\thv[#1,#2,#3]{\left( \begin{array}{c} #1 \\ #2 \\ #3 \end{array} \right)}
\def\twv[#1,#2]{\left( \begin{array}{c} #1 \\ #2 \end{array} \right)}

\def\veff{V_\text{eff}}
\def\vtree{V^{(0)}}
\def\vone{V^{(1)}}

\def\Ooneloop{\mathcal{O}(\textrm{1-loop})}
\def\Otwoloop{\mathcal{O}(\textrm{2-loop})}

\def\HH{{\textcolor{black}{\mathcal{H}}}}
\def\gea{{\textcolor{black}{i}}}
\def\geb{{\textcolor{black}{j}}}
\def\geap{{\textcolor{black}{i'}}}
\def\gebp{{\textcolor{black}{j'}}}
\def\gec{{\textcolor{black}{k}}}
\def\ged{{\textcolor{black}{l}}}
\def\lii{{\textcolor{black}{p}}}
\def\liip{{\textcolor{black}{p'}}}
\def\lij{{\textcolor{black}{q}}}

\def\lik{{\textcolor{black}{r}}}
\def\lil{{\textcolor{black}{s}}}
\def\ligm{{\textcolor{black}{t}}}
\def\lin{{\textcolor{black}{n}}}
\def\lir{{\textcolor{black}{v}}}
\def\lix{{\textcolor{black}{x}}}
\def\liy{{\textcolor{black}{y}}}
\def\liz{{\textcolor{black}{z}}}
\def\liu{{\textcolor{black}{u}}}
\def\heI{{\textcolor{black}{\mathcal{P}}}}
\def\heJ{{\textcolor{black}{\mathcal{Q}}}}
\def\heK{{\textcolor{black}{\mathcal{R}}}}
\def\heL{{\textcolor{black}{\mathcal{S}}}}

\def\heM{{\textcolor{black}{\mathcal{T}}}}
\def\heN{{\textcolor{black}{\mathcal{N}}}}
\def\heX{{\textcolor{black}{\mathcal{X}}}}
\def\heY{{\textcolor{black}{\mathcal{Y}}}}

\newcommand\lights[1]{{\color{black} {#1}}}

\newcommand{\eps}{\delta}

\def\Mm{\ensuremath{M\xspace}}
\def\Ml{\ensuremath{\zeta\xspace}}

\begin{document}

\begin{titlepage}

\begin{flushright}
OU-HET-984
\end{flushright}
\begin{center}

\vspace{1cm}

{\LARGE \bf Matching renormalisable couplings:} 
\vskip 0.3cm
{\LARGE \bf simple schemes and a plot}

\vspace{1cm}

{\Large Johannes~Braathen,$^{\!\!\!\,a,b}$~ 
Mark~D.~Goodsell$^{\,a}$~ and Pietro~Slavich$^{\,a}$}

\vspace*{5mm}

{\sl ${}^a$ Laboratoire de Physique Th\'eorique et Hautes Energies (LPTHE),\\ UMR 7589,
Sorbonne Universit\'e et CNRS, 4 place Jussieu, 75252 Paris Cedex 05, France.  }
\vspace*{2mm}\\
{\sl ${}^b$ Department of Physics, Osaka University, Toyonaka, Osaka 560-0043, Japan. }
\end{center}
\symbolfootnote[0]{{\tt e-mail:}}
\symbolfootnote[0]{{\tt braathen@het.phys.sci.osaka-u.ac.jp}}
\symbolfootnote[0]{{\tt goodsell@lpthe.jussieu.fr}}
\symbolfootnote[0]{{\tt slavich@lpthe.jussieu.fr}}

\vspace{0.7cm}

\abstract{We discuss different choices that can be made when matching a general high-energy theory -- with the restriction that it should not contain
heavy gauge bosons -- onto a general renormalisable effective field theory at one loop, with particular attention to the quartic scalar couplings and Yukawa couplings. This includes a generalisation of the counterterm scheme that was found to be useful in the case of high-scale/split supersymmetry, but we show the important differences when there are new heavy scalar fields in singlet or triplet representations of $SU(2)$. We also analytically compare our methods and choices with the approach of matching pole masses, proving the equivalence with one of our choices. We outline how to make the extraction of quartic couplings using pole masses more efficient, an approach that we hope will generalise beyond one loop. We give examples of the impact of different scheme choices in a toy model; we also discuss the MSSM and give the threshold corrections to the Higgs quartic coupling in Dirac gaugino models.}

\vfill

\end{titlepage}

\tableofcontents

\setcounter{footnote}{0}

\section{Introduction}
\label{SEC:INTRODUCTION}

\setcounter{footnote}{2}

In the absence of clear collider signals of new particles, there has been much recent interest in constraining deviations from the Standard Model (SM) in terms of effective operators. This approach to the ``Standard Model Effective Field Theory'' has primarily been interested in higher-dimensional operators that encode new effective interactions, for example recent work on calculating these in general theories can be found in \cite{Henning:2014wua,Drozd:2015rsp,delAguila:2016zcb,Boggia:2016asg,Henning:2016lyp,Ellis:2016enq,Fuentes-Martin:2016uol,Zhang:2016pja,Ellis:2017jns,Summ:2018oko,Bakshi:2018ics}. However, there is also important information that can be extracted by matching the \emph{renormalisable} couplings of the SM. In particular, this is an increasingly important approach to calculating the Higgs mass from a top-down theory, providing a more accurate calculation than a fixed-order one once new particles that couple to the Higgs are above a few TeV. It is the only approach to constraining the Higgs mass in split supersymmetry \cite{ArkaniHamed:2004fb,Giudice:2004tc,ArkaniHamed:2004yi} where new physics could be around $100-10^5$ TeV \cite{Giudice:2011cg,Arvanitaki:2012ps}; high-scale supersymmetry \cite{Hall:2009nd,Giudice:2011cg,Degrassi:2012ry,Buttazzo:2013uya,Bagnaschi:2014rsa} where it could be around $10^7-10^9$ TeV; the FSSM \cite{Benakli:2013msa,Benakli:2015ioa} where it could be as high as the GUT/Planck scale, etc. Moreover, there is also a parallel effort considering the low-energy theory to be a simple non-supersymmetric extension of the SM such as a Two-Higgs-Doublet Model (THDM) \cite{Haber:1993an,Lee:2015uza,Bagnaschi:2015pwa,Benakli:2018vqz,Bahl:2018jom}, and then it is very interesting to match these theories to new physics at a (much) higher scale. 

With this motivation, we require: (i) the extraction of the renormalisable couplings (gauge couplings, Yukawa couplings and scalar quartic couplings) in the low-energy theory from observables; (ii) renormalisation group equations (RGEs) for the low-energy theory; and (iii) threshold corrections at the matching scale which we shall denote throughout $\Mm$. The RGEs for general renormalisable field theories have been known for some time up to two loop order \cite{Machacek:1983tz,Machacek:1983fi,Machacek:1984zw,Luo:2002ti,Sperling:2013eva,Sperling:2013xqa,Bednyakov:2018cmx,Schienbein:2018fsw} and can be obtained for any model by {\tt SARAH} \cite{Staub:2008uz,Staub:2012pb,Staub:2013tta,Staub:2015kfa} or {\tt PyR@TE} \cite{Lyonnet:2013dna,Lyonnet:2016xiz}, and higher loop orders are available for the SM. On the other hand, for (i) and (iii) the information is less complete: when the low-energy theory is the SM, the Higgs mass is used to extract the running quartic coupling, and the extraction of all couplings can be performed at two loop order (with some three- or four-loop corrections known), e.g.~in \cite{Degrassi:2012ry,Buttazzo:2013uya,Martin:2014cxa,Kniehl:2015nwa,Kniehl:2016enc}, but for general models in {\tt SARAH} it can be done only at one loop order, with two-loop corrections to the Higgs mass in the  limit of vanishing electroweak gauge couplings \cite{Goodsell:2014bna,Goodsell:2015ira,Braathen:2017izn}. Furthermore, threshold corrections to the Higgs quartic coupling have been computed explicitly for some models or scenarios such as split/high-scale supersymmetry, up to full one-loop plus leading two-loop  order\cite{Giudice:2011cg,Draper:2013oza,Bagnaschi:2014rsa,Vega:2015fna,Bagnaschi:2017xid}, and even recently up to leading three-loop order in \cite{Harlander:2018yhj}. These corrections are implemented in public codes for the Higgs mass calculation such as  {\tt SusyHD}~\cite{Vega:2015fna}, {\tt MhEFT}~\cite{Lee:2015uza}, {\tt FlexibleSUSY}~\cite{Athron:2017fvs} and {\tt FeynHiggs}~\cite{Hahn:2013ria,Bahl:2016brp,Bahl:2017aev}. The codes {\tt FlexibleEFTHiggs} \cite{Athron:2016fuq} and {\tt SARAH} \cite{Staub:2017jnp} also allow one-loop matching of a general theory to the SM as the low-energy theory via matching of pole masses. Finally, the code {\tt MatchingTools}~\cite{Criado:2017khh} matches two general theories to each other, but only at the tree level.

While it is vital to reduce the error in the extraction of the top Yukawa couplings and strong gauge coupling, the need for precision in the extraction of low-energy parameters and especially matching is particularly important for quartic couplings, which are well-known to be highly sensitive to quantum corrections, as stressed e.g.~in \cite{Braathen:2017jvs}. The purpose of the running to high scales in the bottom-up approach is to constrain the scale of new physics or investigate the (scale of) instabilities of the potential, and these depend logarithmically on the scale, thus the scale depends exponentially on small differences in the low-energy parameters. 

In this work, we shall instead be interested in the top-down approach and provide all of the ingredients to match the renormalisable couplings between two generic theories where the high-energy theory contains no heavy gauge bosons (this generalises the most interesting well-known examples) elucidating \emph{the various choices that can be made}. In section \ref{SEC:GENERAL} we describe three general approaches to obtaining the matching conditions at one loop, of which we shall develop the one that is most suited to be generalised beyond one loop. However, our main focus will be on certain important further details:

\bigskip

\begin{enumerate}
\item Mixing between heavy and light states is inevitable in models with additional Higgs doublets, and then there are quantum corrections to the mixing angle(s). This has been investigated in the case of one extra doublet \cite{Bagnaschi:2014rsa,Wells:2017vla,Bahl:2018jom} and it was found that a judicious choice of counterterms allows the calculation to be simplified (so that the mixing angle $\beta$ is not modified). We show how this can be generalised beyond one additional doublet. 
\item In the presence of heavy $SU(2)$ singlets or triplets, a trilinear coupling with two light Higgs fields is possible, and then the quartic coupling receives a correction at tree level when integrating out the heavy states. The presence of trilinear couplings with two light Higgs scalars moreover leads to infra-red divergences in the amplitudes which cancel in the threshold corrections: we explicitly show how these cancel and how they can be simply dealt with. 
\item In the presence of gauge singlets, tadpoles are generated before electroweak symmetry breaking. We describe four different approaches to dealing with them.
\item We show that the threshold corrections to the Higgs quartic, under the assumption that there are no heavy gauge bosons, are independent of gauge couplings\footnote{Note that, in the discussion of a general field theory, ``gauge couplings'' refers strictly to the interactions of the gauge bosons. In supersymmetric theories some of the scalar and Yukawa interactions may be related to the gauge couplings, but for the sake of our discussion they are treated just like all other scalar and Yukawa interactions.} at one loop, which is not immediately obvious. 
\item It is clear that cubic scalar couplings in the low energy theory should be at most of the order of the mass scale of the low-energy theory, which we denote \Ml. However, if we insist on including such couplings that do not decouple as we take $\Ml \rightarrow 0$ then we find that we must include higher-dimensional operators to cancel the infra-red divergences. We describe this explicitly in section \ref{SEC:HIGHERDIM}.
\item As a result of points (1) and (4) we give, in section \ref{SEC:COUNTERTERMS}, what we believe is the simplest possible prescription for matching general scalar quartic couplings. 
\item As mentioned above, an alternative approach to matching quartic or Yukawa couplings when the low-energy theory is the SM is to match pole masses in the two theories. However, given that there are different possible choices for parameter definitions when we perform a ``conventional'' matching calculation, it is not immediately obvious how to compare the definitions in the two approaches (i.e.~to know what we actually obtain from the pole-matching calculation!). This has been seen in the case of high scale/split SUSY in \cite{Bagnaschi:2014rsa,Wells:2017vla,Bahl:2018jom}, where the pole mass calculation gives a result equivalent to the ``counterterm'' approach to the angle $\beta$, which we define in section \ref{SEC:MATCHING}. In section \ref{SEC:POLEMATCHING} we derive the matching conditions for a general high-energy theory using the pole matching approach, and show the correspondence with the EFT calculation.
\item As a result of the derivation in  section \ref{SEC:POLEMATCHING}, we propose in section \ref{sec:efficient} a simple and explicitly infra-red safe prescription for matching Higgs quartic couplings where we only need to evaluate two-point scalar amplitudes.
\end{enumerate} 
Our approach to matching is illustrated with examples of the MSSM and Dirac gaugino models in section \ref{SEC:EXAMPLES}, and we investigate the impact of our counterterm choice in a toy model in section \ref{SEC:TOYMODEL}. We then describe the effect of \emph{fermion mixing} on matching Yukawa couplings in section \ref{SEC:YUKAWAS}, before concluding in section \ref{SEC:CONCLUSIONS}. The appendices contain our notation, the general results for threshold corrections, and specific results for Dirac gaugino models.

\section{Deriving the matching conditions}
\label{SEC:GENERAL}

In this paper we are interested in corrections to scalar quartic couplings in general renormalisable field theories, the effect of mixing of scalars, and gauge (in)dependence of the results. It turns out that in the body of the text we only explicitly need to refer to pure scalar interactions, and some interactions of scalars with gauge bosons.  We will work in terms of real scalars, which we denote as $\{\Phi_\gea\}$ -- with indices $\{\gea, \geb, \gec, ...\}$ -- in our high-energy theory, and as $\{\phi_\lii\}$ --  with indices drawn from $\{\lii,\lij,\lik,\lil,\lix,\liy\}$ -- in the low-energy theory.\footnote{Note that we will also use indices $\{\lii, \lij,\lik,\cdots\}$ for states of the high-energy theory that can be identified as light and therefore correspond to states in the low-energy theory. } The gauge bosons -- which appear in both the high- and low-energy theories, since we shall not consider the case of integrating out heavy gauge bosons -- are denoted as $A_\mu^a$ with indices $\{a,b,c,d\}$. Then the  Lagrangian terms of the high-energy theory (HET) that are relevant to the matching conditions are
\begin{align}
\lag_\text{HET} \supset& - t_\gea \Phi_\gea - \frac{1}{2} m_\gea^2 \Phi_\gea^2 - \frac{1}{6} a_{\gea\geb\gec} \Phi_\gea \Phi_\geb \Phi_\gec - \frac{1}{24} \lt_{\gea\geb\gec\ged} \Phi_\gea \Phi_\geb \Phi_\gec \Phi_\ged  + g^{a \gea\geb} A^a_\mu \Phi_\gea \partial^\mu \Phi_\geb,
\label{EQ:HETBasis}\end{align}
while the effective low-energy theory contains
\begin{align}
  \lag_\text{EFT} \supset& 
 - \frac{1}{24} \lambda_{\lii\lij\lik\lil} \phi_\lii \phi_\lij \phi_\lik \phi_\lil  + g^{a \lii\lij} A^a_\mu \phi_\lii \partial^\mu \phi_\lij.
\label{EQ:GENEFT}\end{align}
Since the gauge group is unbroken in each case, the couplings $g^{a\lii\lij}$ are proportional to the group generators (in a real representation). The full set of our conventions (and loop functions) is given in appendix \ref{APP:CONVENTIONS}, but it should be emphasised that we take all purely scalar couplings -- i.e.~$a_{\gea\geb\gec}$ and $\lambda_{\gea\geb\gec\ged}$ -- to be fully symmetric under the exchange of indices, and the $g^{a\gea\geb}$ couplings to be antisymmetric under the exchange $\gea\leftrightarrow\geb$. Note also that we can assume without any loss of generality that we are working with scalars defined in the mass-diagonal basis.

We shall treat the above fields as fluctuations around their values at the minimum of the potential. Since we are assuming that no gauge groups are broken before electroweak symmetry breaking, the only fields that may obtain an expectation value in the HET are gauge singlets. If we start in some basis where the fields have expectation values $\{v_\gea\}$ then to obtain $\bra \Phi_\gea \ket = 0$ we should make the shift 
\begin{align} 
t_\gea &\rightarrow t_\gea + m_{\gea}^2 v_\gea+ \frac{1}{2} a_{\gea\geb\gec} v_\geb v_\gec + \frac{1}{6} \lt_{\gea\geb\gec\ged} v_\geb v_\gec v_\ged, \quad\nn\\
m_{\gea}^2 \delta_{\gea\geb} &\rightarrow m_{\gea}^2 \delta_{\gea\geb} + a_{\gea\geb\gec} v_\gec + \frac{1}{2}\lt_{\gea\geb\gec\ged} v_\gec v_\ged, \nn\\
 a_{\gea\geb\gec} &\rightarrow a_{\gea\geb\gec} +  \lt_{\gea\geb\gec\ged} v_\ged,
\label{EQ:removesingletvevs}\end{align}
and then diagonalise the mass terms again.

We shall match our two theories together at some scale \Mm, assuming that all ``heavy'' fields have masses of this order, and take the mass scale of our low-energy theory to be $\Ml \ll \Mm$. We shall have in mind that this hierarchy can be of more than one-loop order, but in any case since we are only matching at one loop we can treat masses that are suppressed by one loop compared to the scale \Mm ~-- i.e.~all $m_{\lii\lij}^2$ -- as effectively zero. For example, taking the SM as low-energy theory, $\Ml \sim v \sim m_h$. Then it is convenient to take the limit $\Ml \rightarrow 0$ in the loop functions for the final expressions, as terms of order $\Ml$ would lead to corrections to the result suppressed by powers of $v/\Mm$, i.e.~equivalent to higher-dimensional operators. 

While the SM contains no cubic scalar couplings prior to electroweak symmetry breaking, a general low-energy theory (involving, e.g.~electroweak triplets or singlets) could contain them. However, as mentioned in the introduction, for consistency of the theory we must require 
$$a_{\lii\lij\lik} \sim \Ml.$$
One way to see this is just from unitarity considerations \cite{Veneziano:1972rs,Goodsell:2018tti}, and implies that we must include \emph{higher-dimensional operators} in the theory. Another perspective is that if we allow cubic scalar couplings in the low energy theory then we must add a finite set of higher-dimensional operators to cancel infra-red divergences; we discuss this in section \ref{SEC:HIGHERDIM}.

In the presence of singlets, tadpole terms $-t_\lii\phi_\lii$ may also appear in the low-energy theory Lagrangian. However, once again these must at least be linear in the light mass scale
$$ t_\lii\sim\Ml.$$
Consequently, when we take the limit $\Ml \rightarrow 0$ terms with either $t_\lii$ or $a_{\lii\lij\lik}$ vanish, and we have therefore excluded them from equation (\ref{EQ:GENEFT}). 

We shall now briefly review three methods of deriving the matching conditions between these two theories.

\subsection{Diagrammatic}

The conventional approach to matching theories is to compare Feynman diagram calculations. The approach in the next subsection (using path integrals) corresponds to calculating 1PI diagrams, but at the expense of obtaining a non-canonically normalised low-energy theory. If we want to insist that our low-energy theory has canonical kinetic terms, and want to match directly using diagrams, then the obvious and essentially only approach is to match \emph{$S$-matrix elements} in the two theories. The simplest way to do this is to take $\Ml \rightarrow 0$ first, making sure that the pole masses (not just the tree-level masses) of all the light particles are also set to zero, and then matching the results in the two theories as the total external momentum is taken to zero.

\subsection{Effective action: path integral approach}
\label{sec:CDE}

The other intuitive approach to matching effective theories comes from the Wilsonian picture: we want to integrate out the ``heavy'' degrees of freedom $\Phi_H$ and be left with only the ``light'' ones $\phi_L$, so on the one hand, in the absence of mixing, we write \cite{Henning:2014wua,Drozd:2015rsp}
\begin{align}
e^{iS_\text{EFT} [\phi_L]} =& \int D\Phi_H e^{i S[\phi_L, \Phi_H]} \nn\\
=&\, e^{i S[\phi_L, \Phi_H^c[\phi_L]]} \exp \bigg[ - \frac{1}{2} \mathrm{Tr}\log \bigg( - \frac{\delta^2 S}{\delta \Phi_H^2}\Bigg|_{\Phi_H=\Phi_H^c} \bigg)\bigg],
\end{align}
where $\Phi_H^c$ is defined by the relation 
\begin{align} 
 \frac{\delta S}{\delta\Phi_H}\Bigg|_{\Phi_H=\Phi_H^c}=0\,,
\end{align}
and can be expressed in terms of $\phi_L$.  This means that we write 
\begin{align}
S_{\rm EFT} [\phi_L]= S_{\rm EFT}^{\rm tree} + S_{\rm EFT}^{\rm 1-loop} = S[\phi_L, \Phi_H^c[\phi_L]] + \frac{i}{2} \mathrm{Tr}\log \bigg( - \frac{\delta^2 S}{\delta \Phi_H^2}\Bigg|_{\Phi_H=\Phi_H^c} \bigg).
\label{EQ:BasicCDE}\end{align}
On the other hand, in the presence of mixing between light and heavy states the problem of integrating out heavy degrees of freedom has also been addressed \cite{delAguila:2016zcb,Henning:2016lyp,Boggia:2016asg,Ellis:2016enq,Fuentes-Martin:2016uol,Zhang:2016pja,Ellis:2017jns}; writing 
\begin{align}
\twomat[\frac{\delta^2 S}{\delta \Phi_H^2},\frac{\delta^2 S}{\delta  \phi_L \delta\Phi_H}][ \frac{\delta^2 S}{\delta  \phi_L\delta \Phi_H}, \frac{\delta^2 S}{\delta \phi_L^2}] \equiv \twomat[\Delta_H,X_{HL}][X_{LH}, \Delta_L]
\end{align}
we can write \cite{Fuentes-Martin:2016uol}
\begin{align}
S_{\rm EFT}^{\rm 1-loop} =\frac{i}{2} \log \det ( \Delta_H - X_{HL} \Delta^{-1}_{L} X_{LH} )\Big|_{\rm hard}.
  \label{EQ:SEFT}\end{align}
Here ``hard'' means that the integral over loop momentum should be split up into ``hard'' and ``soft'' pieces via the method of regions, and the ``soft'' pieces should be discarded. This neatly avoids infra-red divergences, which must come from the ``soft'' part of the integrals (where the loop momentum is small).

\subsection{Effective action: equations of motion method}
\label{sec:effectiveaction}

Since we shall be interested in this work in large separations of scale between the low- and high-energy theories, we restrict to only \emph{renormalisable} operators,  and this means we shall (almost always) only need the kinetic terms and couplings up to quartic order. Moreover, our focus shall be on the different choices (of parameters, renormalisation schemes etc.) that are possible, and we want a method that makes these transparent. We also want a technique that will generalise (in future work) beyond one-loop order. Such an approach is given by simply evaluating the effective action up to quartic order for a general theory, and then integrating out the ``heavy'' fields using the equations of motion, matching the terms onto the equivalent ones in the low-energy theory. We define the effective action for the full high-energy theory as $S_{HET}[\Phi]$ and recall that it is the generating function of one-particle-irreducible diagrams; in momentum space it is 
\begin{align}
S_{HET} &= i \sum_{n=2}^\infty \frac{1}{n!} \int d^d p_1 ... d^d p_n \Phi_{\lights{i_1}} (p_1) \cdots \Phi_{\lights{i_n}} (p_n) \,\Gamma^{(n)}_{\lights{i_1},\ldots,\lights{i_n}} (p_1, \ldots, p_n)\delta^4\Big(\sum p_i\Big)
\end{align}
which we expand as a series in $p_i/\Mm$, so that $\Gamma^{(n)}_{\lights{i_1},\ldots,\lights{i_n}} (p_1, \ldots, p_n) = \Gamma^{(n)}_{\lights{i_1},\ldots,\lights{i_n}} (0,\ldots,0) + \mathcal{O} (\frac{p_i^2}{\Mm^2})$ and we can write, in the basis after the shifts (\ref{EQ:removesingletvevs}):
\begin{align}
S_{HET} =& \int d^4x \Gamma[\Phi] + ...\\
  \Gamma[\Phi] \equiv&  - \frac{1}{2}  Z_{\gea\geb}\Phi_\gea \partial_\mu \partial^\mu \Phi_\geb - (t_\gea + \delta t_\gea) \Phi_\gea - \frac{1}{2} (m_\gea^2 \delta_{\gea\geb} + \delta m_{\gea\geb}^2) \Phi_\gea \Phi_\geb - \frac{1}{6} (a_{\gea\geb\gec} + \delta a_{\gea\geb\gec})\Phi_\gea \Phi_\geb \Phi_\gec \nn\\
  &- \frac{1}{24} (\lt_{\gea\geb\gec\ged} + \delta \lt_{\gea\geb\gec\ged}) \Phi_\gea \Phi_\geb \Phi_\gec \Phi_\ged. \nn
\end{align}
We work in a minimal subtraction scheme ($\msbar$ or $\drbar$) where the counterterms have already been absorbed in the above; the (finite) quantities $\delta t_\gea$, $\delta m_{\gea\geb}^2$, $\delta a_{\gea\geb\gec}$, $\delta \lt_{\gea\geb\gec\ged}$ are the first through fourth derivatives of the loop correction to the \emph{renormalised} effective potential. 
This is valid to any loop order required, the appropriate corrections being included in the ``couplings.''  We write the quartic coupling in the high-energy theory with a tilde to distinguish it from the quartic in the low-energy theory (no such distinction is necessary for the cubic couplings). 
We then compute  
\begin{align}
0=&\ \frac{\delta \Gamma [\Phi]}{\delta \Phi_\gea}
\label{EQ:EOM}
\end{align}
for the heavy fields and reinsert the results into our expanded effective action. 
To obtain the same result for the effective action as from equation (\ref{EQ:SEFT}) we should expand the scalar mass term as a (diagonal) tree-level piece plus a perturbation
and expand the resulting effective action to one-loop order. First, however, if there are heavy singlet fields, then denoting their indices with an italic capital $\{\heI,\heJ,\heK,\heL\}$, they may have a non-vanishing tadpole before electroweak symmetry breaking and so: 
\begin{enumerate}
\item In some favourable cases a discrete symmetry, which is broken at the same time as electroweak symmetry (or not at all), forbids such a tadpole (such as in e.g.~the $\mathbb{Z}_2$-symmetric singlet-extension of the SM or the $\mathbb{Z}_3$-symmetric NMSSM in the unbroken phase). 
\item We may have the freedom to adjust the tree-level tadpole term $t_\heI$ already in the basis of equation (\ref{EQ:HETBasis}) so that the total tadpole including quantum corrections is zero, without needing to make any shifts of the form (\ref{EQ:removesingletvevs}). This is the case if we specify the high-energy theory by just a matching scale and the dimensionless parameters, for example if we scan over supersymmetric models without specifying a mediation mechanism. 
\item We can assume that the tadpole equation is satisfied at tree level (so that $t_\heI =0$). Then we solve (\ref{EQ:EOM}) treating $\delta t_\heI$ as a one-loop perturbation of the tree-level tadpole condition. I.e. since we have $\bra \Phi_\heI \ket = 0$ at tree level with all non-singlet field expectation values set to zero, the solution to (\ref{EQ:EOM}) is
  $$
\Phi_\heI = - \frac{1}{m_{\heI}^2} \delta t_\heI + \mathcal{O}(\Phi_\gea^2) + \mathcal{O}(2\ \mathrm{loops})
$$
which effectively means shifting
\begin{equation} 
    \delta m_{\gea\geb}^2 \rightarrow \delta m_{\gea\geb}^2 - \frac{\delta t_\heI}{m_\heI^2} \,a_{\heI\gea\geb}, \quad \delta a_{\gea\geb\gec} \rightarrow \delta a_{\gea\geb\gec} - \frac{\delta t_\heI}{m_\heI^2}\, \lt_{\heI\gea\geb\gec} .
  \label{EQ:MyVS}    \end{equation}
In this way we can compute around the tree-level vacuum; in the case that the tree-level expectation value is small or vanishing -- in the basis (\ref{EQ:HETBasis}) before any shifts -- this option would appear to be the most appropriate choice.
\item We can assume that the tadpole equation is satisfied at loop level (so that $t_\heI + \delta t_{\heI} =0$) after making shifts of the form (\ref{EQ:removesingletvevs}). In so doing, we can trade a different dimensionful parameter for each singlet tadpole equation, order by order in perturbation theory. 
  This is the standard approach in pole mass calculations, where the typical choice is to eliminate mass-squared parameters, but this is the most complicated from the EFT point of view because we want to fix the masses in order to perform the matching. The tree-level tadpole equations for the singlets in the basis \emph{before the shifts} (\ref{EQ:removesingletvevs}) read
\begin{align}
m_{\heI}^2 v_\heI = &\ - t_\heI   - \frac{1}{2} a_{\heI\heJ\heK}  v_\heJ v_\heK - \frac{1}{6} \lt_{\heI\heJ\heK\heL} v_\heJ v_\heK v_\heL 
\label{EQ:StandardVS}\end{align}
 and, for the typical choice of adjusting the diagonal terms $m_{\heI}^2$, the one-loop mass shift becomes
  \begin{align}
\delta m_{\heI\heJ}^2 \rightarrow \delta m_{\heI\heJ}^2 - \frac{\delta t_\heI}{v_\heI}\,\delta_{\heI\heJ},
\end{align}
where the tadpole $\delta t_\heI $ is computed at the minimum of the potential.
Note that if we have a case where $v_\heI =0$ for all $\heI$ then this approach reduces to option 2: we shall throughout assume when we refer to option 4 that the expectation values of all the singlets concerned are non-vanishing in the original basis.
\end{enumerate}
We shall henceforth assume that one of these choices has been made and the parameters adjusted accordingly; note that in the path integral method the choice made is implicitly our number $3$, since the tadpole equations are chosen to be satisfied at tree level only.

In section \ref{SEC:COUNTERTERMS} we shall explore alternatives, but persisting for now with the simplest possible approach -- which we shall in the following refer to as the ``perturbative masses'' approach -- we now split the fields \emph{at tree level} into (all of the) heavy ones with upper-case indices and light ones with lower-case indices. We then integrate out the heavy fields, to one-loop order and including only renormalisable operators, and obtain a new Lagrangian for the high-energy theory -- $\lagr_\text{eff} [\Phi]$ -- written entirely in terms of fields $\{\Phi_\lii\}$ that have counterparts in the low-energy theory:
\begin{align} 
\label{EQ:integrateout_heavyfields}
\lagr_\text{eff} [\Phi]=&  -\frac{1}{2} Z_{\lii\lij} \Phi_\lii  \partial_\mu \partial^\mu \Phi_\lij
+ \frac{1}{8  m_\heI^2} a_{\heI\lij\lik}(a_{\heI\lil\ligm} + 2 \delta a_{\heI\lil\ligm}) \Phi_\lij \Phi_\lik \Phi_\lil \Phi_\ligm  \nn\\
&  - \frac{\delta m_{\heI\heJ}^2 }{8 m_\heI^2 m_\heJ^2} a_{\heI\lij\lik} a_{\heJ\lil\ligm} \Phi_\lij \Phi_\lik \Phi_\lil \Phi_\ligm  - \frac{1}{24} (\lt_{\lii\lij\lik\lil} + \delta \tilde{\lambda}_{\lii\lij\lik\lil}) \Phi_\lii \Phi_\lij \Phi_\lik \Phi_\lil ~,
\end{align}
where we have omitted the mass terms and the trilinear couplings for the light fields as we assume that they all vanish in the limit $\Ml\to0$. In this approach, there is a tree-level shift of the quartic coupling of the theory from integrating out heavy fields in the presence of trilinears of the form $a_{\heI\lij\lik}$, see the fist term that multiplies $\Phi_\lij \Phi_\lik \Phi_\lil \Phi_\ligm$ in the equation above. In the path integral approach this is included in $S[\phi_L, \Phi^c_H[\phi_L]]$ of (\ref{EQ:BasicCDE}), while the one-loop corrections stemming from these terms appear  via the term $ X_{HL} \Delta^{-1}_{L} X_{LH} $ in (\ref{EQ:SEFT}). 

To complete the matching, we need to identify the above effective action with the equivalent expression computed in the low-energy theory, which means also rescaling the kinetic terms: we make the mapping
\begin{align}
  \Phi_\gea = U_{\gea\lii} \phi_\lii + U_{\gea\heI} \phi_\heI \equiv \delta_{\gea\lii}\phi_\lii + \delta U_{\gea\lii} \phi_\lii+\delta_{\gea\heI}\phi_\heI + \delta U_{\gea\heI} \phi_\heI,
  \label{defU}
\end{align}
where $\phi_\gea$ are now split into light $\{\phi_\lii\}$ and heavy $\{\phi_\heI\}$ fields, and we can throw away the heavy fields as they are already integrated out. 
In the two theories we have
\begin{align}
Z_{\gea\geb}^H = \delta_{\gea\geb} + \delta Z_{\gea\geb}^H, \qquad Z_{\gea\geb}^L = \delta_{\gea\geb} + \delta Z_{\gea\geb}^L
\end{align}
where the indices $H,L$ indicate whether they are computed in the high- or low-energy theory. It turns out, however, that there is more than one way to make this identification, depending on our choice of counterterms, and we will describe these choices in sections \ref{SEC:MATCHING} and \ref{SEC:COUNTERTERMS}. For now we need just give the general formula, expanded up to one-loop order, for the quartic term $\lambda_{\lii\lij\lik\lil} $ in the effective low-energy theory:
\begin{align}
\lambda_{\lii\lij\lik\lil} =&\ \lt_{\lii\lij\lik\lil} + \delta \lt_{\lii\lij\lik\lil} - \delta \lambda_{\lii\lij\lik\lil}\nn\\
 &+ \bigg[- \frac{1}{8 m_\heI^2}  a_{\heI\lii\lij} a_{\heI\lik\lil}- \frac{1}{4 m_\heI^2}  a_{\heI\lii\lij} \delta a_{\heJ\lik\lil}  + \frac{1}{8 m_\heI^2 m_\heJ^2} \delta m^2_{\heI\heJ}  a_{\heI\lii\lij} a_{\heJ\lik\lil}\nn\\
&\quad\,\,\,\  + \frac{1}{6} \delta U_{\heK\lii} (\lt_{\heK\lij\lik\lil}- \frac{3}{m^{2}_{\heI}}a_{\heI\heK\lij} a_{\heI\lik\lil} )  + \frac{1}{6} \delta U_{\liip\lii} (\lt_{\liip\lij\lik\lil}- \frac{3}{m^{2}_{\heI}} a_{\heI\liip\lij} a_{\heI\lik\lil} )  +  (\lii\lij\lik\lil)\bigg]. 
\label{EQ:GeneralMatching}
\end{align}
Here $\delta \lambda_{\lii\lij\lik\lil}$ denotes the corrections to the light quartic \emph{in the low-energy theory}, so just consisting of light degrees of freedom (if we use the approach of equation (\ref{EQ:SEFT}) then $\delta \lambda_{\lii\lij\lik\lil} = 0$). $(\lii\lij\lik\lil)$ stands for all $24$ permutations of the indices $\{\lii,\lij,\lik,\lil\}$, counting even the cases that the indices are identical -- hence for one light field the matching would be
$$
\lambda_{1111} = \lt_{1111} - \frac{3}{m_\heI^2 } a_{\heI11}^2 + ...
  $$
  We give results for all of the relevant generic expressions for one-loop corrections to the effective action  in appendix \ref{APP:THRESHOLDS} (see also e.g.~\cite{Camargo-Molina:2016moz}). In the next section we shall discuss the cancellation of infra-red divergences and derive an expression for the matrix $U$.

\section{Mixing and Matching}
\label{SEC:MATCHING}

In this section we shall discuss the effects of infra-red safety and gauge dependence of the matching, and also derive the matrices $\delta U$ that encode the effects of mixing of the light and heavy degrees of freedom, employing the ``perturbative masses'' approach; in section \ref{SEC:COUNTERTERMS} we shall show an alternative.

\subsection{Infra-red safety}
\label{sec:IRsafety}

If we compute the shifts with small or vanishing masses for the ``light'' fields, then the corrections $\delta \lambda$ will contain large/divergent logarithms of the form $\log \frac{m_\lii^2}{M^2}$, $m_\lii$ being the light masses and $M$ the mass scale of heavy states, at which the matching is performed. Clearly these should cancel against the corresponding corrections in the high-energy theory, so that the resulting shift is infra-red finite. In the case that the theory contains no couplings of the form $a_{\heI\lii\lij}$ or $a_{\lii\lij\lik}$ the infra-red divergent corrections in $\delta \tilde{\lambda}$ are identical to those in $\delta \lambda$ and so the subtraction is straightforward. On the other hand, once we allow for these other types of coupling the cancellation of infra-red divergences becomes more subtle.

All-light trilinear scalar couplings $a_{\lii\lij\lik}$ are forbidden in the SM by the gauge symmetries, so in order to have such a coupling the low-energy theory would need additional scalars, but as we described in section \ref{SEC:GENERAL} we must forbid couplings $a_{\lii\lij\lik}$ in any model. However, in the presence of couplings $a_{\heI\lii\lij}$ (which, for the low-energy theory being the SM, means the high-energy theory contains either heavy singlets or triplets)  we generate a difference between $\tilde{\lambda}$ and $\lambda$ at tree level. This means that subtracting the low-energy $\delta \lambda_{\lii\lij\lik\lil}$ from the high-energy $ \delta \lt_{\lii\lij\lik\lil}$ is not entirely trivial, as we shall see below. The low-energy amplitude $\delta \lambda_{\lii\lij\lik\lil}$ coming from scalar loops is given by
\begin{align}
\delta \lambda_{\lii\lij\lik\lil} \supset& \ \frac{\kappa}{16} \lambda_{\lii\lij\lix\liy} \lambda_{\lik\lil\lix\liy} P_{SS} (m_\lix^2, m_\liy^2) + (\lii\lij\lik\lil)\,,
\end{align}
where the sum over $\lix, \liy$ is over all light scalars. $\kappa$ is a loop factor defined in eq.~(\ref{EQ:loopfactor}). $P_{SS}$ is defined with our other loop functions in appendix \ref{APP:CONVENTIONS}; as we take $\Ml\rightarrow 0$  it diverges. We will not write here the contributions from fermions, because there is no tree-level shift to the Yukawa couplings; we show in appendices \ref{app:yukawas} and \ref{APP:noIRdiv_fermions} that the cancellation of infra-red divergences in the fermionic contributions to Yukawa and quartic couplings is straightforward. 

In the high-energy theory, there will be an identical contribution to $\delta \tilde{\lambda}_{\lii\lij\lik\lil}$, but 
\begin{align}
  \lambda_{\lii\lij\lik\lil} = \tilde{\lambda}_{\lii\lij\lik\lil} - \bigg( \frac{1}{8m_\heI^2} a_{\heI\lii\lij} a_{\heI\lik\lil} + (\lii\lij\lik\lil)\bigg) + \Ooneloop\label{EQ:lightlambda}
\end{align} 
and on the other hand the corrections $\delta a_{\heI\lii\lij}, \delta m_{\heK\heL}^2$ both contain additional infra-red divergent pieces. Clearly these divergences must cancel, and after a little tedious algebra (which we present in appendix \ref{APP:CANCELIR}) it can be shown that indeed they do. This then motivates using infra-red safe loop functions $\ov{P}_{SS}(x,y)$, $\ov{C}_0(x,y,z)$, $\ov{D}_0 (x,y,z,u)$ given in the appendix, which can be defined in terms of one of:
\begin{itemize}
\item Subtracting an infra-red divergent piece and taking the limit as $\Ml \rightarrow 0$, e.g.
  $$ \ov{P}_{SS} (0,0) \equiv \lim_{x\rightarrow 0} \bigg[ P_{SS} (x,x) - \log \frac{x}{M^2} \bigg]= 0\,.$$
\item Taking the loop integral to only be over the ``hard'' momenta, as described in equation (\ref{EQ:SEFT}).
\item Regularising the infra-red divergences using dimensional regularisation and discarding the divergent terms $\propto \frac{1}{\epsilon_{IR}}$. 
\end{itemize}
We shall then write the infra-red safe shifts as $\ov{\delta} \lt_{\lii\lij\lik\lil}$, $\ov{\delta} a_{\heI\lii\lij}$, $\ov{\delta} m_{\heI\heJ}^2$ etc. 
All three definitions above do not necessarily give the same result: there is some potential ambiguity about the first method, because we can always add a constant piece to the subtraction term. However, once we subtract the contribution from amplitudes containing purely light fields, such as the term ${\delta} \lambda_{\lii\lij\lik\lil}$ in eq.~(\ref{EQ:GeneralMatching}), then the difference is unambiguous. It is then convenient to take $\ov{\delta} \lambda_{\lii\lij\lik\lil} =0,$ which is indeed the result in dimensional regularisation, but may be confusing to some readers.

Finally, we shall see in the next section that we must compute $\delta m^2_{\heI\lii}$ and $\delta Z_{\lii\lij}$, which in principle could contain infra-red divergences. However, a divergence that is not trivially equal to the same contribution in the low-energy theory could only appear from a scalar loop, and the absence of the offending terms at one loop is guaranteed by forbidding couplings of the form $a_{\lii\lij\lik}$. Hence we need make no distinction between $\delta m^2_{\heI\lii} $ and $\db m^2_{\heI\lii}$, etc.

\subsection{Mixing}
\label{sec:mixing}

Here we shall derive the most obvious choice for the matrix $\delta U$. 
 Noting that the fields in the high-energy theory have kinetic terms $ \frac{1}{2} (1 + \delta Z^H)_{\gea\geb} \partial_\mu \Phi_\gea \partial^\mu \Phi_\geb$, and in the low-energy theory $ \frac{1}{2} (1 + \delta Z^L)_{\lii\lij} \partial_\mu \phi_\lii \partial^\mu \phi_\lij $, we can make the identification
\begin{align}
\Phi ~=~ (1 + \delta Z^H)^{-1/2} R \,(1+ \delta Z^L)^{1/2}\phi ~\equiv~ U\,\phi~.
\label{EQ:heavylightrelation}\end{align}
 To ensure that the transformation $U$ connecting the high-energy and low-energy fields is invertible, we retain the ``dummy'' heavy fields $\{\phi_\heI\}$ in the low-energy set, see eq.~(\ref{defU}), and define $(\delta Z^L)_{\lii\heI} = (\delta Z^L)_{\heI\lii} = (\delta Z^L)_{\heI\heJ} = 0$ (note that the fields $\{\phi_\heI\}$ can eventually be disregarded, as the general matching condition for $\lambda_{\lii\lij\lik\lil}$ in eq.~(\ref{EQ:GeneralMatching}) depends only on $U_{\heI\lii}$ and $U_{\lii\lij}$). Here $R = 1 + \delta R$ is a unitary rotation, which we are free to introduce as it leaves the kinetic terms unchanged. Taking the masses of the heavy fields to be diagonal at tree level and expanding this just to one-loop order we obtain
\begin{align}
\label{EQ:mixing_pertapproach}
\delta U =& - \frac{1}{2} \Delta Z + \delta R, \qquad \Delta Z \equiv \delta Z^H - \delta Z^L.
\end{align}
With this transformation the kinetic terms will have the correct normalisation, but we must also choose $\delta R$ to eliminate the mass-mixing between heavy and light states: assuming that we have diagonalised the masses at tree level (in the end, we only require that we diagonalise the heavy masses and remove light-heavy mixing) we have
\begin{align}
0 &= (- \frac{1}{2} \Delta Z_{\lii\heI} + \delta R_{\lii\heI}) m_\lii^2 +  (- \frac{1}{2} \Delta Z_{\heI\lii} + \delta R_{\heI\lii})m_\heI^2 + \delta m^2_{\heI\lii} 
\end{align}
where $\delta m^2_{\heI\lii} = \Pi_{\heI\lii}(0)$, 
which, for vanishing light-scalar masses, leads to 
\begin{align}
 \delta R_{\lii\heI} = - \delta R_{\heI\lii} & = \frac{\delta m^2_{\lii\heI}}{m_\heI^2} - \frac{1}{2} \Delta Z_{\lii\heI} 
\end{align}
and so
\begin{align}
U_{\lii\heI}= \frac{\delta m^2_{\lii\heI}}{m_\heI^2}- \Delta Z_{\lii\heI}, \qquad U_{\heI\lii} = -\frac{\delta m^2_{\heI\lii}}{m_\heI^2}.
\label{EQ:NaiveUIj}\end{align}
On the other hand, we have the freedom whether or not to diagonalise the \emph{mass} terms of the low-energy theory. We can set $\delta R_{\lii\lij} =0$ so that
\begin{align}
U_{\lii\lij} &= \delta_{\lii\lij} -  \frac{1}{2} \Delta Z_{\lii\lij}, 
\label{EQ:SimpleUij}
\end{align}
and we will then have mass terms for the light fields of
\begin{align}
\mathcal{L}_\text{EFT} \supset - \frac{1}{2} ( m_{\lii\lij}^2 +  \ov{\delta} m_{\lii\lij}^2  -  m_{\lii\lik}^2 \Delta Z_{\lik\lij} ) \phi_\lii \phi_\lij ,
\end{align}
where we allow now for non-diagonal masses for the light fields at tree level. 
Since we are neglecting all terms of order \Ml\ in our calculations this is not a problem: it may be more desirable to calculate all these terms from the high-energy theory and then diagonalise the light fields only \emph{after} electroweak symmetry breaking. On the other hand, if we want to diagonalise our light fields at zero expectation value for the Higgs field then we require an extra rotation component in $\delta U_{\lii\lij}$: we would have
$$
\delta U_{\lii\lij} \rightarrow -\frac{\delta m_{\lii\lij}^2}{m_\lii^2 - m_\lij^2} + \frac{m_\lij^2 \Delta Z_{\lii\lij}}{m_\lii^2 - m_\lij^2}. 
$$
Note that at two loops we would necessarily take the tree-level ``light'' masses to be diagonal and of one-loop order, but it is still not necessary to perform this extra diagonalisation before electroweak symmetry breaking. Alternatively, we can add finite counterterms for these masses to ensure that they are zero -- and then we can simply use (\ref{EQ:SimpleUij}) again.

Up to one loop this gives
\begin{empheq}[box=\fbox]{align}
\label{EQ:MatchingPerturbative}
\lambda_{\lii\lij\lik\lil}  =&\,\lt_{\lii\lij\lik\lil} + \ov{\delta} \lt_{\lii\lij\lik\lil}  \nn\\
& + \bigg[- \frac{1}{8m^{2}_{\heI}} a_{\heI\lii\lij} a_{\heI\lik\lil}- \frac{1}{4m^{2}_{\heI}}  a_{\heI\lii\lij}\ov{\delta} a_{\heI\lik\lil}  + \frac{1}{8 m_\heI^2} \ov{\delta} m^2_{\heI\heJ} \frac{1}{m_\heJ^2} a_{\heI\lii\lij} a_{\heJ\lik\lil}\nn\\
& \quad\,\ \ - \frac{1}{6} \frac{\ov{\delta} m^2_{\heI\lii}}{m_\heI^2} \big(\lt_{\heI\lij\lik\lil}- \frac{3}{m^{2}_{\heJ}}a_{\heI\heJ\lij} a_{\heJ\lik\lil} \big) \nn\\
& \quad\,\ \ - \frac{1}{12} \Delta Z_{\lii\liip} \big(\lt_{\liip\lij\lik\lil}- \frac{3}{m^{2}_{\heI}}a_{\heI\liip\lij} a_{\heI\lik\lil} \big)  +  (\lii\lij\lik\lil)\bigg]. 
\end{empheq}
The term on the third line generalises the shift in rotation angle in Two-Higgs-doublet models observed e.g.~in \cite{Bagnaschi:2014rsa,Wells:2017vla,Bahl:2018jom}. Complete expressions for the different terms in the above equation are given in appendix \ref{APP:THRESHOLDS}.

\subsection{Gauge dependence}
\label{sec:gaugedep}

\begin{figure}[t]\centering
  \begin{tabular}{ccc}
\includegraphics[height=0.15\textheight]{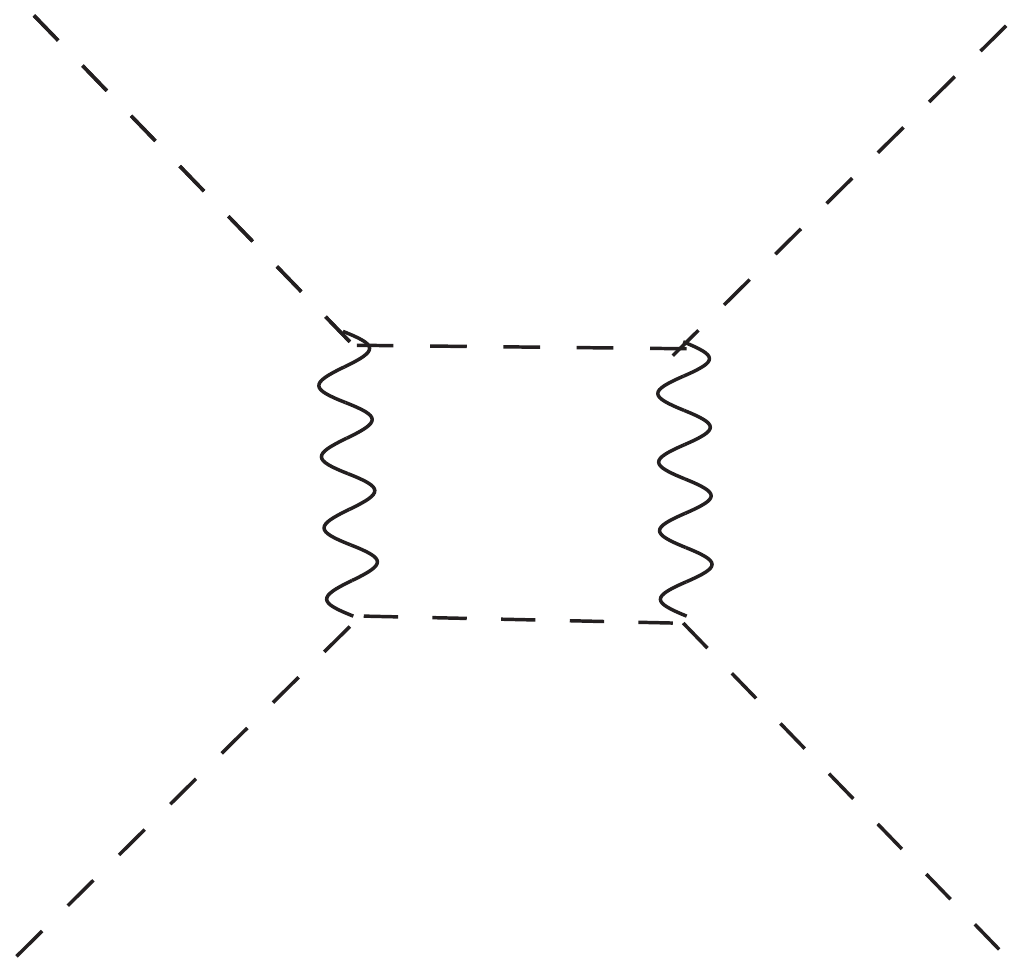} &  \includegraphics[height=0.15\textheight]{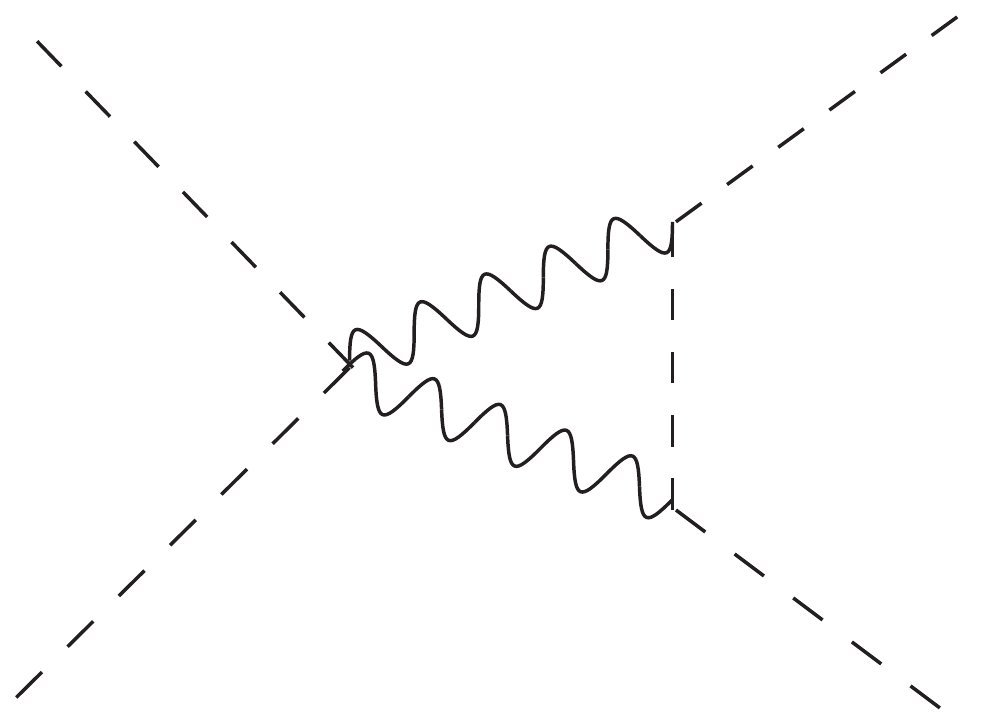}  &  \includegraphics[height=0.15\textheight]{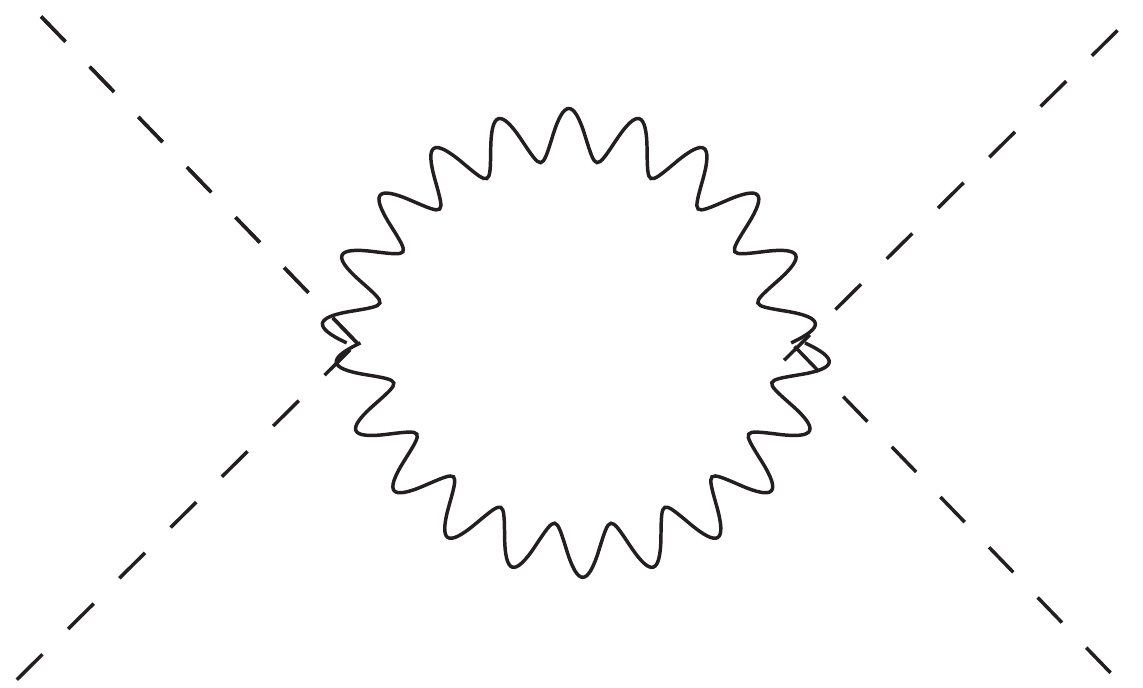}
  \end{tabular}
  \begin{tabular}{cc}
 \includegraphics[height=0.15\textheight]{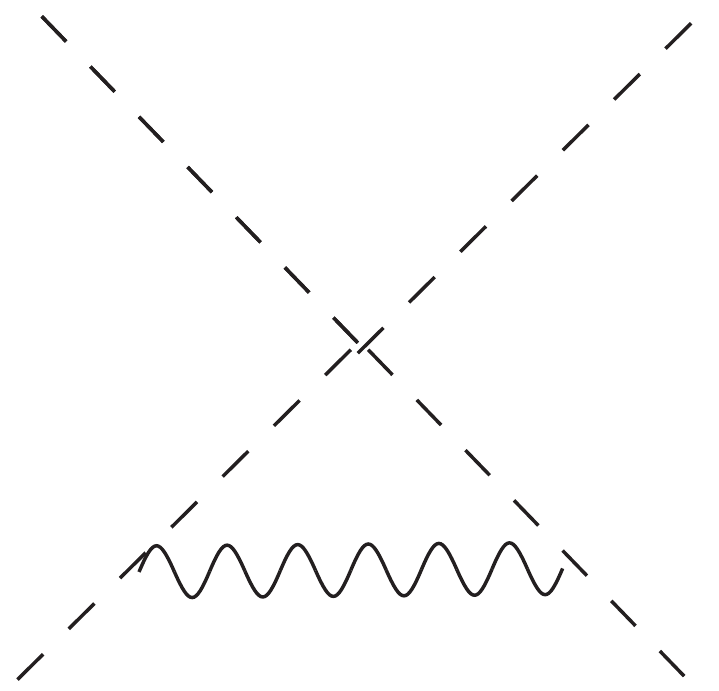}  &   \includegraphics[height=0.15\textheight]{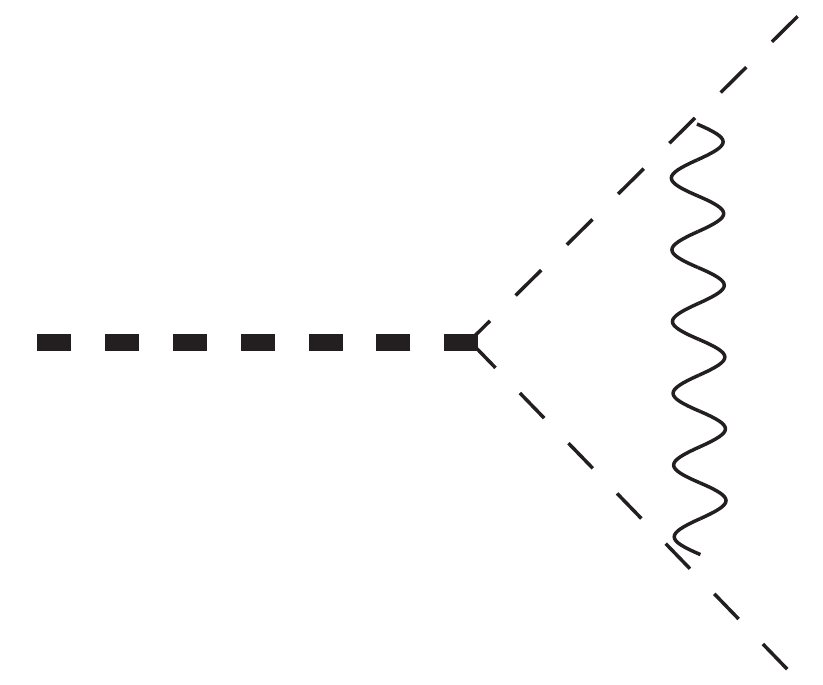} 
  \end{tabular}
  \caption{Gauge-dependent diagrams with only light scalars/gauge bosons in the loops, which must contribute zero after infra-red regulation; heavy fields are denoted with a thick line.}
  \label{FIG:LightGauge}
\end{figure}

\begin{figure}[t]\centering
  \begin{tabular}{ccc}
    \includegraphics[width=0.25\textwidth]{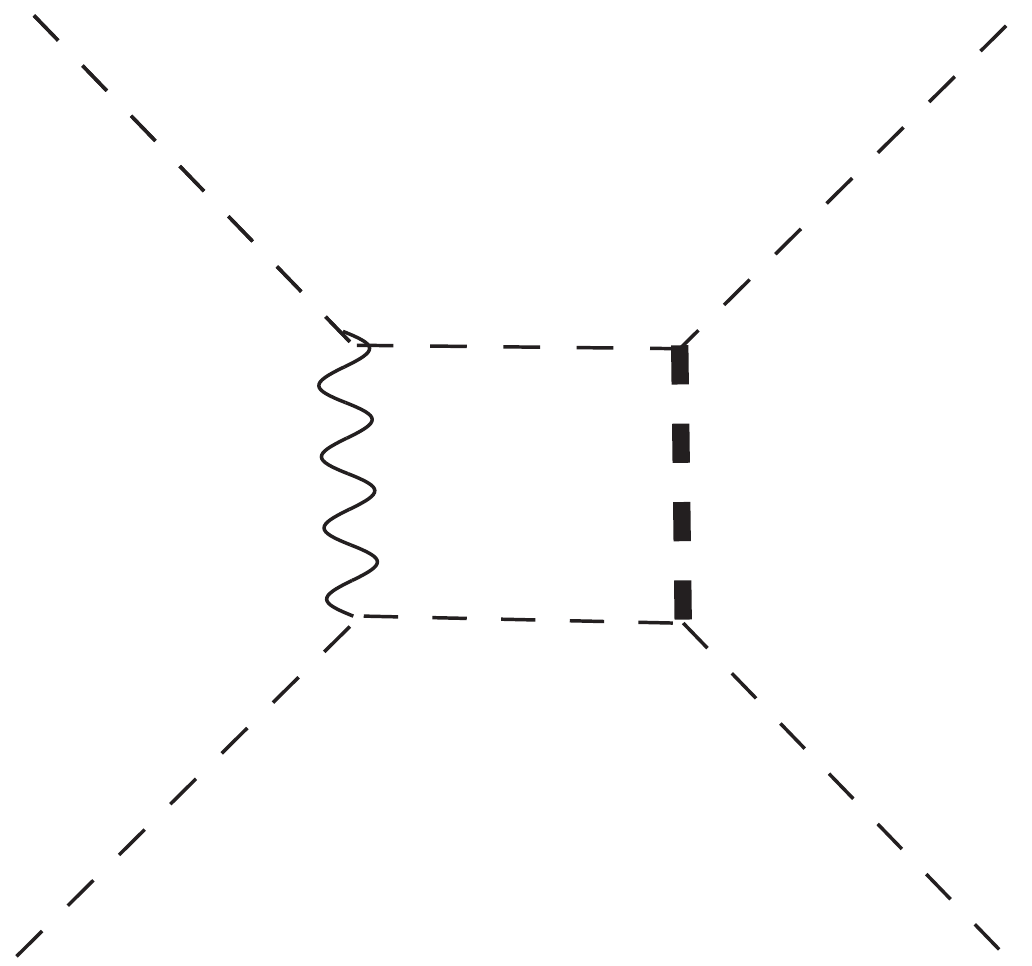} &   \includegraphics[width=0.25\textwidth]{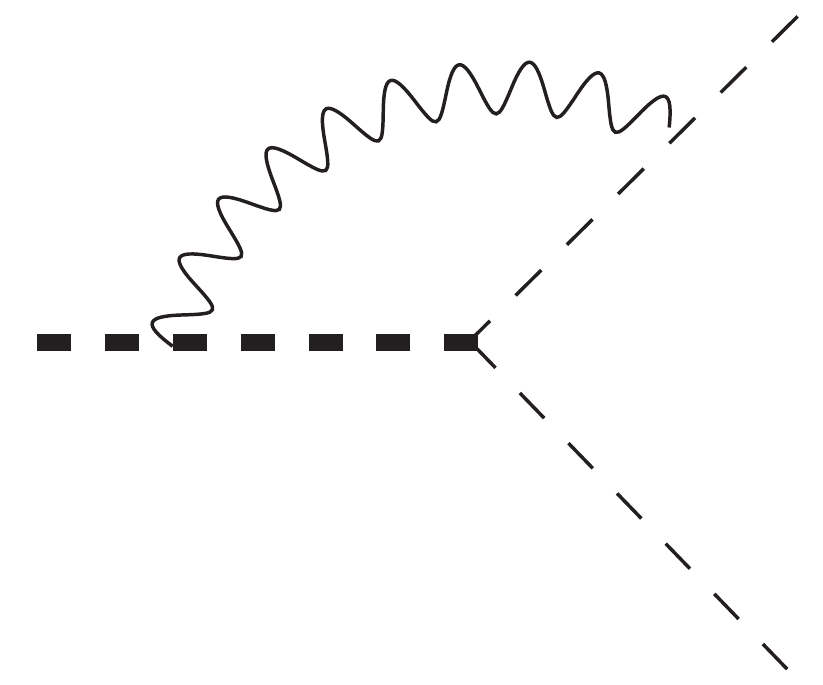} & \raisebox{0.5\height}{ \includegraphics[width=0.25\textwidth]{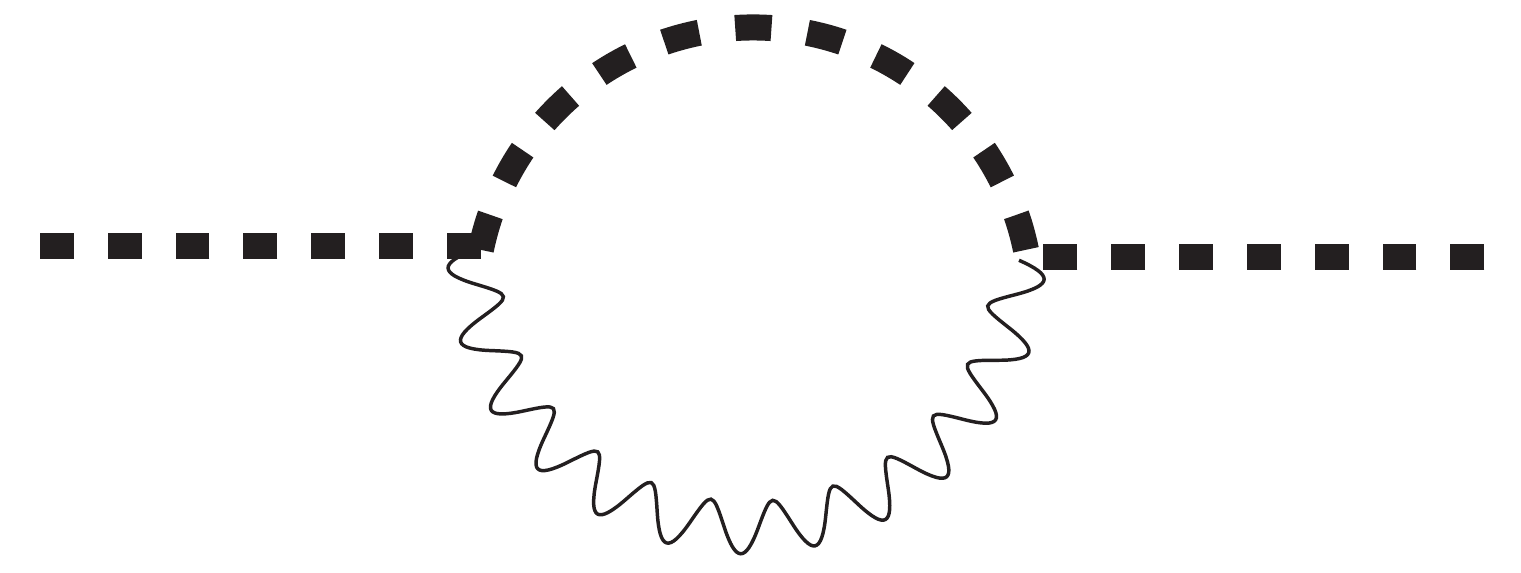}}
  \end{tabular}
  \caption{Gauge-dependent diagrams with one heavy scalar in the loop; heavy fields are denoted with a thick line.}
  \label{FIG:HeavyGauge}
\end{figure}

Since we take all gauge groups to be unbroken in the limit $\Ml \rightarrow 0$, we may expect that gauge couplings ought to induce no net contribution to $\lambda_{\lii\lij\lik\lil}$. Indeed, if there are no trilinear couplings in the theory, then this is immediately obvious: the gauge contributions to $\delta \lt_{\lii\lij\lik\lil}$ and $\delta \lambda_{\lii\lij\lik\lil}$ are identical in this case, because the unbroken gauge interactions cannot mix heavy and light fields and certainly the corrections of quartic order in the gauge couplings -- i.e. the first row of diagrams in figure \ref{FIG:LightGauge} -- must always be equal.\footnote{On the other hand, there \emph{is} a difference if we compute the corrections in different schemes; if we match a theory in the $\drbar $ scheme onto a theory in the $\msbar$ one then there is a shift to the quartic couplings of quartic order in the gauge couplings, see e.g.~\cite{Martin:1993yx,Summ:2018oko} for general formulae.} For corrections of quadratic order in the gauge couplings, the second row of diagrams in figure \ref{FIG:LightGauge} all contain only massless/light fields in the loops, and so we expect them not to contribute. However,  once we include trilinear couplings, there are diagrams such as those given in figure \ref{FIG:HeavyGauge} which are individually non-zero after infra-red regulation, and so it is possible that there could be some residual dependence on the gauge couplings. However, this cancels out, as we show below.

The individual infra-red safe contributions (it is straightforward to show that the infra-red divergences cancel) are
\begin{align}
\kappa^{-1}  \ov{\delta}_{g^2} \lt_{\lii\lij\lik\lil} =& - \frac{1}{2}\xi g^{a\lii\ligm} g^{a\lij\lir} a_{\ligm\heI\lik} a_{\heI\lir\lil} \ov{C}_0 (m_\ligm^2,m_\heI^2,m_\lir^2) + (\lii\lij\lik\lil) = - \frac{1}{2}\xi g^{a\lii\ligm} g^{a\lij\lir} a_{\ligm\heI\lik} a_{\heI\lir\lil} \frac{A_0 (m_\heI^2) }{m_\heI^4} + (\lii\lij\lik\lil)\nn\\
  =& \,-\frac{1}{8} \xi g^2 C_2 (\heI) a_{\heI\lik\lil} a_{\heI\lii\lij}  \frac{A_0 (m_\heI^2) }{m_\heI^4} + (\lii\lij\lik\lil),  \nn\\
 \kappa^{-1} \ov{\delta}_{g^2} a_{\heI\lii\lij} =&  -\xi g^{a\heI\heM} g^{a\lij\lin} a_{\heM \lin\lii} \ov{P}_{SS} (m_\heM^2, m_\lin^2)- \xi g^{a\heI\heM} g^{a\lii\lin} a_{\heM \lin\lij} \ov{P}_{SS} (m_\heM^2, m_\lin^2) \nn\\
=& \,\xi g^{a\heI\heM} g^{a\heM\heN} a_{\heN\lii\lij} \frac{A_0 (m_\heM^2) }{m_\heM^2} = -\xi g^2 C_2 (\heI) a_{\heI\lii\lij} \frac{A_0 (m_\heI^2) }{m_\heI^2},   \nn\\
\kappa^{-1}  \ov{\delta}_{g^2} m^2_{\heI\heJ} =& - \xi g^{a\heI\heM} g^{a\heJ\heM} A_0 (m_\heM^2) =  -\xi \delta_{\heI\heJ} g^2 C_2 (\heI) A_0 (m_\heI^2),
\end{align}
where $g^2$ is the relevant gauge coupling and $C_2(\heI)$ is the quadratic Casimir of the corresponding representation of heavy field $\heI$. For each term we have used gauge invariance to simplify the expressions. We see that all of these contributions are proportional to the gauge-fixing parameter $\xi$, which tells us that the total contribution must vanish; this would not have been obvious if we had worked in the Feynman gauge (but of course would be in the Landau gauge!).  Indeed, combining the above contributions as in eq.~(\ref{EQ:MatchingPerturbative}) gives
\begin{align}
 \ov{\delta}_{g^2} \lt_{\lii\lij\lik\lil} - \bigg[  \frac{1}{4 m_\heI^2} a_{\heI\lik\lil} \ov{\delta}_{g^2} a_{\heI\lii\lij} -  \frac{1}{8 m_\heI^2 m_\heJ^2} a_{\heI\lik\lil} a_{\heJ\lii\lij} \ov{\delta}_{g^2} m_{\heI\heJ}^2 + (\lii\lij\lik\lil) \bigg]=& \ 0\,.
\end{align}
Hence we can indeed neglect gauge contributions at one loop, as there is no gauge contribution to $\delta m_{\lii\heI}^2$ and $\delta_{g^2} Z^H = \delta_{g^2} Z^L$. However, it is important to note that we require all of the separate pieces together in order to cancel the gauge dependence, which will be relevant in section \ref{SEC:COUNTERTERMS}.

\subsection{Trilinear couplings and higher-dimensional operators}
\label{SEC:HIGHERDIM}

We end this section by considering the case of non-vanishing trilinear couplings between light states, i.e.~$a_{\lii\lij\lik}\neq0$. Such couplings result in new divergent diagrams compared to the case considered in appendix \ref{APP:CANCELIR} (where we demonstrate the cancellation of all IR divergences when $a_{\lii\lij\lik}=0$). Indeed, considering the different contributions appearing in equation (\ref{EQ:MatchingPerturbative}), one can observe that several divergent terms in the high-energy part of the matching do not seem to cancel out with any term in the low-energy part, namely
\begin{align}
\label{EQ:extra_div}
 \kappa\,\bigg\{&-\frac{1}{2}\lt_{\lii\lij\lix\heY}a_{\lik\lix\liz}a_{\lil\heY\liz}\frac{P_{SS}(m^2_\lix,m^2_\liz)}{m_{\heY}^2}+\frac{1}{2}a_{\lii\heX\heY}a_{\lij\heX\liz}a_{\lik\heY\liu}a_{\lil\liz\liu}\frac{P_{SS}(m^2_\liz,m^2_\liu)}{m_{\heX}^2m_{\heY}^2}\nn\\
 &-\frac{1}{4}a_{\heI\lii\lij}\lt_{\heI\lik\lix\liy}a_{\lil\lix\liy}\frac{P_{SS}(m_\lix^2,m_\liy^2)}{m_{\heI}^2}+\frac{1}{2}a_{\heI\lii\lij}a_{\heI\heX\liy}a_{\lik\heX\liz}a_{\lil\liy\liz}\frac{P_{SS}(m_\liy^2,m_\liz^2)}{m_{\heI}^2m_{\heX}^2}\nn\\
 &-\frac{1}{12}a_{\heI\lix\liy}a_{\lii\lix\liy}\bigg[\lt_{\heI\lij\lik\lil}-\frac{3}{m_{\heJ}^2}a_{\heI\heJ\lij}a_{\heJ\lik\lil}\bigg]\frac{P_{SS}(m_\lix^2,m_\liy^2)}{m_{\heI}^2}\nn\\
 &-\frac{1}{2}a_{\lii\heX\liy}a_{\lij\heX\liz}a_{\lik\liy\liu}a_{\lil\liz\liu}\frac{P_{SS}(m_\liz^2,m_\liu^2)}{m_{\heX}^4}+(\lii\lij\lik\lil)\bigg\}\,.
\end{align}
Note that all these terms involve one, or two, trilinear couplings between light scalars. Moreover, one may observe that these remaining terms are all proportional to a $P_{SS}$ loop function -- in some cases this $P_{SS}$ being obtained from the expansion of a $C_0$ or $D_0$ function -- while it can be shown that divergent terms with $C_0(m_x^2,m_y^2,m_z^2)$ or $D_0(m_x^2,m_y^2,m_z^2,m_u^2)$ -- with all masses being light -- do cancel out. 

If we reason with orders of magnitude, it is natural to assume that couplings $a_{\heI\lij\lik}$ and $a_{\heI\heJ\lik}$ are of the order of a heavy mass, say $M$, times numerical factors of $\mathcal{O}(1)$. From this we can easily see that all of the above terms are of order $a_{\lii\lij\lik}/M$ (and even $(a_{\lii\lij\lik}/M)^2$ for the last one). As we could expect $a_{\lii\lij\lik}$ to be of the order of a light mass (e.g.~$m_{\lii}\sim\Ml$), it would seem natural that the above terms be suppressed at least as $\mathcal{O}(\Ml/M)$ -- and therefore also go to zero in the limit $\Ml\to0$. The finite part of the matching is then exactly the same as that obtained previously. 

However, one can still want to understand what happens if the trilinear couplings between light states are not of the order of $\Ml$. Having very large trilinear couplings in the low-energy theory could potentially cause a breakdown of perturbativity and/or unitarity, as well as expectation values in the low-energy theory of the order of the heavy masses. Nevertheless, it is actually still possible in such a case to cancel all of the IR divergences, by taking into account higher-dimensional operators. 

More specifically, one can deduce from the form of the divergent terms in equation~(\ref{EQ:extra_div}) that the required new operators are a dimension-5 operator $c_5^{\lii\lij\lik\lil\ligm}\phi_\lii\phi_\lij\phi_\lik\phi_\lil\phi_\ligm$ and a dimension-6 operator $k_6^{\lii\lij\lik\lil}\phi_\lii\phi_\lij\partial_\mu\phi_\lik\partial^\mu\phi_\lil$ (a correction to the kinetic term of the scalars). The former will cancel out with the first three lines of eq.~(\ref{EQ:extra_div}), while the latter compensates the last remaining term. 

\subsubsection{Higher-dimensional operators in a toy model}
\label{sec:Higherd}

\begin{figure}
 \includegraphics[width=\textwidth]{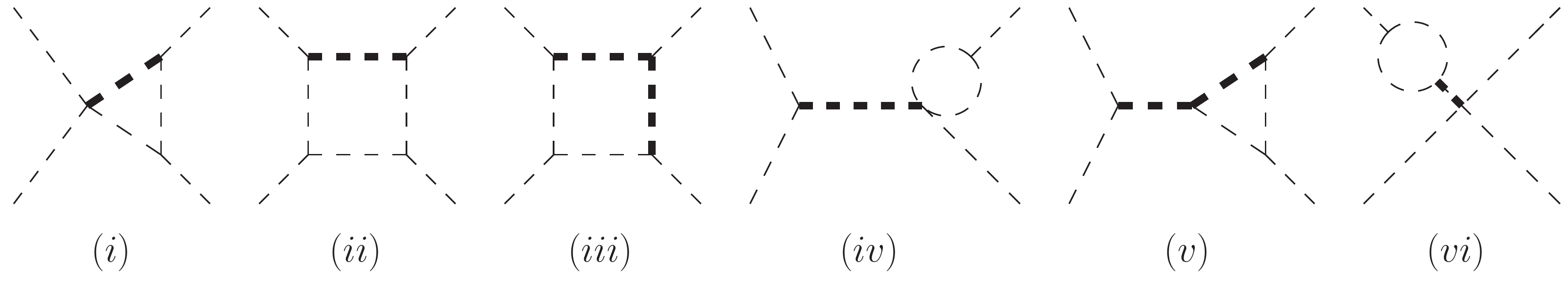}
 \caption{Types of diagrams that contribute to the high-energy part of the matching of quartic couplings and that contain divergences only regulated when including higher-dimensional operators in the low-energy theory. Bold lines denote heavy propagators. }
 \label{FIG:noncancelleddiags}
\end{figure}

To illustrate how to address the additional terms of eq.~(\ref{EQ:extra_div}), it will prove useful to first consider a simple toy example, with only two scalars -- one light $L$ and one heavy $H$ -- in the high-energy theory. The Lagrangian of such a model reads
\begin{align}
 \mathcal{L}_\text{HET}\supset&-\frac12 m_L^2L^2-\frac12 m_H^2H^2-\frac16a_{LLL}L^3-\frac12a_{LLH}L^2H-\frac12a_{LHH}LH^2-\frac16a_{HHH}H^3\nn\\
&-\frac{1}{24}\tilde\lambda_{LLLL}L^4-\frac16\lt_{LLLH}L^3H-\frac{1}{4}\tilde\lambda_{LLHH}L^2H^2-\frac16\lt_{LHHH}LH^3-\frac{1}{24}\tilde\lambda_{HHHH}H^4
\end{align}
Performing the one-loop matching of the quartic coupling $\lambda^{LLLL}$ as previously, we obtain 
\begin{align}
 \lambda_{LLLL}+\delta\lambda_{LLLL}=&\,\lt_{LLLL}-\frac{3}{m_H^2}(a_{LLH})^2+\delta\lt_{LLLL}-\frac{6a_{LLH}}{m_H^2}\delta a_{LLH}+\frac{3(a_{LLH})^2}{m_H^4}\delta m_{HH}^2\,.
\end{align}
The IR-divergent terms left on the right-hand side of the matching are then
\begin{equation}
\label{EQ:divterms_toymodel}
 -\kappa\bigg[20\frac{a_{LLH}a_{LLL}\lt_{LLLH}}{m_H^2}-30\frac{a_{LHH}(a_{LLH})^2a_{LLL}}{m_H^4}+12\frac{(a_{LLH})^2(a_{LLL})^2}{m_H^4}\bigg]P_{SS}(m_L^2,m_L^2)\,,
\end{equation}
and correspond to the types of diagrams shown in figure~\ref{FIG:noncancelleddiags} (recall that $\kappa$ is the loop factor). One can expect the divergent terms in equation~(\ref{EQ:divterms_toymodel}) to correspond to diagrams in the EFT similar to those in figure~\ref{FIG:noncancelleddiags}, but with the dashed lines corresponding to heavy propagators contracted to points. In particular, the low-energy diagrams corresponding to diagrams $(i)$ and $(iii)-(vi)$ in figure~\ref{FIG:noncancelleddiags} will involve a coupling between five scalars, while the diagram corresponding to $(ii)$ will involve a dimension-6 coupling between four scalars, suppressed by $m_H^4$. We define the corresponding operators as 
\begin{equation}
 \mathcal{L}_\text{EFT}\supset-\frac{1}{5!}c_5^{LLLLL}L^5-\frac14k_6^{LLLL}L^2(\partial_\mu L)^2
\end{equation}

\begin{figure}
\centering
 \includegraphics[width=.9\textwidth]{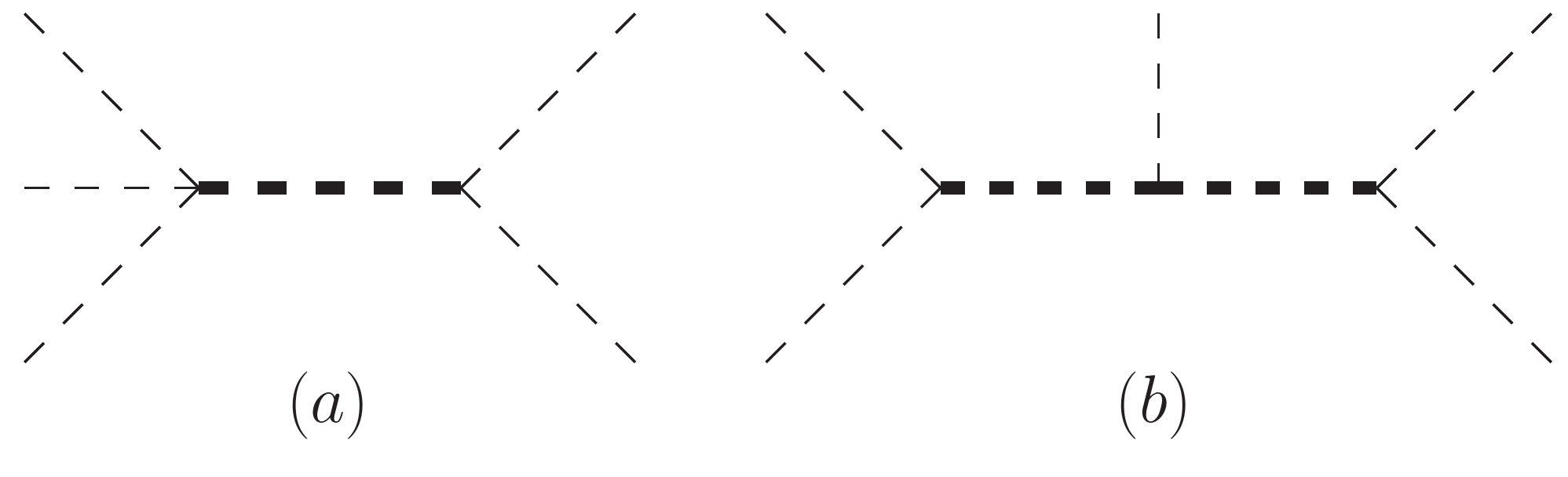}
 \caption{Diagrams in the high-energy theory that contribute to the tree-level expression of $c_5^{LLLLL}$ (or in general of $c_5^{\lii\lij\lik\lil\ligm}$). Bold lines denote heavy propagators. }
 \label{FIG:dim5op_diags}
 
 \includegraphics[width=.5\textwidth]{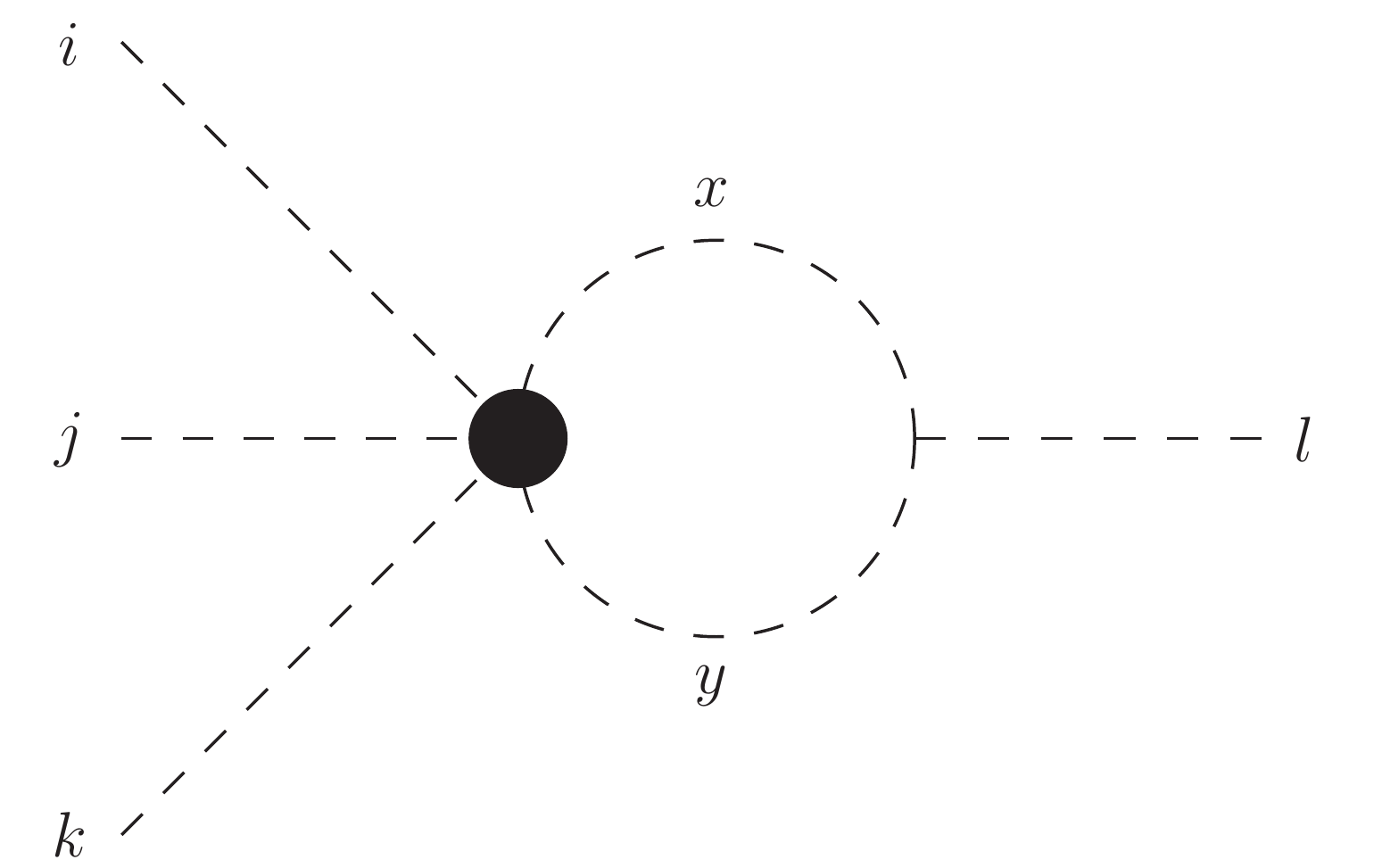}
 \caption{New diagram in the low-energy theory part of the matching of the scalar quartic coupling, involving a dimension-5 scalar operator. }
 \label{FIG:dim5}
 
 \includegraphics[width=.35\textwidth]{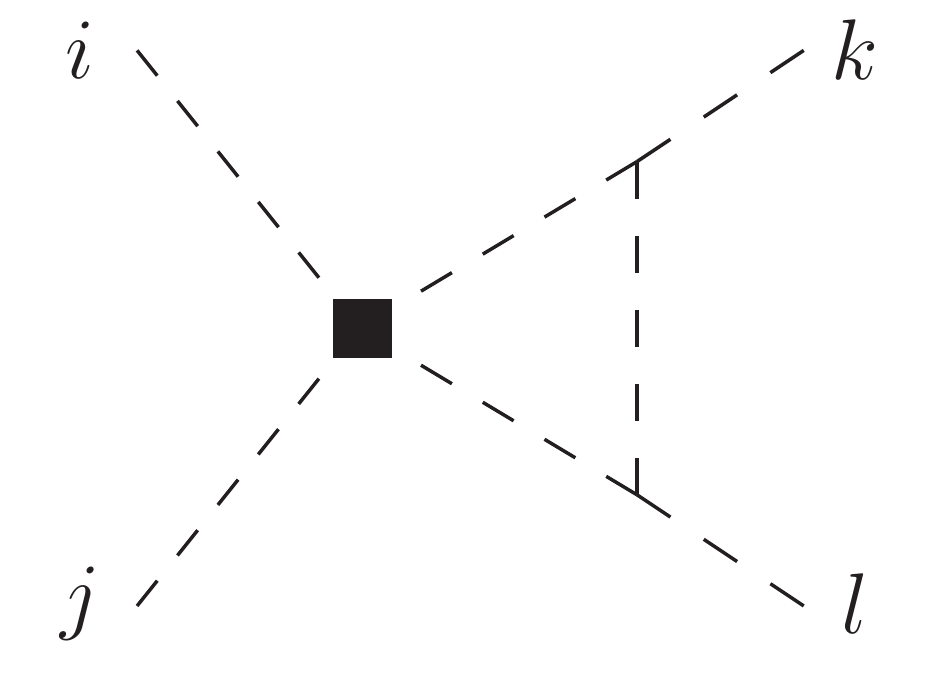}
 \caption{New diagram involving a dimension-6 operator -- denoted by a black square -- in the low-energy side of the one-loop matching.}
 \label{dim6diag}
\end{figure}

Before deriving the expressions of these two higher-dimensional operators, it is important to note that they will only appear in one-loop diagrams in the low-energy part of the matching and therefore it will suffice for the discussion at hand here to obtain their tree-level expressions. For the dimension-5 coupling, two different sorts of diagrams contribute to its tree-level expression, as shown in figure~\ref{FIG:dim5op_diags}. We find 
\begin{equation}
\label{dim5toymodel}
 c_5^{LLLLL}=-10\frac{a_{LLH}\lt_{LLLH}}{m_H^2}+15\frac{a_{LHH}(a_{LLH})^2}{m_H^4}\,.
\end{equation}

The dimension-6 operator is obtained from similar diagrams as the tree-level threshold corrections to $\lambda_{LLLL}$, but taking the second order in the $p^2/m_H^2$ expansion of the heavy propagator. Finally, we find
\begin{equation}
\label{dim6toymodel}
 k_6^{LLLL}=-\frac{2}{m_H^4}(a_{LLH})^2\,.
\end{equation}

One can then compute the new contributions to $\delta\lambda_{LLLL}$ arising from diagrams involving $c_5^{LLLLL}$ and $k_6^{LLLL}$ -- shown respectively in figures~\ref{FIG:dim5} and \ref{dim6diag} -- and one finds 
\begin{align}
 \kappa^{-1}\delta\lambda_{LLLL}\supset 2c_5^{LLLLL}a_{LLL}P_{SS}(m_L^2,m_L^2)+6k_6^{LLLL}(a_{LLL})^2\big[P_{SS}(m_L^2,m_L^2)-m_L^2C_0(m_L^2,m_L^2,m_L^2)\big]\,.
\end{align}
Note that the last term within parentheses in the above expression is regular in the limit $m_L\to0$. Using the tree-level expressions in eqs.~(\ref{dim5toymodel}) and~(\ref{dim6toymodel}), we can rewrite these contributions as
\begin{align}
 \kappa^{-1}\delta\lambda_{LLLL}\supset&-\bigg[20\frac{a_{LLH}a_{LLL}\lt_{LLLH}}{m_H^2}-30\frac{a_{LHH}(a_{LLH})^2a_{LLL}}{m_H^4}\bigg]P_{SS}(m_L^2,m_L^2)\nn\\
 &-12\frac{(a_{LLH})^2(a_{LLL})^2}{m_H^4}\big[P_{SS}(m_L^2,m_L^2)+\text{reg.}\big]\,,
\end{align}
which exactly correspond to the divergent terms in equation~(\ref{EQ:divterms_toymodel}). 

\subsubsection{Discussion of the dimension-5 operator for a general theory}
We now investigate the corresponding higher-order operators in the context of the matching of generic theories. 
We denote these two couplings as $c_5^{\lii\lij\lik\lil\ligm}$ and $k_6^{\lii\lij\lik\lil}$, respectively, and we define the corresponding Lagrangian terms as
\begin{equation}
\label{EQ:higherdimops}
 \lagr_\text{EFT}\supset-\frac{1}{5!}c_5^{\lii\lij\lik\lil\ligm}\phi_\lii\phi_\lij\phi_\lik\phi_\lil\phi_\ligm-\frac{1}{4}k_6^{\lii\lij\lik\lil}\phi_\lii\phi_\lij\partial_\mu\phi_\lik\partial^\mu\phi_\lil\,.
\end{equation}
For both operators, we will first derive their (tree-level) expression in terms of couplings of the high-energy theory, before showing how their inclusion allows to cancel all remaining IR divergences in the one-loop matching of the quartic coupling. 

We must first derive the tree-level expression of the new dimension-5 scalar $c_5^{\lii\lij\lik\lil\ligm}$ operator that is generated in the low-energy theory, by repeating the matching of effective actions done in equation~(\ref{EQ:integrateout_heavyfields}). Keeping now terms with five scalars, the Lagrangian of the high-energy theory contains
\begin{align}
 \lagr_\text{eff}[\Phi]\supset&\ \frac{1}{12m_{\heI}^2}a_{\heI\lii\lij}\lt_{\heI\lik\lil\ligm}\Phi_\lii\Phi_\lij\Phi_\lik\Phi_\lil\Phi_\ligm-\frac{1}{8m_{\heI}^2m_{\heJ}^2}a_{\heI\lii\lij}a_{\heI\heJ\lik}a_{\heJ\lil\ligm}\Phi_\lii\Phi_\lij\Phi_\lik\Phi_\lil\Phi_\ligm+\cdots
\end{align}
Identifying this with the definition of $c_5$ in eq.~(\ref{EQ:higherdimops}), and symmetrising the indices, we obtain for the tree-level matching of $c_5$
\begin{align}
c_5^{\lii\lij\lik\lil\ligm}=&-\frac{1}{12m_{\heI}^2}a_{\heI\lii\lij}\bigg[\lt_{\heI\lik\lil\ligm}-\frac{3}{2m_{\heJ}^2}a_{\heI\heJ\lik}a_{\heJ\lil\ligm}\bigg]+(\lii\lij\lik\lil\ligm)\,.
\end{align}

The dimension-5 operator gives rise to a new type of diagrams, shown in figure~\ref{FIG:dim5}, contributing to the low-energy part of the matching of the quartic couplings. The additional terms in $\delta \lambda_{\lii\lij\lik\lil}$ read
\begin{align}
\label{EQ:dim5_general_correction}
 \kappa^{-1}\delta \lambda_{\lii\lij\lik\lil}\supset&\ 
 \frac{1}{12}c_5^{\lii\lij\lik\lix\liy}a_{\lix\liy\lil}P_{SS}(m_\lix^2,m_\liy^2)+(\lii\lij\lik\lil)\nn\\
 =&-\frac{5}{6m_{\heI}^2}a_{\heI\lii\lij}a_{\lix\liy\lil}\bigg[\lt_{\heI\lik\lix\liy}-\frac{3}{2m_{\heJ}^2}a_{\heI\heJ\lik}a_{\heJ\lix\liy}\bigg]P_{SS}(m_\lix^2,m_\liy^2)+(\lii\lij\lik\lil)\,.
\end{align}
Using permutations of indices, it can be shown that the terms in equation~(\ref{EQ:dim5_general_correction}) match exactly the three first lines of equation~(\ref{EQ:extra_div}). 
Only the last term 
\begin{equation*}
 -\frac{1}{2}a_{\lii\heX\liy}a_{\lij\heX\liz}a_{\lik\liy\liu}a_{\lil\liz\liu}\frac{P_{SS}(m_\liz^2,m_\liu^2)}{m_{\heX}^4}+(\lii\lij\lik\lil)\,,
\end{equation*}
is left, and has to be cancelled out by the dimension-6 operator defined in equation~(\ref{EQ:higherdimops}). 

\subsubsection{Discussion of the dimension-6 operator for a general theory}
We start by deriving the (tree-level) matching condition for this coupling, using once again the equation of motion for the heavy fields, and we have
\begin{align}
 \mathcal{L}_\text{eff}[\Phi] &\supset\frac18\frac{a_{\heI\lii\lij}a_{\heI\lik\lil}}{m_{\heI}^4}\partial_\mu\big(\Phi_\lii\Phi_\lij\big)\partial^\mu\big(\Phi_\lik\Phi_\lil\big)+\cdots
 =\frac12\frac{a_{\heI\lii\lik}a_{\heI\lij\lil}}{m_{\heI}^4}\Phi_\lii\Phi_\lij\partial_\mu\Phi_\lik\partial^\mu\Phi_\lil+\cdots
\end{align}
Matching the Lagrangians of the high- and low-energy theory at tree level, we obtain
\begin{align}
\label{dim6_matching}
 k_6^{\lii\lij\lik\lil}=-\frac{2}{m_{\heI}^4}a_{\heI\lii\lik}a_{\heI\lij\lil}\,.
\end{align}
%
%
In turn, we find the following Feynman rule for the dimension-6 coupling
\begin{figure}[h]
\centering
 \begin{tikzpicture}
  \draw[dashed] (-1,-1) -- (1,1);
  \draw[dashed] (-1,1) -- (1,-1);
  \fill (-.1,-.1) -- (-.1,.1) -- (.1,.1) -- (.1,-.1) -- (-.1,-.1);
  \draw[->] (-.9,-1.1) -- (-.4,-.6);
  \draw[->] (-.9,1.1) -- (-.4,.6);
  \draw[->] (.9,-1.1) -- (.4,-.6);
  \draw[->] (.9,1.1) -- (.4,.6);
  \node at (-.65,1.3) {$p_1$};
  \node at (-.65,-1.3) {$p_2$};
  \node at (.65,1.3) {$p_3$};
  \node at (.65,-1.3) {$p_4$};
  \node at (-1.2,1.2) {$i$};
  \node at (-1.2,-1.2) {$j$};
  \node at (1.2,1.2) {$k$};
  \node at (1.2,-1.2) {$l$};
  \node at (5.5,0) {$=+ik_6^{\lii\lij\lik\lil}p_3\cdot p_4=-\displaystyle{\frac{i}{2}}k_6^{\lii\lij\lik\lil}\big[(p_3-p_4)^2-p_3^2-p_4^2\big]\,.$};
 \end{tikzpicture}
\end{figure}

The new type of diagrams in the low-energy theory that will cancel the remaining divergence is shown in figure~\ref{dim6diag}.
From these, we have a contribution to $\delta \lambda_{\lii\lij\lik\lil}$ 
\begin{align}
\label{dim6_contribution}
 \kappa^{-1}\delta\lambda_{\lii\lij\lik\lil}\supset&\ \frac{1}{4}k_6^{\lii\lij\lix\liy}a_{\lik\lix\liz}a_{\lil\liy\liz}\big(P_{SS}(m_\liy^2,m_\liz^2)-m_\lix^2C_0(m_\lix^2,m_\liy^2,m_\liz^2)\big)+(\lii\lij\lik\lil)\nn\\
                                           =&-\frac{1}{2m_{\heI}^4}a_{\heI\lii\lix}a_{\heI\lij\liy}a_{\lik\lix\liz}a_{\lil\liy\liz}\big(P_{SS}(m_\liy^2,m_\liz^2)-m_\lix^2C_0(m_\lix^2,m_\liy^2,m_\liz^2)\big)+(\lii\lij\lik\lil),
\end{align}
where for the second line, we used the matching in equation (\ref{dim6_matching}). The first term within parentheses in the equation (\ref{dim6_contribution}) exactly cancels the last divergent term in equation (\ref{EQ:extra_div}), while the second term is regular in the limit $m_\lix^2\rightarrow0$.

\section{Non-minimal counterterm approach}
\label{SEC:COUNTERTERMS}

In the derivation of section \ref{sec:mixing}  we used the running ($\msbar$ or $\drbar$) parameters of the high-energy theory as inputs. However, we may prefer to define our matching scale in terms of the loop-corrected masses and mixings, for example in high-scale/split SUSY cases we can typically choose to adjust $m_{H_u}^2, m_{H_d}^2, B_\mu$ and also the $Z$-counterterms. In other words, we can allow new \emph{finite corrections to the counterterms}, which we denote $\delta_{ct} Z,\,  \delta_{ct} m^2$:\footnote{We \emph{do not} include the divergent parts of the counterterms in $\delta_{ct} Z,\, \delta_{ct} m^2$ as they have already been implicitly subtracted.}
\begin{align}
\Gamma [\Phi] =& \frac{1}{2} \big(\delta_{\gea\geb} - \Pi_{\gea\geb}^\prime (0) + \delta_{ct} Z_{\gea\geb}\big) \partial_\mu \Phi_\gea \partial^\mu \Phi_\geb - \frac{1}{2} \big(m^2_{\gea  }\delta_{\gea\geb} + \Pi_{\gea\geb}(0) + \delta_{ct} m_{\gea\geb}^2\big) \Phi_\gea \Phi_\geb + ...
\end{align}
where we made use of the relations $\delta Z_{\gea\geb} = -\Pi_{\gea\geb}^\prime (0)$ and  $\delta m^2_{\gea\geb} = \Pi_{\gea\geb} (0)$ to avoid confusion with the counterterms, and the prime on $\Pi'_{ij}$ denotes the derivative of $\Pi_{ij}$ with respect to the external momentum. We then make the definition
\begin{align}
U &\equiv N^H R N^L
\end{align}
where $N^H, N^L$ are not (necessarily) unitary, and $R$ is unitary -- in section \ref{SEC:MATCHING} we had $N^{H} = 1 - \frac{1}{2} \delta Z^H, N^L = 1 + \frac{1}{2} \delta Z^L$ (see equation~(\ref{EQ:heavylightrelation})). Wanting to diagonalise the masses of the heavy states, and ensure all fields have correctly normalised kinetic terms we have
\begin{align}
  N^H_{\geap\gea} \big( \delta_{\geap\gebp} - \Pi_{\geap\gebp}^{HET\,\prime} (0) + \delta_{ct} Z_{\geap\gebp}\big)N_{\gebp\geb}^H &= \delta_{\gea\geb}\,, \nn\\
R_{\gec\heI} N^H_{\geap\gec} \big( m^2_{\geap}  \delta_{\geap\gebp}+  \Pi_{\geap\gebp}^{HET}(0) + \delta_{ct} m_{\geap\gebp}^2\big) N_{\gebp\ged}^{H}R_{\ged\geb} &\equiv M_\heI^2 \delta_{\heI\geb}\,.
\end{align}
However, with the intention of using top-down information on the new dimensionless couplings appearing in the HET (since we cannot fix them from the bottom up, and since they are often given by e.g.~unification or symmetry relationships such as the relationship between the Higgs quartic coupling and the gauge couplings in supersymmetric models) we should maintain the use of $\msbar$ (or $\drbar$) values for them, and so we should set $\delta_{ct} Z =0$ as before; this also precludes additional finite counterterms for the cubic and quartic terms. Then at one-loop order
$$R = 1 + \delta R, \quad R^T = 1 - \delta R, \quad U = 1 - \frac{1}{2} \Delta Z + \delta R$$
and (dropping the HET on the self-energies when it is unambiguous)
\begin{align}
M_\heI^2 &= m^2_{\heI} + m^2_{\heI} \Pi^\prime_{\heI\heI} (0)+ \Pi_{\heI\heI} (0) + (\delta_{ct} m^2)_{\heI\heI}, \nn\\
0&= \frac{1}{2} \Pi_{\heI\geb}^\prime (0) [m^2_\heI + m^2_\geb] + \Pi_{\heI\geb} (0) + \delta_{ct} m^2_{\heI\geb} + \delta R_{\heI\geb} [m^2_\heI - m^2_\geb],
\end{align}
with no summation on repeated indices. 
We then have some freedom to choose our mass counterterms to adjust  $\delta R$. We could choose $\delta m_{\gea\geb}^2$ so that $\delta R_{\gea\geb} =0$, but then we still have off-diagonal contributions to $U$ from the wave-function renormalisation. The most expedient choice seems to be to eliminate the terms $U_{\heI\lij}$ (but not $U_{\lii\heJ}$) via
\begin{align}
\delta_{ct} m^2_{\heI\lij} =& - \Pi_{\heI\lij} (0) ~~\Rightarrow~~
\delta U_{\heI\lij} = - \frac{m_\lij^2}{m_\heI^2 - m_\lij^2} \Pi_{\heI\lij}^\prime (0)\,.
\label{EQ:MAINCTCHOICE}\end{align}
For the light masses, the above equations do not determine  $\delta R_{\lii\lij}$, and so we can take it to be zero and work in the flavour basis (of course, any unitary rotation of the fields is equivalent).\footnote{Since the light masses need tuning to remain small, we see that we should either adjust the tree-level masses order by order in perturbation theory, or take $\delta_{ct} m^2_{\lii\lij} = - \Pi_{\lii\lij}^{HET} (0) + \Pi_{\lii\lij}^{LET} (0).$}
In other words, once we set the light masses to zero, 
\begin{align}
U_{\heI\lij} &= 0\,, \qquad U_{\lii\lij} = \delta_{\lii\lij} -\frac{1}{2} \Delta Z_{\lii \lij}\,.
\label{EQ:NoMixingU}\end{align}
This generalises the result for two Higgs doublets in \cite{Bahl:2018jom}.
Hence in the non-minimal counterterm approach, we can eliminate the mixing term between the heavy and light states. The result is:
\begin{align}
\lambda_{\lii\lij\lik\lil}  =&\ \lt_{\lii\lij\lik\lil} + \ov{\delta} \lt_{\lii\lij\lik\lil} \nn\\
& + \bigg[- \frac{1}{8m_\heI^2} a_{\heI\lii\lij} a_{\heI\lik\lil}- \frac{1}{4m_\heI^2}  a_{\heI\lii\lij} \ov{\delta} a_{\heI\lik\lil}  + \frac{1}{8 m_\heI^2} (\db m^2_{\heI\heJ} + \delta_{ct} m^2_{\heI\heJ})\frac{1}{m_\heJ^2} a_{\heI\lii\lij} a_{\heJ\lik\lil}\nn\\
&\quad\ \,\ - \frac{1}{12} \Delta Z_{\lii\liip} \big(\lt_{\liip\lij\lik\lil}- \frac{3}{m^{2}_{\heI}}a_{\heI\liip\lij} a_{\heI\lik\lil} \big)  +  (\lii\lij\lik\lil)\bigg]. 
\label{EQ:MasterCT}\end{align}
We then still have the choice of counterterm for $\delta_{ct} m_{\heI\heJ}^2$. If we take
\begin{align}
  \delta_{ct} m_{\heI\heJ}^2 =& 0
\label{EQ:DeltaM0}
\end{align}
then we can evaluate (\ref{EQ:MasterCT}) with $g\rightarrow 0$ in all of the loop corrections, but at the expense of including $\ov{\delta} m_{\heI\heJ}^2$. On the other hand, we in principle also have the freedom to take $\delta_{ct} m^2_{\heI\heJ} = - \delta m^2_{\heI\heJ} $ \emph{but}, as we have seen in the previous sections, this will spoil the cancellation of infra-red divergences \emph{and} gauge dependence. Indeed, if we set $\delta_{ct} m^2_{\heI\heJ} = - \db m^2_{\heI\heJ}$ then we still have the problem of gauge invariance, and so the simplest possible choice is
 \begin{align}
\delta_{ct} m^2_{\heI\heJ} = - \db m^2_{\heI\heJ}\bigg|_{g \rightarrow 0}.
\label{EQ:DeltaMdb} \end{align}
The mass-squared quantities defined using this counterterm choice do not have a simple physical interpretation (they correspond neither to a running mass nor a pole mass) but nevertheless will be the quantities that appear as the expansion parameters, and could potentially be identified with the matching scale; we also expect that it should be these quantities that are most useful beyond one-loop order, but we leave the investigation of that to future work.

In this way, when calculating the quartic coupling in the low-energy theory, we should take
\begin{empheq}[box=\fbox]{align}
\lambda_{\lii\lij\lik\lil}  =\lt_{\lii\lij\lik\lil}& \nn\\
 -\bigg[& \frac{1}{8m_\heI^2} a_{\heI\lii\lij} a_{\heI\lik\lil} -\frac{1}{24}\ov{\delta} \lt_{\lii\lij\lik\lil}+ \frac{1}{4m_\heI^2}  a_{\heI\lii\lij} \ov{\delta} a_{\heI\lik\lil} 
+ \frac{1}{12} \Delta Z_{\lii\liip} \big(\lt_{\liip\lij\lik\lil}- \frac{3}{m_\heI^2} a_{\heI\liip\lij} a_{\heI\lik\lil} \big)\nn\\
&+  (\lii\lij\lik\lil)\bigg]_{g\rightarrow 0}. 
\label{EQ:SIMPLEST}\end{empheq}
This is one of the main results of this work: we have a prescription that eliminates mixing between light and heavy degrees of freedom that can be applied in any model. 

It can also be convenient to write the above explicitly for the case of the low-energy theory being the SM, and where the (neutral component of the) Higgs field $H$  is complex with interactions
\begin{align}
  \mathcal{L}_{EFT} \supset& - \frac{\lambda_{HH}^{HH}}{4} |H|^4 , \quad  \mathcal{L}_{HET} \supset - \frac{\lt_{HH}^{HH}}{4} |H|^4 - \frac{1}{2} a_{\heI HH} \Phi_\heI H^2 - \frac{1}{2} a_{\heI}^{HH} \Phi_\heI \ov{H}^2 -a_{\heI H}^{H} \Phi_\heI |H|^2 . 
\end{align}
Note that when working with complex fields  we use lowered indices for a given field and raised indices for its complex conjugate. We obtain for the matching 
\begin{align}
\lambda_{HH}^{HH}  =\lt_{HH}^{HH}& -\frac{2}{m_\heI^2} (a_{\heI H}^H)^2 -\frac{1}{m_\heI^2} a_{\heI HH} a_{\heI}^{HH}   \nn\\
 +\bigg[&\ov{\delta} \lt_{HH}^{HH}   - \frac{4}{m_\heI^2}  a_{\heI H}^H \ov{\delta} a_{\heI H}^H -\frac{1}{m_\heI^2}  ( a_{\heI HH} \ov{\delta} a_{\heI}^{HH} + a_{\heI}^{HH} \ov{\delta}a_{\heI HH}  ) \nn\\
&-2\Delta Z_H^H (\lt_{HH}^{HH} - \frac{2}{m_\heI^2} (a_{\heI H}^H)^2 - \frac{1}{m_\heI^2} a_{\heI HH} a^{HH}_\heI) \bigg]_{g\rightarrow 0}. 
\end{align}

As a coda to this discussion, we note that another counterterm choice that is available is to use \emph{pole masses} for  the heavy states. This would have the advantage that standard expressions could be used to define the counterterms, and it would avoid the problems of infra-red divergences (at least at one loop) and gauge dependence because the pole mass is a well-defined quantity. Furthermore, if the heavy states had masses not outside experimental reach (for example coloured superpartners around 2 TeV) then we would be using physically measurable quantities. However, from an effective field theory point of view this choice is less practical, because (1) there would not be cancellations between the counterterms and the terms in the effective potential (which are evaluated at zero external momentum); (2) the loop functions containing external momenta become much more complicated at one loop, and the full set is not known analytically at two. 

Finally, a more extreme counterterm choice would be to use pole masses for all states, both light and heavy, without taking the limit $\Ml \rightarrow 0$. This would technically remove the problem of infra-red divergences, but replace it with a practical one (the computations would become much more cumbersome, with numerically large logarithms, unless the limit of $\Ml \rightarrow 0 $ were taken analytically, when they would reduce to the expressions above).

\section{Comparison with the pole matching approach}
\label{SEC:POLEMATCHING}

As mentioned in the introduction, an alternative approach to matching quartic couplings in effective theories is to match the pole masses of the light scalar fields; this method has recently been advocated as an efficient matching technique in \cite{Athron:2016fuq}. This is only really tractable when the low-energy theory has scalars that do not mix with each other, and so if we assume that the low-energy theory is the SM (or any extension thereof without any additional scalars/gauge bosons), then there is only one physical scalar mass, and then there is only one equation to solve:
\begin{align}
\frac{1}{2}\lambda_{SM} v^2_{SM} + \Delta M^2_{SM} (m_h^2) =&\ m_0^2 + \Delta M^2_{HET} (m_h^2),
\label{EQ:BASICPOLEMATCHING}\end{align}
where: we define the quartic term in the SM Higgs potential as $\frac{\lambda_{SM}}{4} |H|^4\,$; $m_0$ is the tree-level Higgs boson mass in the high-energy theory; $\Delta M^2_{SM} (m_h^2)$ and $\Delta M^2_{HET} (m_h^2)$ denote the one-loop corrections to the Higgs mass in the SM and in the high-energy theory, respectively, computed with external momentum equal to the Higgs pole mass $m_h^2\,$. Since we work in the broken phase of the theory, the value for $\lambda_{SM}$ extracted in this way will be correct up to subleading terms of order $v^2/\Mm^2$. This approach has the advantage of requiring only two-point functions, at the expense of requiring numerical cancellations between large terms. Given that we described several choices in sections \ref{SEC:GENERAL}, \ref{SEC:MATCHING} and \ref{SEC:COUNTERTERMS}, it is interesting and important to compare this calculation with our traditional EFT approach so that we understand the results obtained via the pole matching method.

To extract $\lambda_{SM}$, we can next perform a double expansion in $v$ as well as loop order, neglecting subleading terms, because the EFT approach will only capture the leading terms in the expectation values. So we write the threshold corrections for all parameters $g_i$ as 
\begin{align}
g_i^{SM} = g_i^{HET} + \Delta_{g_i} (g, \lambda, v, ...)\,.
\end{align}
To extract the quartic coupling $\lambda$, we need thresholds for all parameters that appear at tree-level in the equation (\ref{EQ:BASICPOLEMATCHING}), which consists only of $\lambda$ and $v$. The other important parameters of the SM are then the gauge and Yukawa couplings; the threshold corrections to these are only needed for running (or e.g.~for supersymmetric relationships) in the high-energy theory but not for the extraction of $\lambda$ at one-loop order (whereas at two-loop order they are required). Nevertheless, the one-loop gauge threshold corrections are given in \ref{app:gaugebosons} and those to the Yukawas in \ref{app:yukawas}; alternatively the Yukawa couplings can be extracted by pole-mass matching of the quarks/leptons, under the assumption that the couplings are real and diagonal.

To match $v$, we can match the pole mass of the $Z$-boson and use the relation
\begin{align}
m_Z^2 = \frac{1}{4} (g_Y^2 + g_2^2) v^2 + \Pi_{ZZ} (m_Z^2).
\end{align}
Then clearly we need $\Pi_{ZZ}$ and thresholds to $g_Y$ and $g_2$ to determine the shift to $v$. So then
\begin{align}
 v_{SM}^2 =&\ v_{HET}^2+ \frac{4}{g_Y^2 + g_2^2} \bigg[ \Pi_{ZZ}^{HET} (m_Z^2)- \Pi_{ZZ}^{SM} (m_Z^2) - \frac{1}{4} v^2 (\Delta_{g_Y^2} + \Delta_{g_2^2} ) \bigg]
\label{EQ:vshift}\end{align}
Now we can take the $v=0$ expressions in $\Delta_{g_i^2}$ because they already have a prefactor of $v^2$. As we show in appendix \ref{app:gaugethresholds}, these are given by 
\begin{align}
\label{EQ:gaugethresholds}
\Delta_{g_Y^2} + \Delta_{g_2^2}\Big|_{v=0} &= 
(g_Y^2 + g_2^2) \bigg[ \Pi_{ZZ}^{HET\,\prime} (0)-\Pi_{ZZ}^{SM\,\prime} (0) \bigg]_{v=0}~.
\end{align}
For the self-energies we need to expand them to order $v^2$, which is equivalent to order $p^2$:
\begin{align}
\Pi_{ZZ} (m_Z^2) = \Pi_{ZZ}(0)\Big|_{v=0} + v^2 \bigg[ \partial_{v^2} \Pi_{ZZ}(0) + \frac{g_Y^2 + g_2^2}{4} \Pi_{ZZ}^\prime (0) \bigg]_{v=0} + ... 
\end{align}
This then yields
\begin{align}
  v_{SM}^2 =&\ v_{HET}^2+ \frac{4}{g_Y^2 + g_2^2} \bigg[ \Pi_{ZZ}^{HET}(0) - \Pi_{ZZ}^{SM}(0) \bigg] + \mathcal{O}(v^4).
  \label{vexp}
\end{align}
In other words, we do not need the momentum dependence of the gauge-boson self-energies. Note also that the self-energies in eq.~(\ref{vexp}) are not expanded in $v^2$, in order to retain the correct dependence on the ${\cal O}(v^2)$ terms. 
Under the assumption that there are no heavy gauge bosons being integrated out, we need only consider heavy fermions and scalars in the above, and the resulting shift in $v$ is ultimately independent of the gauge couplings.

Armed with this, we would now like to use the pole mass approach to obtain the \emph{most efficient} way of extracting the EFT matching condition for $\lambda$, which means that we are interested in an ultimately infra-red safe expression (i.e.~containing no large logarithms) and valid up to leading order in an expansion in $v$ -- recall that $v$ is of order $\Ml$. Then the tree-level Higgs mass $m_0^2$ is of order $\Ml^2$, and we see 
\begin{align}
\Delta M^2 (m_0^2) = \Delta M^2 (0) + m_0^2 \Pi_{hh}^\prime (0) + \mathcal{O} (\Ml^4).
\end{align}
Next, we need to solve the relation (\ref{EQ:BASICPOLEMATCHING}). This gives:
\begin{align}
  \lambda_{SM} = \frac{2}{v_{HET}^2} \bigg[& m_0^2 \bigg(1 + [\Pi^{HET\,\prime}_{hh} (0)- \Pi^{SM\,\prime}_{hh} (0)]\bigg)  - \frac{m_0^2}{m_Z^2} [\Pi_{ZZ}^{HET}(0) - \Pi_{ZZ}^{SM}(0)]  \nn\\
  & + [\Delta M^2_{HET} (0)- \Delta M^2_{SM} (0)] \bigg],
\label{EQ:polemass1}\end{align}
Here $m_0^2$ is a function of $v_{HET}$ (which is defined in terms of the $Z$ mass) but we could alternatively express the quantities on both sides of equation (\ref{EQ:BASICPOLEMATCHING}) in terms of $v_{SM}$ which would yield the same result. 
Now $\Delta M^2_{SM}(0)$ depends on $\lambda_{SM}$, so we will need to solve the above relationship recursively (at one loop this is one recursion). We know from the previous sections that in general there will be a \emph{tree-level} difference between $\lambda_{SM}$ and the quartic self-coupling of the Higgs in the high-energy theory, and it is interesting to see how this arises.

First we need to divide the fields into \emph{three} types: the index $1$ for the Higgs, greek letters $\{\alpha \}$ for heavy (doublet) fields that mix with the Higgs at zero expectation value, and capital roman letters $\{I,J,K,L\}$ for heavy scalar fields that mix with the Higgs only \emph{after} EWSB\footnote{Elsewhere we use $\{I,J,K,L\}$ for fermions: in this section we do not have explicit fermion indices so there should hopefully be no confusion.}; in the previous (and subsequent) sections we had $\{\alpha \} \subset \{\heI\}, \{I,J,K,L \} \subset \{\heI\}$ as we did not need to explicitly distinguish between the set $\{\alpha\}$ and $\{I,J,K,L\}$, but in this approach it becomes important. As before, though, we do not need to explicitly discuss fields that never obtain an expectation value or mix with the Higgs (such as squarks and fermions etc.). Then the allowed scalar couplings (under gauge symmetries) are
\begin{align}
 \{ a_{I 11}, a_{I 1\alpha}, a_{I \alpha \beta}, a_{IJK},\lt_{1 1 1 1}, \lt_{\alpha 1 1 1}, \lt_{\alpha \beta 1 1}, \lt_{\alpha \beta \gamma 1}, \lt_{\alpha \beta \gamma \delta}, \lt_{IJ 11}, \lt_{IJ\alpha 1}, \lt_{IJ\alpha \beta}, \lt_{IJKL}\}.
\end{align}
In the pole-mass approach, the expectation values of fields are usually treated as fixed, with chosen dimensionful parameters being fixed by the tadpole equations order by order in perturbation theory.  We can rotate all of the \emph{doublets} so that only one has an expectation value and
$$v_\alpha =0\,.$$
In any theory of many Higgs doublets, such as the MSSM/THDM, this choice corresponds to the so-called ``Higgs basis'', but in the presence of singlets/triplets this is no longer the case, since the expectation values of the latter cannot be just rotated away. This basis is not commonly used in the practical calculation but it will greatly simplify our analysis, in particular because non-alignment effects only appear at higher order in $v_1$.

Now we wish to derive $\Delta M_{HET}^2(0)$, which can be obtained by taking derivatives of the effective potential, and expand it to order $\mathcal{O} (v_1^2)$. First we split the one-loop effective potential into a supertrace over heavy and light fields:
\begin{align}
 \label{eq:Vheavy}
V^{(1)} \equiv&\ \ov{V}^{(1)} + V^{(1)}_{IR}, \qquad \ov{V}^{(1)} \equiv \sum_{i \in \rm heavy\ fields} \frac{1}{4} (-1)^{2s_i} (2s_i+1) m_i^4 ( \blog m_i^2 - 3/2)
\end{align}
(where $s_i$ is the spin of the field). $\ov{V}^{(1)}$ is regular as $v_1 \rightarrow 0$,  whereas $V^{(1)}_{IR} $ has infra-red divergences in its second and higher derivatives. Hence we next expand only the derivatives of $\ov{V}^{(1)} $ in the tadpole equations as a series in $v_1$ and $v_I$. To this end we define
\begin{align}
\Delta V_{i_1 \cdots i_n} \equiv \left. \kappa \frac{\partial^n \ov{V}^{(1)}}{\partial \Phi_{i_1} \cdots \partial \Phi_{i_n}} \right|_{\Phi_\gea=0, v_\gea =0}, \qquad \Delta V_{i_1 \cdots i_n}^{IR} \equiv \left. \kappa \frac{\partial^n V^{(1)}_{IR}}{\partial \Phi_{i_1} \cdots \partial \Phi_{i_n}} \right|_{\Phi_\gea = v_\gea, v_1 \ne 0}.
\end{align}
We work in the basis after any shifts of the parameters (\ref{EQ:removesingletvevs}) -- but, crucially, such shifts are made before electroweak symmetry breaking, so in all cases $v_I$ will be nonzero but small after electroweak symmetry is broken. 
In this notation the expansions of the tadpole equations become:
\begin{align}
  0=&\ \Delta V_{1}^{IR} +  m_{11}^2 v_1 + a_{I 11} v_1 v_I + \frac{1}{6} \lt_{1111} v_1^3 + \frac{1}{2} \lt_{11IJ} v_1 v_I v_J+ \Delta V_{11} v_1 + \Delta V_{I 11} v_I v_1 + \frac{1}{6} \Delta V_{1111} v_1^3 + ...   \nn\\
  0=&\ \Delta V_{\alpha}^{IR} +  m_{\alpha 1}^2 v_1 + a_{I\alpha 1} v_1 v_I + \frac{1}{6} \lt_{\alpha 111} v_1^3 + \frac{1}{2} \lt_{\alpha 1 IJ} v_1 v_I v_J + \Delta V_{\alpha 1} v_1 + \Delta V_{I \alpha 1} v_I v_1 + \frac{1}{6} \Delta V_{\alpha 111} v_1^3 + ... \nn\\
  0=&\ \Delta V_{I}^{IR}+ t_I + \Delta V_I + m_{IJ}^2 v_J + \frac{1}{2} a_{IJK} v_J v_K + \frac{1}{2} a_{I 11} v_1^2 + \frac{1}{2}\lt_{IJ 11} v_J v_1^2 + \frac{1}{6} \lt_{IJKL} v_J v_K v_L\nn\\
  &+ \Delta V_{IJ} v_J + \frac{1}{2} \Delta V_{IJK} v_J v_K + \frac{1}{2} \Delta V_{I 11} v_1^2 + \frac{1}{2}\Delta V_{IJ 11} v_J v_1^2 + \frac{1}{6} \Delta V_{IJKL} v_J v_K v_L + ...
\end{align}
Although we have not expanded the derivatives of $V_{IR}$, we know that $V^{(1)}_{IR} \sim \mathcal{O}(\Ml^4)$ so $\Delta V_{1}^{IR} \sim \mathcal{O}(\Ml^3) $  and $\Delta V_{11}^{IR} \sim \mathcal{O}(\Ml^2),$ so we will not need to.
We must now understand how to treat the expectation values $v_I$ -- recall that these are really the differences between the singlet expectation values and their values at  $v_1 =0$. 
Firstly we can solve the third equation for 
\begin{align}
m_I^2 v_I =& - \bigg[ \frac{1}{2} a_{I11} v_1^2 + \frac{1}{2} \Delta V_{I11} v_1^2 + \Delta V_{IJ} v_J + \sum_{J \ne I} m_{IJ}^2 v_J + t_I + \Delta V_I +\Delta V_{I}^{IR} + ... \bigg] ,
\label{EQ:HeavyTadpole}\end{align}
where we write $m_I^2 \equiv m_{II}^2$, singling out the diagonal element, since at tree level and for $v_1 =0$ we take $m_{IJ}^2$ to be diagonal, and thus $m_{IJ}^2$ is of subleading order for $I \ne J$. Now, depending on our treatment, we have 
$$t_I  =\left\{ \begin{array}{cl}0 & \mathrm{Possibilities\ 1\ and\ 3} \\ -\Delta V_I  & \mathrm{Possibilities\ 2\ and\ 4} \end{array}\right. $$
In other words, recalling that $ \Delta V_{I}^{IR} \sim \mathcal{O}(\Ml^3) $ we find that 
$$ v_I =  - \frac{1}{m_I^2} (t_I + \Delta V_I) + \mathcal{O}(v_1^2/\Mm) + \Otwoloop.$$
Note that for triplets $t_I = \Delta V_I = 0$ whatever the option. This means there is no ambiguity in the definition of $v_{HET} = v_1$ in (\ref{EQ:vshift}), since the corrections to the $Z$ mass from triplets will be of subleading order compared to that from doublets and can be neglected.

 Now we must consider the mass matrices, and perturbatively determine them both to one-loop order and to order $v_1^2$. Since we are interested in $\Delta M^2_{HET}(0)$ we just need the second derivative of the effective potential
\begin{align}
\mathcal{M}^2_{\gea\geb} \equiv \frac{\partial^2 }{\partial v_\gea \partial v_\geb} [ V^{(0)} + \kappa V^{(1)}] 
\label{EQ:MatchingMassSquared}  \end{align}
  so we have
  \begin{align}
  (\mathcal{M}^2)_{11} &= m_{11}^2 + \Delta V_{11} + \Delta V_{11}^{IR} + \frac{1}{2} \lt_{11IJ} v_I v_J + (a_{I11} + \Delta V_{I11}) v_I + \frac{1}{2} (\lt_{1111} + \Delta V_{1111} ) v_1^2 + ...  \nn\\
    &= \frac{1}{3} (\lt_{1111} + \Delta V_{1111}) v_1^2 + \Delta V_{11}^{IR} - \frac{1}{v_1} \Delta V_{1}^{IR} +\mathcal{O}(v_1^3), 
  \end{align}
as we would expect. For the other doublets,   
\begin{align}
  (\mathcal{M}^2)_{\alpha 1} &= \frac{1}{3} (\lt_{\alpha 111} + \Delta V_{\alpha 111}) v_1^2 + \Delta V_{\alpha 1}^{IR}  -\frac{1}{v_1} \Delta V_{\alpha}^{IR}  + ...
  \end{align}
  which will not contribute to the mass of the lightest eigenvalue at order $v_1^2 \sim \Ml^2$, and so they can be neglected, as claimed.
  Finally,  
  \begin{align}
   (\mathcal{M}^2)_{I1} &= (a_{I11}+ \Delta V_{I11}) v_1 + \Delta V_{I1}^{IR}+ \lt_{IJ 11} v_J v_1 +... \nn\\
  (\mathcal{M}^2)_{IJ} &= m_I^2 \delta_{IJ} +  \Delta V_{IJ} + \Delta V_{IJ}^{IR} + a_{IJK} v_K + \mathcal{O}(v_1)
\end{align}
Then the result for the mass shift (after performing a double expansion in $v$ and loop order) is
\begin{align}
m_0^2 +  \Delta M_{HET}^2(0) =&\  (\mathcal{M}^2)_{11} -  \frac{((\mathcal{M})_{I1}^2)^2}{m_I^2} + \frac{(\mathcal{M})_{I1}^2 (\mathcal{M})_{J1}^2 }{m_I^2 m_J^2} (\Delta V_{IJ} + \Delta V_{IJ}^{IR} + a_{IJK} v_K) + \mathcal{O}(v_1^3) + \Otwoloop \nn\\
  =&\  v_1^2 \bigg[\frac{1}{3} (\lt_{1111} + \Delta V_{1111}) - \frac{a_{I11}^2}{m_I^2} - \frac{2 a_{I11} \Delta V_{I11}}{m_I^2}  +  \frac{a_{I11} a_{J11} \Delta V_{IJ}}{m_I^2 m_J^2} \bigg]\nn\\
& - \frac{v_1^2}{m_K^2}  (t_K + \Delta V_K)  \bigg[ - \frac{2a_{I11} \lt_{IJ11}}{m_I^2} + \frac{a_{I11} a_{J11} a_{IJK}}{m_I^2 m_J^2} \bigg]\nn\\
& + \Delta V_{11}^{IR} - \frac{1}{v_{1}} \Delta V_{1}^{IR} - \frac{2 a_{I11} v_1 \Delta V_{I1}^{IR}}{m_I^2} +  \frac{a_{I11} a_{J11} v_1^2  \Delta V_{IJ}^{IR}}{m_I^2 m_J^2} \nn\\
&+ \mathcal{O}(v_1^3) + \Otwoloop.
\end{align}
The equivalent expression in the low-energy theory is of course just
\begin{align}
\frac{1}{2} \lambda_{SM} v_{SM}^2 + \Delta M_{SM}^2 (0) &= \frac{1}{3} v_{SM}^2 \lambda_{1111} + \Delta V_{11}^{SM} - \frac{1}{v_{SM}} \Delta V_{1}^{SM},
\end{align}
where $\Delta V_{1}^{SM}, \Delta V_{11}^{SM} $ are the first and second derivatives of the Standard Model effective potential, and it is important to note that they are \emph{not equal} to $\Delta V_{1}^{IR}, \Delta V_{11}^{IR} $ in the presence of $a_{I11}$ terms. This leads to 
\begin{align}
    \lambda_{SM} =&\ \frac{2}{3} \lambda_{1111} \,.
\end{align}
Next we can see that 
\begin{align}
\Delta V_{11}^{IR} - \frac{1}{v_{1}} \Delta V_{1}^{IR} - \frac{2 a_{I11} v_1 \Delta V_{I1}^{IR}}{m_I^2} +  \frac{a_{I11} a_{J11} v_1^2  \Delta V_{IJ}^{IR}}{m_I^2 m_J^2} - \Delta V_{11}^{SM} + \frac{1}{v_1} \Delta V_{1}^{SM} = \mathcal{O}(v_1^3) + \Otwoloop.
\end{align}
and so we can identify 
the derivatives of the one-loop contribution to the effective potential with corrections to the different couplings, i.e.
\begin{align}
 \Delta V_{1111}  \rightarrow \db \lt_{1111}\,,\qquad\Delta V_{I11}\rightarrow\db a_{I11}\,,\qquad\Delta V_{IJ} \rightarrow \db m_{IJ}^2\,.
\end{align}
The result for matching $\lambda_{1111}$ becomes
\begin{align}
  \lambda_{1111} =&\ \lt_{1111} - 3\frac{a_{I11}^2}{m_I^2} + \db \lt_{1111}  - \frac{6 a_{I11} \db a_{I11}}{m_I^2} + 3 \frac{a_{I11} a_{J11} \db m^2_{IJ}}{m_I^2 m_J^2} \nn\\
& + (t_K + \Delta V_K)  \bigg[ \frac{6a_{I11} \lt_{IJ11}}{m_I^2 m_K^2 }-  \frac{3 a_{I11} a_{J11} a_{IJK}}{m_I^2 m_J^2 m_K^2} \bigg] \nn\\
  & - \bigg(\lt_{1111} - 3\frac{a_{I11}^2}{m_I^2}\bigg) \bigg( \Delta Z_{11} + \frac{1}{m_Z^2} [\Pi_{ZZ}^{HET}(0) - \Pi_{ZZ}^{SM}(0)]\bigg) .
\label{EQ:FullMatchingComparison}\end{align}
Next, it can be shown that\footnote{For example, this can be shown either explicitly at one loop using the expressions in the appendix, or using the Ward identities ($29$) and ($30$) from \cite{Denner:1994xt}; see also $(3.10)$ from \cite{Bohm:1986rj} and D$.11$ from \cite{Belanger:2003sd}.}
\begin{align}
 \frac{1}{m_Z^2} [\Pi_{ZZ}^{HET}(0) - \Pi_{ZZ}^{SM}(0)] = \Delta Z_{11} + \mathcal{O} (v^2)  
\label{EQ:PiZequivalence}\end{align}
and we conclude that the pole-mass calculation is equivalent to the EFT calculation \emph{with the counterterm choices} (\ref{EQ:MAINCTCHOICE}) for the heavy-light mixing and (\ref{EQ:DeltaM0}) for the heavy masses.  This agrees with the result found in the MSSM in \cite{Wells:2017vla} where it was found that the classic PBMZ calculation of the Higgs mass \cite{Pierce:1996zz} yields a result equivalent to including a counterterm for the rotation angle between the fields such as used in \cite{Bagnaschi:2014rsa}. 

Furthermore, we find that it is straightforward to make a connection between the pole mass matching and the EFT approach for the treatment of the singlet expectation values: the second line in equation (\ref{EQ:FullMatchingComparison}) vanishes for options 2 or 4 for the singlet tadpoles, and gives exactly the shifts (\ref{EQ:MyVS}) for option 3, where $t_K =0$. This was not necessarily obvious, since the definitions are subtly different (in the pole matching procedure, the conditions are specified at $v_1 \ne 0$). Note that the treatment of the singlet tadpoles in the pole mass matching approach is commonly chosen to be option 4.

\subsection{Efficient computation of the matching}
\label{sec:efficient}

Since it is typically simpler to compute two-point functions, it is to be expected that the pole-mass matching procedure should be easier to implement than a conventional calculation. However, there remains the problem of efficiently subtracting the large logarithmic terms. The above derivation shows us that the calculation (\ref{EQ:polemass1}) can be simplified to
\begin{align}
  \lambda_{SM} = \frac{2}{v_{HET}^2} \bigg[& m_0^2 (1 - 2 \Delta Z_{11}) + \hat{\Delta} M^2_{HET} (0) \bigg],
\label{EQ:EFFICIENT}\end{align}
where $\Delta Z_{11} $ is computed at zero external momentum with all light masses  set to zero; the second term is defined with a hat to mean that we drop all terms which contain \emph{only} light masses, and for remaining terms (of the type $P_{SS} (m_\heI^2, m_\lii^2)$, etc.) we set all logarithms of light masses $\blog m_\lii^2 \rightarrow 0$. Furthermore, we can also set the gauge contributions to zero.

However, we must also take care with the gauge dependence in the presence of heavy triplet scalars (such as in Dirac gaugino models). In that case, if we set the gauge contributions to zero in the matching, then we must also set them to zero in the heavy tadpole relationship between $m_I^2$ and $v_I$ (\ref{EQ:HeavyTadpole}) -- otherwise we will reintroduce gauge dependence into the result.

\section{Examples}
\label{SEC:EXAMPLES}

\subsection{Pole matching in the MSSM}

The calculation in section \ref{SEC:POLEMATCHING} is perhaps couched in unfamiliar terms, so it is useful to present the standard example of split or high-scale supersymmetry, where the MSSM scalars are heavy and, when integrated out, yield a scalar sector that is just that of the SM, so ideal for application of the pole matching procedure.

The relevant part of the scalar sector consists of two complex fields $H_u^0, H_d^0$ that mix and have as potential prior to electroweak symmetry breaking
\begin{align}
V^{(0)} &= (m_{H_u}^2 + |\mu|^2) |H_u^0|^2 + (m_{H_d}^2 + |\mu|^2) |H_d^0|^2 - (B_\mu H_u^0 H_d^0 + h.c.) + \frac{g_Y^2 + g_2^2}{8} (|H_u^0|^2 - |H_d^0|^2)^2.
\label{EQ:mssm_pot}
\end{align}
After electroweak symmetry breaking we give expectation values to both fields of $\bra H_u^0 \ket \equiv \frac{v}{\sqrt{2}} \sin \beta,  $ $\bra H_d^0 \ket \equiv \frac{v}{\sqrt{2}} \cos \beta,  $ and we take CP to be conserved so that the neutral SM Higgs boson comes from the mixing of the scalar components. Solving the one-loop tadpole equations for $m_{H_u}^2, m_{H_d}^2$ the tree-level Higgs mass matrix for the real components $h_{u,d} \equiv \sqrt{2} \mathrm{Re}(H_{u,d}^0)$, writing $t_\beta \equiv \tan \beta$ etc., is 
\begin{align}
\mathcal{M}^2_{0} =&  \twomat[B_\mu t_\beta + \frac{1}{4} (g_Y^2 + g_2^2) v^2 c_\beta^2, - B_\mu - \frac{1}{8} (g_Y^2 + g_2^2) v^2 s_{2\beta}][- B_\mu - \frac{1}{8} (g_Y^2 + g_2^2) v^2 s_{2\beta}, \frac{B_\mu}{ t_\beta} + \frac{1}{4} (g_Y^2 + g_2^2) v^2 s_\beta^2]. 
\end{align}
The contributions of the heavy particles to the one-loop corrections to the mass matrix at zero external momentum can be obtained from the derivatives of $\ov{V}^{(1)}\!$, see eq.~(\ref{eq:Vheavy}), computed at the minimum of the potential:
\begin{align}\label{EQ:LeadingMH2MSSM}
  \left[\hat\Delta \mathcal{M}^2(0)\right]_{ij} ~=~
  \kappa\,\frac{\partial^2 \ov{V}^{(1)}}{\partial h_i \partial h_j} \bigg|_\text{min}
  -~~\kappa\,\frac{\delta_{ij}}{v_i}
\frac{\partial \ov{V}^{(1)}}{\partial h_i} \bigg|_\text{min},~~~~\quad\text{($i,j=u$ or $d$)}.
\end{align}
In order to obtain the correction to the quartic coupling from
eq.~(\ref{EQ:EFFICIENT}), we expand to $\mathcal{O}(v^2)$ the corrections to the mass matrix. To this effect, we define a derivative along the direction of $h$
\begin{equation}
 \frac{\partial}{\partial h}\equiv s_\beta \frac{\partial}{\partial h_u}+ c_\beta \frac{\partial}{\partial h_d}~,
\end{equation}
and we obtain:
\begin{align}\label{EQ:LeadingMH2MSSMexp}
  \hat\Delta \mathcal{M}^2(0) =&\twomat[-V_{ud} t_\beta, V_{ud}][V_{ud}, - \frac{V_{ud}}{t_\beta}] +\frac{v^2}{3}\twomat[V_{ddhh}, V_{udhh}][V_{udhh},V_{uuhh}]+\frac{v^2}{6}\twomat[-V_{udhh}t_\beta, V_{udhh}][V_{udhh},-\frac{V_{udhh}}{t_\beta}]+ \mathcal{O}(v^4),
\end{align}
where we introduced the following abbreviations for the $\mathcal{O}(v^0)$ parts of the derivatives of $\ov{V}^{(1)}$:
\begin{align}
V_{ij} \equiv \kappa\frac{\partial^2 \ov{V}^{(1)}}{\partial h_i \partial h_j} \bigg|_\text{min}^{v=0},~~~~~\quad V_{ijhh} \equiv \kappa\frac{\partial^4 \ov{V}^{(1)}}{\partial h_i \partial h_j (\partial h)^2} \bigg|_\text{min}^{v=0},~~~~\quad\text{($i,j=u$ or $d$)}.
\end{align}
Note that, in the above, all contributions with an odd number of derivatives of $\ov{V}^{(1)}$ vanish for $v=0$ because the one-loop corrections to the potential only contain terms with even powers of the Higgs fields.
Rotating the combined mass matrix to the Higgs basis,  we see that both the $\mathcal{O}(v^0)$ and the second of the $\mathcal{O}(v^2)$ terms in the one-loop corrections will cancel out in the correction to the Higgs mass. From the remaining $\mathcal{O}(v^2)$ terms from eq.~(\ref{EQ:LeadingMH2MSSMexp}), we find that the correction to the Higgs boson mass in the MSSM is
\begin{align}
  \hat\Delta M_{HET}^2(0)=&\ \frac{v^2}{3}\bigg[s_\beta^2 V_{uuhh}+2s_\beta c_\beta V_{udhh}+c_\beta^2 V_{ddhh}\bigg]+ \mathcal{O}(v^4)\nn\\
                     =&\ \frac{v^2}{3}\bigg(s_\beta \frac{\partial}{\partial h_u} + c_\beta \frac{\partial}{\partial h_d}\bigg)^4 \ov{V}^{(1)}+ \mathcal{O}(v^4).
\end{align}
Inserting this into equation (\ref{EQ:EFFICIENT}), and noting that in the MSSM there are no trilinear couplings involving only the Higgs bosons, we find
\begin{align}
\lambda_{SM} &= \frac{2}{3}( \lt_{1111}  + \ov{\delta} \lt_{1111})  - \frac{4}{3} \Delta Z_{11} \lt_{1111}  + \mathcal{O}(v^2/\Mm^2)
\end{align}
where
\begin{align}
  \lt_{1111} &= \frac{\partial^4 V^{(0)}}{\partial h^4}
  = \frac{3}{4} c_{2\beta}^2 (g_Y^2 + g_2^2), \\
\ov{\delta} \lt_{1111}  &=       \frac{\partial^4 \ov{V}^{(1)}}{\partial h^4} + \Delta_{\mathrm{reg}} \lt_{1111}.\nn
\end{align}
$ \Delta_{\mathrm{reg}}\lt_{1111}$ is a shift due to changing between the $\drbar$ and $\msbar$ schemes, given e.g.~in \cite{Bagnaschi:2014rsa} or the general expressions in \cite{Martin:1993yx}. We have checked that, when using the general formulae in the appendix for the self-energies and derivatives of the one-loop effective potential, we can reproduce the matching condition from \cite{Bagnaschi:2014rsa} -- after accounting for the different definitions of the electroweak gauge couplings in the tree-level part.

The above illustrates the equivalence between the pole-matching procedure and the EFT calculation for the MSSM matching to the SM, and is much simpler than an explicit term-by-term derivation in e.g.~\cite{Athron:2016fuq}.

\subsection{Dirac gauginos}

In the context of matching a heavy theory onto the SM, Dirac gaugino models are particularly interesting because they contain both singlet and triplet scalars, which are the most general possibilities for the presence of a coupling $a_{I11}$ at $\mathcal{O}(\Ml^0) $ with a SM doublet: $SU(2)$ gauge invariance forbids other representations (although in the most general case we would also be allowed triplets carrying hypercharge $\pm 1$). Moreover, in many scenarios a hierarchy between the singlet/triplet states and the Higgs is natural, which comes from a large Dirac gaugino mass, so an EFT approach to the Higgs mass calculation is particularly appropriate. Indeed first attempts were made in this direction in \cite{Benakli:2013msa,Benakli:2015ioa,Benakli:2018vqz}; in \cite{Benakli:2013msa,Benakli:2015ioa} a Dirac-gaugino model was matched onto the SM -- without (most) threshold corrections -- while in \cite{Benakli:2018vqz} the Minimal Dirac Gaugino Supersymmetric Standard Model (MDGSSM) and Minimal R-symmetric Supersymmetric Standard Model (MRSSM) were matched onto the THDM, giving one-loop threshold corrections in the limit that the Dirac gaugino masses were small. Here we shall consider the one-loop threshold corrections of the MDGSSM matching onto the SM plus higgsinos in the limit that the Dirac gaugino masses are large. 

Using the conventions and choices of \cite{Benakli:2018vqz} where we take an approximate R-symmetry to hold, the theory consists of the MSSM superfields plus additional adjoint chiral superfields, namely a (complex) singlet $\mathbf{S}$, a triplet of $SU(2)$ $\mathbf{T} $ and an octet of $SU(3)$ $\mathbf{O}$, all having no hypercharge, and superpotential 
\begin{eqnarray}
W_{\text{Higgs}} =  \mu \,  \mathbf{H_u} \cdot \mathbf{H_d}+ \lambda_S \mathbf{S} \, \mathbf{H_u} \cdot \mathbf{H_d} + 2 \lambda_T \, \mathbf{H_d} \cdot \mathbf{T} \mathbf{H_u} \label{W_Higgs}
\end{eqnarray} 
in addition to the usual Yukawa coupling terms. These are supplemented by standard soft terms
\begin{align}
\lagr_{\rm standard\ soft} =& -m_{H_u}^2 |H_u|^2 - m_{H_d}^2 |H_d|^2 - B_{\mu} (H_u \cdot H_d + \text{h.c}) \ \\ 
 & - \bigg(\frac{1}{\sqrt{2}} t_S S + h.c.\bigg) - m_S^2 |S|^2 - 2 m_T^2 \text{tr} (T^{\dagger} T)\nn\\
 &- \frac{1}{2} B_S \left(S^2 + h.c\right)-  B_T\left(\text{tr}(T T) + h.c.\right)  - m_O^2 |O|^2 - B_O\left(\text{tr}(O O) + h.c.\right)\,, \nn
\end{align} 
as well as \emph{supersoft} operators $m_{Di }\theta^\alpha$ for Dirac masses
\begin{align}
\int d^2\theta \left[  \sqrt{2} \, m_{DY} \theta^\alpha \mathbf{W}_{1\alpha} \mathbf{S} + 2 \sqrt{2} \, m_{D2}\theta^\alpha \text{tr} \left( \mathbf{W}_{2\alpha} \mathbf{T}\right)  +  2 \sqrt{2} \, m_{D3}\theta^\alpha \text{tr} \left( \mathbf{W}_{3\alpha} \mathbf{O}\right)\! \right]\,,
\end{align} 
where $\mathbf{W}_{i\alpha}$ are the gauge field-strength superfields. 
We shall take for simplicity $\mu \ll \Mm \sim m_{Di} \sim \sqrt{B_\mu}$, which also requires $t_S \ll \Mm^3 $, and assume that CP is conserved. We shall also neglect any trilinear soft terms such as $S^3, ST^2$ (even though these are not forbidden by any symmetry) both for simplicity, and  because they are typically found to be very small in gauge mediation scenarios \cite{Benakli:2016ybe}.

As stated above, this model has almost all of the interesting ingredients that differentiate it from the MSSM in the matching: the singlet $S$ and the triplet scalars $T$ split into scalar and pseudoscalar pieces 
\begin{align}
S = \frac{1}{\sqrt{2}} (v_S + S_R + i S_I), \qquad T^a = \frac{1}{\sqrt{2}} (T_{P}^a + i T_M^a) 
\end{align}
with masses 
\begin{align}
m_{SR}^2 &= m_S^2 + B_S + 4 m_{DY}^2 , \qquad\ m_{SI}^2 = m_S^2 - B_S, \\
m_{TP}^2 &= m_T^2 + B_T + 4 m_{D2}^2 , \qquad m_{TM}^2 = m_T^2 - B_T. 
\end{align}
The neutral scalar component of the triplet $T_P^0$ and the scalar component of the singlet $S_R$ can then mix with the light Higgs after electroweak symmetry breaking. Hence both these fields have trilinear couplings with the light Higgs of the form $a_{I11}$: working in terms of complex fields $H \equiv \frac{1}{\sqrt{2}} ( h + i G^0)$ where $h$ is the neutral Higgs and $G^0$ the would-be Goldstone boson (there is no expectation value because we work in the basis before electroweak symmetry breaking), we have
\begin{align}
a^{S_R H}_{H} &= - g_Y m_{DY} c_{2\beta}   \nn\\
a^{T^0_P H}_H &= g_2 m_{D2} c_{2\beta} .                  
\end{align}
The triplet cannot obtain an expectation value before electroweak symmetry breaking. However, while at tree level we can take the singlet to have no expectation value, at one loop there is an unavoidable tadpole and the quantum tadpole equation becomes
\begin{align}
0=&\  m_{SR}^2 v_S + t_S  + \delta t_S,                                    
\label{EQ:DiracTadpole}
\end{align}
where as before $t_S$ is the tree-level tadpole (which we are assuming is small). The simplest option to deal with this is to adjust the (supersymmetry-breaking) tadpole term to ensure that $v_S = 0$. Indeed, if we are working in a model where parameters such as the singlet tadpole and sfermion masses are not specified from the bottom up, then this is acceptable. However, in other cases we must choose one of the options 3 or 4 from section \ref{sec:effectiveaction}. If we take option 3 (i.e.~we take $v_S \simeq 0$ to be the VEV of the tree-level potential), then since we neglect $\mu$ the only important cubic coupling is $a_{S_R H}^{\HH} $, where $\HH$ is the neutral component of the heavy Higgs doublet:
\begin{align}
  \mathcal{L} \supset& -g_Y m_{DY} s_{2\beta} S_R ( H \ov{\HH} + \ov{H} \HH) .
\end{align}
Then the mass mixing term becomes
\begin{align}
  (\ov{\delta} m^2)_H^{\HH} \rightarrow &\  (\ov{\delta} m^2)_H^{\HH} + g_Y m_{DY} s_{2\beta} \frac{\delta t_S}{m_{SR}^2} .
\label{EQ:DiracMassShift}\end{align}
This potentially provokes a change in $\tan \beta$. However, this shift is simply absorbed into the counterterm if we use the choice (\ref{EQ:MAINCTCHOICE}). For the shifts to cubic couplings, we note that there is no quartic coupling $\lt_{S_R H}^{HH}$ or $\lt_{S_R HH}^H $, but there is a coupling $\lt_{S_R S_R H}^H = \lambda_S^2$, therefore
\begin{align}
a^{S_R H}_H \rightarrow& - g_Y m_{DY} c_{2\beta} -\lambda_S^2  \frac{\delta t_S}{m_{SR}^2}  
\label{EQ:DiracTrilinearShift}
\end{align}
and finally we find that our expression for the Higgs quartic is
\begin{align}
  \lambda_{HH}^{HH} =&\ (1 - 2 \delta Z_H^H ) \bigg(\frac{1}{2} (g_Y^2 + g_2^2) c_{2\beta}^2 + (\lambda_S^2 + \lambda_T^2) s_{2\beta}^2\bigg)  \nn\\
  &- 2c_{2\beta}^2 \bigg(\frac{g_Y^2 m_{DY}^2}{m_{SR}^2} ( 1 - \frac{\ov{\delta} m_{SR}^2}{m_{SR}^2})  + \frac{g_2^2 m_{D2}^2}{m_{TP}^2} ( 1 - \frac{\ov{\delta} m_{TP}^2}{m_{TP}^2}) \bigg)  \nn\\
                     & +\ov{\delta} \lt_{HH}^{HH} - \frac{4}{m_{SR}^2} a^{S_R H}_{H} \ov{\delta} a^{S_R H}_H -  \frac{4}{m_{TP}^2} a^{T_P^0 H}_{H} \ov{\delta} a^{T_P^0 H}_H \nn\\
  & - 4 \frac{c_{2\beta} \lambda_S^2 g_Y m_{DY} }{m_{SR}^4}\delta t_S - \frac{\kappa}{2} (g_Y^2 + 3 g_2^2 + 2g_Y^2 g_2^2).  
    \label{EQ:DiracMatching}\end{align}
  The final term accounts for the conversion from $\drbar$ to $\msbar$: all of the quantities on the right hand side are expressed in terms of $\drbar$ values.
The expressions for all of the loop quantities are given in appendix \ref{APP:DIRAC}. Note that if we used option 2 from section \ref{sec:effectiveaction} then we would obtain the same result but with $\delta t_S =0$.
On the other hand, if we use option 4 then the above shift (\ref{EQ:DiracMassShift}) in $(m^2)_H^{\HH}$ is automatically transferred into the definition of $\tan \beta$. However, we must treat $v_S$ to be small and non-vanishing, and thus we would need to compute all of the loop functions with modified couplings (which would not affect the quartics, but would affect the cubic couplings, fermion masses etc.). Since we still treat $t_S$ as small, however, we can regard $v_S$ as being of one-loop order, and we obtain exactly the same result as (\ref{EQ:DiracMatching}) once we identify $v_S = - \frac{\delta t_S}{m_{SR}^2}. $

\section{Comparing two approaches to mixing-angle renormalisation}
\label{SEC:TOYMODEL}
A last useful illustration of our results is to compare for a simple toy model the ``perturbative'' and ``non-minimal counterterm'' approaches to the renormalisation of the mixing between light and heavy states. 

We therefore consider a model of 3 scalars, two of them mixing that we call $h_1,\,h_2$ and a third scalar $S$ that does not mix with the other two. We also define two $\mathbb{Z}_2$ symmetries: $\mathbb{Z}_2^A$ under which $h_1,\,h_2$ are charged and $\mathbb{Z}_2^B$ under which only $S$ is charged, i.e.
\begin{equation}
 (h_1,h_2,S)\overset{\mathbb{Z}_2^A}{\longrightarrow}(-h_1,-h_2,S)\quad\text{ and }\quad(h_1,h_2,S)\overset{\mathbb{Z}_2^B}{\longrightarrow}(h_1,h_2,-S)\,.
\end{equation}
With these symmetries, the most general Lagrangian is
\begin{align}
\label{EQ:nondiag_basis}
 \mathcal{L}=&\ \frac{1}{2}\big(\partial^\mu h_i\big)^2+\frac{1}{2}\big(\partial^\mu S\big)^2-\frac{1}{2}m_{ij}^2h_ih_j-\frac{1}{2}m_S^2S^2\nn\\
 &-\frac{1}{24}\lt^{ijkl}h_ih_jh_kh_l-\frac{1}{4}\lt^{ijSS}h_ih_jS^2-\frac{1}{24}\lt^{SSSS}S^4\,.
\end{align}
We define new mass-diagonal states $h,\,H$ and rewrite the Lagrangian as
\begin{align}
 \mathcal{L}=&\ \frac{1}{2}\big(\partial^\mu h\big)^2+\frac{1}{2}\big(\partial^\mu H\big)^2+\frac{1}{2}\big(\partial^\mu S\big)^2-\frac{1}{2}m_h^2h^2-\frac{1}{2}m_H^2H^2-\frac{1}{2}m_S^2S^2\nn\\
 &-\frac{1}{24}\lt^{hhhh}h^4-\frac{1}{6}\lt^{hhhH}h^3H-\frac{1}{4}\lt^{hhHH}h^2H^2-\frac{1}{6}\lt^{hHHH}hH^3-\frac{1}{24}\lt^{HHHH}H^4\nn\\
 &-\frac{1}{4}\lt^{hhSS}h^2S^2-\frac{1}{2}\lt^{hHSS}hHS^2-\frac{1}{4}\lt^{HHSS}H^2S^2-\frac{1}{24}\lt^{SSSS}S^4\,.
\end{align}
We will consider that $H$ and $S$ are heavy fields and we will consider the matching of the quartic coupling $\lambda^{hhhh}$ of the light scalar $h$ in the low-energy theory. 

\subsection{``Perturbative masses'' approach}
We first derive the matching relation for $\lambda^{hhhh}$ in the ``perturbative masses'' approach, as described in section \ref{SEC:MATCHING}. The absence of trilinear couplings in this toy model simplifies greatly the expressions of the matching condition -- see eq.~(\ref{EQ:MatchingPerturbative}) -- and of the different terms contributing to it. Using the general results given in appendix \ref{APP:THRESHOLDS}, we obtain the following IR-safe contributions for the relevant terms
\begin{align}
\label{EQ:terms_pert_matching}
 \delta Z_{ij}&=0\quad\forall i,j\in\{h,H,S\}\,,\nn\\
 \kappa^{-1}\ov\delta m^2_{hH}&=\frac12\lt^{hHHH}A_0(m_H^2)+\frac12\lt^{hHSS}A_0(m_S^2)\,,\nn\\
 \kappa^{-1}\ov\delta \lt^{hhhh}&=\frac{3}{2}(\lt^{hhHH})^2P_{SS}(m_H^2,m_H^2)+\frac{3}{2}(\lt^{hhSS})^2P_{SS}(m_S^2,m_S^2)+3(\lt^{hhhH})^2P_{SS}(0,m_H^2)\,.
\end{align}
The matching condition we find is then
\begin{align}
\label{EQ:match_toymodel_pert}
 \lambda^{hhhh} =\lt^{hhhh}&+\frac{3}{2}\kappa\bigg[(\lt^{hhHH})^2P_{SS}(m_H^2,m_H^2)+(\lt^{hhSS})^2P_{SS}(m_S^2,m_S^2)+2(\lt^{hhhH})^2P_{SS}(0,m_H^2)\bigg]\nn\\
                        &-2\kappa\frac{\lt^{hhhH}}{m_H^2}\bigg[\lt^{hHHH}{A_0(m_H^2)}+\lt^{hHSS}{A_0(m_S^2)}\bigg]\,.
\end{align}

\subsection{``Non-minimal counterterm'' approach}
We may instead choose to use the modified scheme presented in section \ref{SEC:COUNTERTERMS}\footnote{Note that in section \ref{SEC:COUNTERTERMS}, we discussed the choice of counterterm for the heavy masses, however, as there are no trilinear couplings in this model we do not need to worry about this here for the matching condition for the quartic coupling. } to simplify the matching relation by eliminating the mixing term between light and heavy states $\ov\delta m^2_{hH}$ -- see in particular equation~(\ref{EQ:MAINCTCHOICE}). 
In this modified scheme, the one-loop matching condition becomes 
\begin{align}
\label{EQ:match_toymodel_ct}
 \lambda^{hhhh} &=\lt^{hhhh}_\text{c.t.}+\frac{3}{2}\kappa\bigg[(\lt^{hhHH}_\text{c.t.})^2P_{SS}(m_H^2,m_H^2)+(\lt^{hhSS}_\text{c.t.})^2P_{SS}(m_S^2,m_S^2)+2(\lt^{hhhH}_\text{c.t.})^2P_{SS}(0,m_H^2)\bigg]
\end{align}

The subscript ``c.t.'' on the couplings in the high-energy theory indicates that these are computed in this non-minimal counterterm scheme. Indeed the masses and the mixing angle between $h$ and $H$ are modified in the counterterm scheme, which in turn changes the couplings. If the rotation matrix that diagonalises the matrix $m_{\gea\geb}^2$ in the ``perturbative masses'' approach is denoted $R\equiv R(\beta_\text{pert})$ -- i.e.~$R_{\gea\geap} m_{\geap\gebp}^2 R^T_{\gebp\geb}=[\text{diag}(m_h^2,m_H^2)]_{\gea\geb}$ -- then the modified mixing angle is found by diagonalising the matrix
\begin{equation}
 R^T\begin{pmatrix}
     m_h^2 & \ov\delta m^2_{hH}\\
     \ov\delta m^2_{hH} & m_H^2
    \end{pmatrix}
R\,.
\end{equation}
Once we have this modified angle, we can compute couplings in the non-minimal counterterm scheme. Note however that as $\lt_\text{c.t.}^{hhHH}$, $\lt_\text{c.t.}^{hhSS}$, and $\lt_\text{c.t.}^{hhhH}$ only appear in the one-loop correction in eq.~(\ref{EQ:match_toymodel_ct}), the change of scheme for these couplings is only a two-loop effect in the matching -- and only the change in $\lt_\text{c.t.}^{hhhh}$ is relevant at one-loop order.  

\subsection{Numerical example}

\begin{figure}
\centering
 \includegraphics[width=.8\textwidth]{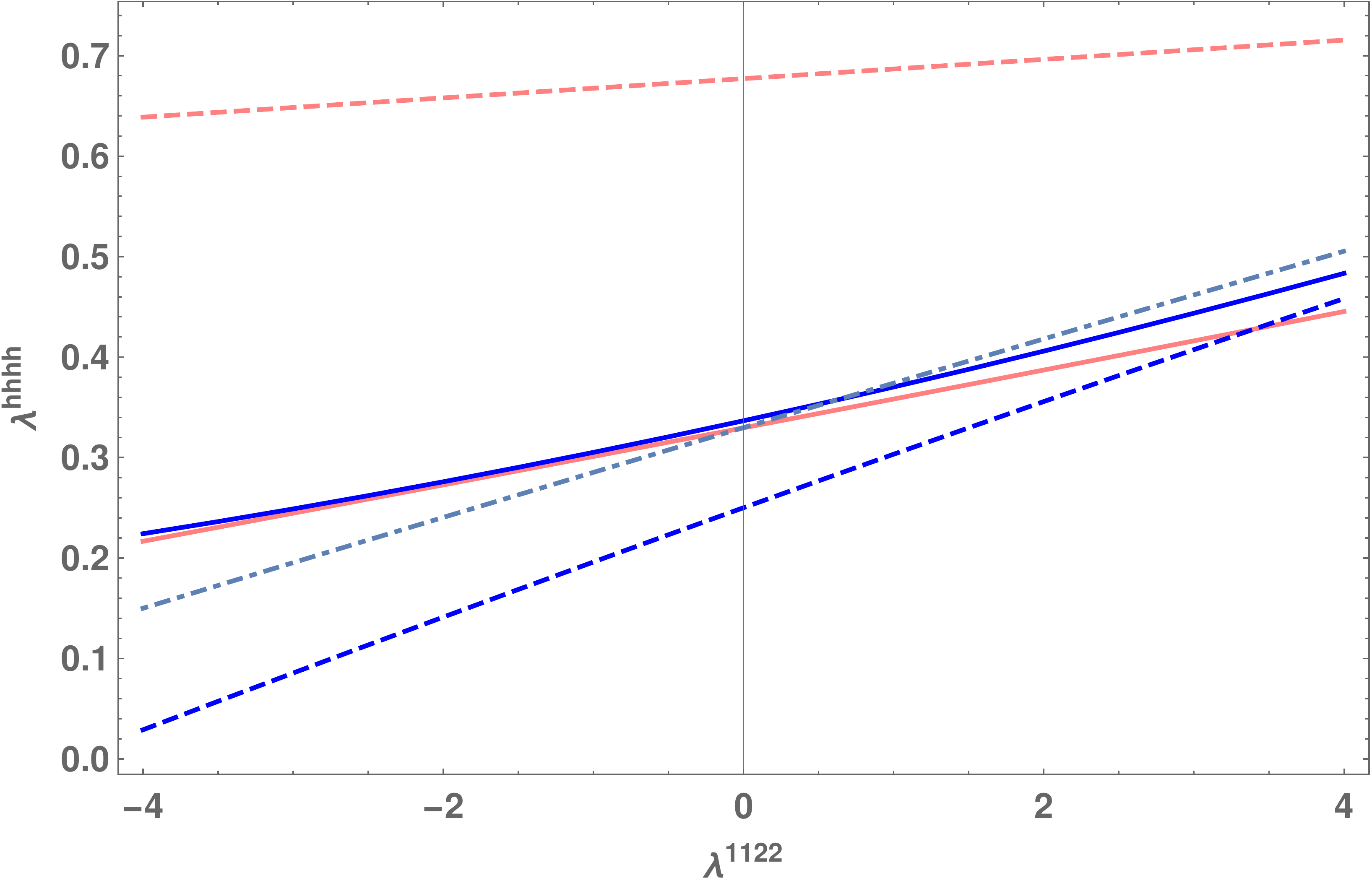}
 \caption{Quartic coupling $\lambda^{hhhh}$ between four light states obtained after integrating out the heavy scalars $H$ and $S$, as a function of $\lambda^{1122}$. Dashed curves show the results obtained at tree level -- i.e.~$\lambda^{hhhh}=\lt^{hhhh}$ -- in the ``perturbative'' (light-red) and ``non-minimal counterterm'' (blue) schemes, while solid and dot-dashed curves are the results at one-loop order in the two schemes -- found using equations~(\ref{EQ:match_toymodel_pert}) and (\ref{EQ:match_toymodel_ct}). For the values in the ``counterterm'' scheme, the (blue) solid and dot-dashed curves differ by the choice of couplings used in the one-loop corrections (see eq.~(\ref{EQ:match_toymodel_ct})): the dot-dashed line corresponds to using couplings computed in the standard ``perturbative'' approach at one loop, and the solid one corresponds to using couplings computed in the modified scheme. The difference between these two choices is formally a two-loop effect. }
 \label{tildelambda_hhhh}
\end{figure}
To compare the different results obtained in the ``perturbative'' and ``non-minimal counterterm'' schemes, we consider the parameter points defined -- in the non-diagonal basis of eq.~(\ref{EQ:nondiag_basis}) -- as 
\begin{align}
\label{EQ:toymodel_param}
 &m_{11}^2=(100\gev)^2,\,m_{12}^2=(400\gev)^2,\,m_{22}^2=(2000\gev)^2,\,m_S^2=(5000\gev)^2,\nn\\
 &\lambda^{1111}=1,\,\lambda^{1112}=2,\,\lambda^{1122}\in[-4,4],\,\lambda^{1222}=1.5,\,\lambda^{2222}=0.5,\nn\\
 &\lambda^{11SS}=0,\,\lambda^{12SS}=3.5,\,\lambda^{22SS}=0\,.
\end{align}
In the above inputs, we have chosen a small value for $m_{12}^2$ with respect to $m_{22}^2$ in order to have a small mixing between $h$ and $H$ at tree level in the ``perturbative'' scheme, and we have also taken large values for $m_S^2$ and $\lambda^{12SS}$ (and thus $\lambda^{hHSS}$) to maximise the effect of the off-diagonal loop-level mixing term proportional to $\ov\delta m_{hH}^2$ in equation~(\ref{EQ:match_toymodel_pert}).
Moreover, from the mass parameters in eq.~(\ref{EQ:toymodel_param}), we can derive the mass eigenvalues to be $m_h=60\gev$ and $m_H^2=2002\gev$, indeed ensuring that our EFT approach of integrating out the heavy mass eigenstate $H$ and the additional scalar $S$ is valid. We will consider that the input values given in equation~(\ref{EQ:toymodel_param}) are defined at renormalisation scale equal to $m_H$.

Figure \ref{tildelambda_hhhh} shows the values that we find for $\lambda^{hhhh}$ respectively in the ``perturbative'' (light-red curves) and the ``non-minimal counterterm'' (blue curves) schemes, at tree level (dashed lines) and one-loop level (solid lines), as a function of the coupling $\lambda^{1122}$ of the non-diagonal basis. At tree level, one can observe a large difference between the quartic couplings obtained in the two schemes.  This can be understood because the mixing between $h$ and $H$ is small at tree level, but the loop-level mixing $\ov\delta m^2_{hH}$ is large, therefore the relative effect of the loop-induced mixing is large and the mixing angle is modified significantly between the two schemes. 

At one-loop, we see that the loop corrections are much larger in the ``perturbative'' scheme than the ``non-minimal counterterm'' scheme; again, this comes from the fact that the loop-level mixing term -- proportional to $\ov\delta m_{hH}^2$ -- is large for the parameter points we considered. However, while the loop corrections differ in magnitude, the one-loop results for $\lambda^{hhhh}$ in the two approaches are close. The differences that appear for increasing $\lambda^{1122}$ can be interpreted as indications of the importance of two-loop corrections. A simple way to estimate the typical size of the two-loop corrections to the matching is to compute the matching relation (\ref{EQ:match_toymodel_ct}) using for the couplings appearing in the one-loop terms the values obtained in the ``perturbative'' scheme -- i.e.~we use equation~(\ref{EQ:match_toymodel_ct}) with $\lt^{hhhh}_\text{c.t.}$, $\lt^{hhHH}$, $\lt^{hhSS}$, and $\lt^{hhhH}$ -- as the difference with using all couplings computed in the ``counterterm'' scheme
is a two-loop effect. 
Doing so, we obtain the dot-dashed curve in figure \ref{tildelambda_hhhh}, which is still close to the result of the ``perturbative masses'' scheme and only differs significantly for large $|\lambda^{1122}|$ -- this indeed confirms missing two-loop corrections as the origin of the difference between the solid curves for $\lambda^{hhhh}$. 

Before ending this section, a final comment is at hand about the choice of inputs and of scheme when integrating out heavy fields. If we had proceeded naively -- or incorrectly -- and had not specified the scheme in which the diagonal-basis couplings are given, or in which they are computed from other inputs (such as in eq.~(\ref{EQ:toymodel_param})), we could have obtained widely different results for $\lambda^{hhhh}$. Indeed for a given value of $\tilde\lambda^{hhhh}$, depending on the scheme that it is considered to be given (or computed) in, the loop corrections that are added to it change drastically -- as we saw in the above.

\section{Threshold corrections to Yukawa couplings}
\label{SEC:YUKAWAS}

Finally\footnote{Scalar trilinears $a_{ijk}$ are of order $\Ml$ and as we have taken the limit $\Ml\to0$, there are no scalar trilinear couplings in our setting and hence no corrections to them either -- note that these corrections could in principle be obtained easily from the results in appendix \ref{app:scalartrilinears} together with the modified loop functions defined in appendix \ref{APP:IRsafe_loopfn}.} we discuss threshold corrections to  Yukawa couplings, which are much simpler than those to quartic scalar couplings. Since we are not considering heavy gauge bosons, there are no contributions to the matching proportional to gauge couplings (as before, provided we use the same renormalisation scheme both above and below the matching scale). However, we must take mixing effects into account:
\begin{align}
y^{IJ\lii}_{EFT} =&\ y^{IJ\lii}_{HET} + \ov{\delta} y^{IJ\lii} + y^{IJ\heK} \delta U_{\heK\lii} - \frac{1}{2} y^{IJ\lik} \delta Z_{\lik\lii} + \bigg[y^{I' J \lii} (\delta U_F)_{I'}^{I} + (I\leftrightarrow J) \bigg].
\label{EQ:MASTERYUKAWA}\end{align}
Here we use capitals $\{I,J\}$ for fermions (see appendix \ref{APP:CONVENTIONS} for all our conventions).
We provide the expressions for $\db y^{IJ\lii}$ in appendix \ref{app:yukawas} and $\delta U_{\heK\lii}$ is given either by (\ref{EQ:NaiveUIj}) in the ``perturbative masses'' approach or $0$ in the ``counterterm'' approach. However, we have so far not discussed \emph{fermion} mixing, which may be important in models e.g.~with heavy top partners, or the FSSM \cite{Benakli:2013msa,Benakli:2015ioa}. The derivation is very similar to the scalar case: we give the corrections to the kinetic and mass terms (in two-component spinor notation)
\begin{align}
\Gamma_{HET} \supset& \ i (\delta Z_F^{HET})_J^I \psi_I \sigma^\mu \partial_\mu \ov{\psi}^J - \bigg(\frac{1}{2} (\delta M^{IJ} + \delta_{ct} M^{IJ})\psi_I \psi_J + h.c. \bigg),
\end{align}
in eq.~(\ref{EQ:FermionSelf}) (or see \cite{Martin:2005ch}), and we can divide the fermions into heavy and light states, then make the identification again $\Delta Z_F \equiv \delta Z_F^{HET} - \delta Z_F^{LET}$ and
\begin{align}
(\psi_I)^{HET} \equiv (U_F)_I^J  (\psi_J )^{EFT} =  (1 - \frac{1}{2} \Delta Z_F + \delta R_F)_I^J \psi_J^{EFT}.
\end{align}
Here $(\Delta Z_F)^\dagger = \Delta Z_F, (\delta R_F)^\dagger = - \delta R_F$.
The difference from scalars is that we should diagonalise the matrix $M_{II'} M^{I' J}$ (note that Dirac fermions do not have diagonal matrices $M_{IJ}$ in two-spinor notation); if we write $\{L, L'\}, \{H, H',H^{\prime \prime}\}$ as indices for light and heavy fermions respectively, then we have
\begin{align}
M_{LL} =&\ M_{LH} = 0, \qquad M_{HH^{\prime \prime}} M^{H^{\prime \prime} H'} = \delta_H^{H'} m_H^2, \nn\\
0 =& - \frac{1}{2} (\Delta Z_F)_{H}^L m_H^2 + (\delta R_F)_{H}^L m_H^2 + (\delta M + \delta_{ct} M)^{LH'} M_{H' H}
\end{align}
which leads to
\begin{align}
(\delta R_F)_{H}^L =&\ \frac{1}{2} (\Delta Z_F)_{H}^L - (\delta M + \delta_{ct} M)^{LH'} \frac{M_{H' H}}{m_H^2} \nn\\
(\delta U_F)_{L'}^{L} =& - \frac{1}{2} (\Delta Z_F)_{L'}^{L} \nn\\
(\delta U_F)_{H}^{L} =& \left\{ \begin{array}{cl} - \delta M^{LH'} \frac{M_{H' H}}{m_H^2} & \mathrm{perturbative\ masses\ approach} \\ 0 & \mathrm{counterterm\ approach} \end{array} \right. .
\end{align}

\section{Outlook}
\label{SEC:CONCLUSIONS}

We have described how to match renormalisable couplings between general theories and explained the different choices that can be made. Our aim is to simplify the calculation of the matching as much as possible, since already at one loop the expressions are rather long; we provide what we expect to be the simplest possible prescription for matching onto the SM using only two-point scalar amplitudes in section \ref{sec:efficient}, and the simplest general prescription in equation (\ref{EQ:SIMPLEST}). 

The logical extension is to pursue our approach(es) at two loops. Beyond one loop, we expect the use of mass counterterms to become more important to simplify the removal of infra-red divergences: in particular, if the hierarchy between \Ml ~and \Mm ~is comparable to or greater than one loop order (so that the scales are highly tuned) then we expect the ``naive perturbative'' approach should break down, because we will not be able to treat the ``light'' states in the loops as massless. Investigating this and its relationship to the Goldstone Boson Catastrophe \cite{Martin:2014bca,Elias-Miro:2014pca,Braathen:2016cqe,Braathen:2017izn} will be the subject of future work.

\section*{Acknowledgements}
We thank Sebastian Pa\ss{}ehr for useful discussions. MDG thanks Florian Staub and Martin Gabelmann for related discussions.
We acknowledge support from French state funds managed by the Agence Nationale de la Recherche (ANR), in the context of the LABEX
ILP (ANR-11-IDEX-0004-02, ANR-10-LABX-63), and MDG and PS acknowledge support from the ANR grant ``HiggsAutomator'' (ANR-15-CE31-0002). 
PS also acknowledges support by  the European Research Council (ERC) under the Advanced Grant “Higgs@LHC” (ERC-2012-ADG 20120216-321133).
During parts of this work, JB was supported by a scholarship from the Fondation CFM and by a fellowship from the Japan Society for the Promotion of Science.

\appendix

\section{Conventions and loop functions}
\label{APP:CONVENTIONS}

We shall work with a theory of scalars, fermions and massless gauge bosons (i.e.~we shall assume for this work that the SM gauge group is not extended). 
The general Lagrangian interaction terms are
\begin{align}
\lag &= \lag_S + \lag_{SF} + \lag_{SV} + \lag_{FV} + \lag_{\rm gauge} + \lag_{S \mathrm{ghost}} .
\end{align}
We use indices $\{i,j,k,l\}$ for general real scalars, $\{I,J,K,L\}$ for Weyl fermions, and $\{a,b,c,d\}$ for gauge bosons. The interactions are
\begin{align}
\lag_S &\equiv - \frac{1}{6} a_{ijk} \Phi_i \Phi_j \Phi_k - \frac{1}{24} \lt_{ijkl} \Phi_i \Phi_j \Phi_k \Phi_l \nn\\
\lag_{SF} &\equiv - \frac{1}{2} y^{IJk} \psi_I \psi_J \Phi_k - \frac{1}{2} y_{IJk} \ov{\psi}^I \ov{\psi}^J \Phi_k \nn\\
\lag_{FV} &\equiv g^{aJ}_I A_\mu^a \ov{\psi}^I \ov{\sigma}^\mu \psi_J \nn\\
\lag_{SV} &\equiv \frac{1}{2} g^{abi} A^a_\mu A^{\mu b} \Phi_i + \frac{1}{4} g^{abij} A^a_\mu A^{\mu b} \Phi_i \Phi_j + g^{aij} A^a_\mu \Phi_i \partial^\mu \Phi_j \nn\\
\lag_{\rm gauge}&\equiv g^{abc} A^a_\mu A^b_\nu \partial^\mu A^{\nu c} - \frac{1}{4} g^{abe} g^{cde} A^{\mu a} A^{\nu b} A_\mu^c A_\nu^d + g^{abc} A^a_\mu c^b \partial^\mu \ov{c}^c \nn\\
\lag_{S \mathrm{ghost}} &\equiv -\frac{1}{2} \xi \hat{g}^{abi} \Phi_i \ov{c}^a c^b .
\end{align}
These differ a little from the conventions of e.g.~\cite{Martin:2003it} because we use the metric $(+,-,-,-)$. With the assumption that the gauge groups are unbroken, $\lag_{S \mathrm{ghost}} =0$. 
The mass terms of fermions are 
$$ \lag_F \supset - \frac{1}{2} M^{IJ} \psi_I \psi_J - \frac{1}{2} M_{IJ} \ov{\psi}^I \ov{\psi}^J$$
where $M^{IJ} = M_{IJ}^*$ is not necessarily diagonal (indeed it cannot be for Dirac fermions) but 
$$ M^{IJ} M_{JK} \equiv \delta^I_K m_I^2.$$

We will also make use of the effective potential $\veff$, which we can expand perturbatively to one-loop order as
\begin{equation}
 \veff=\vtree+\kappa\vone=\vtree+\kappa\big(\vone_S+\vone_F+\vone_V\big)\,,
\end{equation}
where $\vtree$ is the tree-level potential, and $\vone_S$, $\vone_F$, $\vone_V$ are respectively the scalar, fermion, and gauge-boson contributions to the one-loop potential, with the loop factor denoted
\begin{equation} 
\label{EQ:loopfactor}
 \kappa\equiv \frac{1}{16\pi^2}.
\end{equation}

\subsection{One-loop functions}

We shall use loop functions that mostly coincide with those of \cite{Martin:2003qz}:
the one-loop integrals are defined in $d=4-2\epsilon$ dimensions, in terms of Euclidean momenta
\begin{align}
 \mathbf{A}(x)&\equiv C\int\frac{d^dk}{k^2+x}\\
 \mathbf{B}(p^2;x,y)&\equiv C\int\frac{d^dk}{(k^2+x)((p-k)^2+y)}\\
 \mathbf{C_0}(x,y,z)&\equiv C\int\frac{d^dk}{(k^2+x)(k^2+y)(k^2+z)}\\
 \mathbf{D_0}(x,y,z,u)&\equiv C\int\frac{d^dk}{(k^2+x)(k^2+y)(k^2+z)(k^2+u)}
\end{align}
where
\begin{equation}
 C=16\pi^2\frac{\mu^{2\epsilon}}{(2\pi)^d}\,.
\end{equation}
We then define
\begin{align}
Q^2 \equiv 4\pi e^{-\gamma_E} \mu^2, \qquad \blog x \equiv \log x/Q^2.
\end{align}
From these, we use the finite parts, namely
\begin{align}
 A_0(x)&\equiv\lim_{\epsilon\to0}\bigg[\mathbf{A}(x)+\frac{x}{\epsilon}\bigg]=x\big(\llog x-1\big)\nn\\
 B(p^2;x,y)&\equiv\lim_{\epsilon\to0}\bigg[\mathbf{B}(p^2;x,y)-\frac{1}{\epsilon}\bigg]\nn\\
 P_{SS}(x,y)&\equiv-\lim_{\epsilon\to0}\bigg[\mathbf{B}(0;x,y)-\frac{1}{\epsilon}\bigg]=\frac{A_0(x)-A_0(y)}{x-y}=-B(0;x,y)\nn\\
 P_{SS}(x,x)&=\llog x .
\end{align}
The functions $\mathbf{C}_0, \mathbf{D_0}$ are UV-finite, so we can safely take the limit $\epsilon \rightarrow 0$ when there are no IR poles:
\begin{align}
  C_0 (x,y,z) &\equiv \lim_{\epsilon \rightarrow 0} \mathbf{C_0}(x,y,z) = \frac{1}{x-y} \bigg[ P_{SS}(x,z) - P_{SS} (y,z) \bigg] \nn\\
  D_0 (x,y,z,u) &\equiv \lim_{\epsilon \rightarrow 0} \mathbf{D_0}(x,y,z,u) = \frac{1}{y-x} \bigg[ C_0(x,z,u) - C_0 (y,z,u) \bigg].
\end{align}
In the case of coincident masses, one can take the limit as $y\rightarrow x$ in the above. Note that $P_{SS}, C_0, D_0$ are symmetric under permutation of all masses.

Finally, for the kinetic terms we require the derivatives of the $B$ function evaluated at zero external momentum; we denote throughout with a prime the derivative with respect to external momentum squared:
\begin{align}
B^\prime(0;x,y) &\equiv \frac{d}{dp^2} B (p^2;x,y)\bigg|_{p^2=0} = \frac{1}{2(x-y)^3} \bigg[ x^2 - y^2 + 2 xy \log (y/x) \bigg] \nn\\
B^\prime (0;x,x) &= \frac{1}{6x} \nn\\
B^\prime (0;0,x) &= \frac{1}{2x} .
\end{align}

\subsection{Infra-red safe loop functions}
\label{APP:IRsafe_loopfn}

Throughout the text we require infra-red safe loop functions, which can be defined in several ways (as described in section \ref{sec:IRsafety}) but the simplest of which is just using dimensional regularisation. We have
\begin{align}
\ov{P}_{SS} (0,0) &\equiv 0 \equiv \ov{C}_0 (0,0,0 ) \equiv \ov{D}_0 (0,0,0,0 )\nn\\
\ov{C}_0 (0,0,X ) &\equiv   \frac{1}{X} \ov{P}_{SS} (0,X) = \frac{A_0 (X)}{X^2} \nn\\
\ov{D}_0 (0,0,X,Y) &\equiv - \frac{1}{X - Y} \bigg[\ov{C}_0 (0,0,X ) - \ov{C}_0 (0,0,Y ) \bigg] \nn\\
\ov{D}_0 (0,0,X,X) &\equiv \frac{\blog X - 2}{X^2}.  
\label{EQ:IRSAFEDEFINITIONS}\end{align}
If we want to retain the infra-red divergences, we have, noting for example that $\mathbf{B} (0;0,0) = 0 = \frac{1}{\epsilon_{UV}} - \frac{1}{\epsilon_{IR}} + \mathcal{O}(\epsilon)  $:
\begin{align}
P_{SS} (0,0) &= \frac{1}{\epsilon_{IR}} \nn\\
C_0 (0,0,0) &= D_0 (0,0,0,0) = 0 \nn\\
C_0 (0,0,X) &= - \frac{1}{X} \frac{1}{\epsilon_{IR}} + \ov{C}_0 (0,0,X ) \nn\\
D_0 (0,0,X,Y) &= - \frac{1}{X-Y} \bigg[C_0 (0,0,X ) - C_0 (0,0,Y ) \bigg] = - \frac{1}{XY}  \frac{1}{\epsilon_{IR}} + \ov{D}_0 (0,0,X,Y) \nn\\
D_0 (0,0,0,X) &= \frac{1}{X^2} \frac{1}{\epsilon_{IR}} + \ov{D}_0 (0,0,0,X).
\end{align}

\section{One-loop threshold corrections}
\label{APP:THRESHOLDS}

Here we give all of the corrections to all necessary $n$-point functions in the limit of vanishing expectation values and external momenta in a general theory with massless gauge bosons. For the matching procedure we need to compute these in the high-energy theory and use them as described in the body of the paper.  

\subsection{Two-point couplings}

\subsubsection{Scalar self-energies}

The full expressions for scalar self-energies at one loop were given, for example, in \cite{Martin:2003it}. Here we give the zero-momentum limit:
\begin{align}
\kappa^{-1}\Pi_{ij} (0) =& \ \frac{1}{2} \lt_{ijxx} A_0 (m_x^2)  + \frac{1}{2} a_{ixy} a_{jxy} P_{SS} (m_x^2,m_y^2) - \xi g^{aik} g^{ajk} A_0 (m_k^2)\nn\\
&-\mathrm{Re} [ y^{KLi} y_{KLj} ] G (0; m_K^2, m_L^2) -2 \mathrm{Re} [y^{KLi} y^{K'L'j} M_{KK'} M_{LL'}] P_{SS} (m_K^2, m_L^2) ,  \nn\\
\kappa^{-1}\Pi_{ij}^\prime (0) =& -\frac{1}{2}a_{ixy} a_{jxy} B^\prime (0;m^2_x,m^2_y) +g^{aik} g^{ajk}[\frac{1}{2} (\xi + 5) - (3-\xi) \blog m_{k}^2 ] \nn\\
& -\mathrm{Re} [ y^{KLi} y_{KLj} ] G^\prime (0; m_K^2, m_L^2) +2 \mathrm{Re} [y^{KLi} y^{K'L'j} M_{KK'} M_{LL'}] B^\prime (0;m_K^2, m_L^2)  ,
\label{EQ:ScalarSE}\end{align}
where 
\begin{align}
G (p^2; x,y) \equiv&\ (p^2 - x -y ) B (p^2; x,y) + A_0(x) + A_0(y).
\end{align}
We have included the gauge dependent parts, although we do not need them for $\Pi_{ij}^\prime$ because they will be the same in both the low- and high-energy theories.

\subsubsection{Fermion self-energies}

The full expressions for fermion self-energies at one-loop were given, for example, in \cite{Martin:2005ch}. Here we simply state the formulae that we need: the zero-momentum and zero gauge coupling limit contributions to the effective action terms
\begin{align}
\Gamma \supset& \ i (\delta Z_F)_J^I \psi_I \sigma^\mu \partial_\mu \ov{\psi}^J - \bigg(\frac{1}{2} \delta M^{IJ} \psi_I \psi_J + h.c. \bigg),
\end{align}
and we find
\begin{align}
(\delta Z_F)_J^I|_{g\rightarrow 0} &= \kappa y^{IKi} y_{JKi} B_1 (0; m_K^2, m_i^2),\nn\\
\delta M^{IJ} |_{g\rightarrow 0} &= \kappa y^{IKi} y^{JK'i} M_{KK'} P_{SS} (m_K^2, m_i^2),
\label{EQ:FermionSelf}\end{align}
where 
\begin{align}
B_1(0; x,y) =&\ \frac{1}{2} [(x -y) B^\prime (0;x,y) - P_{SS}(x,y) ].
\end{align}

\subsubsection{Gauge-boson self-energies}
\label{app:gaugebosons}

In the absence of heavy gauge bosons, the threshold corrections to gauge boson self-energies come only from fermions and scalars and are given by
\begin{align}
\Delta \Pi_{ab} (p^2) =&\ \Pi_{ab}^{HET} (p^2) -\Pi_{ab}^{EFT} (p^2) = \hat{\Pi}_{ab}^{HET} (p^2) -\hat{\Pi}_{ab}^{EFT} (p^2), \\
\kappa^{-1}\hat{ \Pi}_{ab} (p^2) \equiv&\ 2 g^{aij} g^{bij} \tilde{B}_{22} (p^2; m_i^2, m_j^2)+ 2\mathrm{Re} (g^{aI}_J g^{bI'}_{J'}M_{II'} M^{JJ'}) B (p^2; m_I^2, m_J^2) - g^{aI}_J g^{bJ}_I H_0 (p^2; m_I^2, m_J^2) .\nn 
\end{align}
The hat indicates that the pure gauge parts have already been removed.
The definitions for the functions $H_0$ and $\tilde{B}_{22}$ are the same as those of PBMZ \cite{Pierce:1996zz} and, as they are long, we do not repeat them here. 
The required limits for the general case are
\begin{align}
\tilde{B}_{22} (0; m_i^2, m_j^2) &=  \frac{1}{4} (m_i^2-m_j^2)^2 B^\prime (0;m_i^2,m_j^2).\nn\\
H_0 (0; m_I^2, m_J^2) &= (m_I^2 - m_J^2)^2 B^\prime (0;m_I^2, m_J^2)  +( m_I^2 + m_J^2) P_{SS}(m_I^2, m_J^2) .
\end{align}
These identities can be used to prove (\ref{EQ:PiZequivalence}) at one loop. 
In the limit of an unbroken gauge group, we have 
\begin{align}
 \hat{\Pi}_{ab} (p^2) &= \kappa g^2 \delta^{ab} \bigg[ 2 S_2 (i) \tilde{B}_{22} (p^2; m_i^2, m_i^2)-   S_2 (I) \bigg( 2 m_I^2 B (p^2; m_I^2, m_I^2)+ H_0 (p^2; m_I^2, m_I^2) \bigg)\bigg] , \nn\\
\hat{\Pi}_{ab} (0) &= 0\,, \nn\\
\hat{\Pi}_{ab}^\prime (0) &=\kappa g^2 \delta^{ab} \bigg[ \frac{1}{6} S_2 (i) \blog  m_i^2 + \frac{2}{3} S_2 (I) \blog m_I^2 \bigg],
\end{align}
where $S_2$ is the Dynkin index of the representation of the scalars or fermions, and $g$ is the gauge coupling for the unbroken gauge group. The final expression gives the well-known one-loop corrections to gauge thresholds:
\begin{align}
g_{EFT}^2 \delta_{ab} = g^2_{HET}\delta_{ab} + g^2 \bigg[(\hat{\Pi}^\prime_{ab} (0) )^{HET} - (\hat{\Pi}^\prime_{ab}(0)  )^{EFT} \bigg].
\end{align}

\subsection{Three-point couplings} 

The only relevant three-point couplings that we need are cubic scalar couplings and Yukawa couplings, all those involving gauge bosons just being given by the gauge couplings. 

\subsubsection{Scalar couplings}
\label{app:scalartrilinears}
For a term in the effective action 
\begin{align}
\Gamma \supset - \frac{1}{6} \delta a_{ijk} \Phi_i \Phi_j \Phi_k
\end{align}
recall that we have
\begin{align}
\delta a_{ijk} = \kappa \frac{\partial^3 \vone}{\partial \Phi_i \partial \Phi_j \partial \Phi_k}, \qquad \vone \equiv \vone_S + \vone_F + \vone_V.
\end{align}
Then 
\begin{align}
\frac{\partial^3 \vone_S}{\partial \Phi_i \partial \Phi_j \partial \Phi_k}  =& \  \frac{1}{4} \lt_{ijxy} a_{kxy} P_{SS} (m_x^2,m_y^2) + \frac{1}{6} a_{kxy} a_{jyz} a_{izx}C_0 (m_x^2,m_y^2,m_z^2) + (ijk)\,, \\
\frac{\partial^3 \vone_F}{\partial \Phi_i \partial \Phi_j \partial \Phi_k} =& - \frac{2}{3} \mathrm{Re}( y^{IJi} y^{J'Kj} y^{K'I'k} M_{II'} M_{JJ'} M_{KK'}) C_0(m_I^2,m_J^2,m_K^2) \nn\\
& -2 \mathrm{Re} (y^{IJi} y^{J'Kj} y_{KIk} M_{JJ'}) F_3 (m_I^2,m_J^2,m_K^2) + (ijk)\,,
\label{EQ:threeptcontrib_fermion}
\end{align}
where
\begin{align}
 F_3 (m_I^2,m_J^2,m_K^2) \equiv & \lim_{\epsilon\to0}\bigg[C\int d^dk (-k^2)\left(\prod_{\mathcal{I}\in\{I,J,K\}}\frac{1}{k^2+m_\mathcal{I}^2}\right)+\frac1\epsilon\bigg]\nn\\
 =&\ m_I^2 C_0(m_I^2,m_J^2,m_K^2) + P_{SS} (m_J^2,m_K^2) .
\end{align}
Note that the function $F_3$ could also have been written in a form where it is manifestly symmetric under the exchange of any two of its arguments. It only has an infra-red divergence for all three arguments vanishing, so we can define
\begin{align}
\ov{F}_3 (0,0,0) =0; \qquad \ov{F}_3 (x,y,z) = F_{3} (x,y,z) \qquad (x,y,z) \ne (0,0,0). 
\end{align}

Finally for contributions from massless gauge bosons:
\begin{align}
\frac{\partial^3 \vone_V}{\partial \Phi_i \partial \Phi_j \partial \Phi_k} &=- \frac{1}{2} \xi g^{ail} g^{ajm} a_{lmk} P_{SS}(m_l^2, m_m^2) + (ijk) + \mathcal{O}(g^3).
\end{align}
The $\mathcal{O}(g^3) $ pieces automatically cancel between high- and low-energy theories, so we do not include them. 

\subsubsection{Yukawa couplings}
\label{app:yukawas}

With our assumptions of having no heavy gauge bosons, the only vertex corrections to Yukawa couplings come from triangle diagrams with scalars and fermions in the loop.
The result is that 
\begin{align}
\Gamma \supset& - \frac{1}{2}\delta y^{IJi} \psi_I \psi_J \Phi_i , \nn\\
\kappa^{-1} \ov{\delta} y^{IJi} = &\ M_{KK'} a_{ijk} y^{KIk} y^{K'Jj}C_0 (m_j^2, m_k^2, m_K^2) - y^{KIk} y^{K'Jk} y_{KK'i} \ov{F}_3 (m_k^2, m_K^2, m_{K'}^2)  \nn\\
& + M_{KL} M_{K'L'} y^{LL'i} y^{KIk} y^{K'Jk} C_0(m_k^2, m_K^2 ,m_{K'}^2) .
\end{align}
To find the matching condition, we need to supplement this with corrections from the scalar and fermion self-energies and insert them all in equation (\ref{EQ:MASTERYUKAWA}). Note that the infra-red divergences are much more simply tamed than in the pure scalar couplings: the difference between the HET and EFT is automatically infra-red safe and corresponds just to replacing $F_3 \rightarrow \ov{F}_3$ in the HET. The other two terms are always infra-red safe, because the first term can only diverge for the case of $j,k$ both light fields, so the coupling $a^{ijk}$ must vanish when $i$ is also light; while the last term has mass prefactors that vanish for light fermions.

\subsection{Four-point couplings}

For a term in the effective action 
\begin{align}
\Gamma \supset - \frac{1}{24} \delta \lt_{ijkl} \Phi_i \Phi_j \Phi_k \Phi_l
\end{align}
recall that we have
\begin{align}
\delta \lt_{ijkl} = \kappa \frac{\partial^4 \vone}{\partial \Phi_i \partial \Phi_j \partial \Phi_k \partial \Phi_l}, \qquad \vone \equiv \vone_S + \vone_F + \vone_V.
\end{align}
Then
\begin{align}
\frac{\partial^4 \vone_S}{\partial \Phi_i \partial \Phi_j \partial \Phi_k \partial \Phi_l}
=&  \ \frac{1}{16} \lt_{ijxy} \lt_{klxy} P_{SS} (m_x^2,m_y^2) + \frac{1}{4} \lt_{ijxy} a_{kyz} a_{lzx} C_0 (m_x^2,m_y^2,m_z^2)\nn\\
&- \frac{1}{8} a_{ixy} a_{jyz} a_{kzu} a_{lux} D_0(m_x^2,m_y^2,m_z^2,m_u^2)  + (ijkl). 
\end{align}

\begin{align}
\label{EQ:fourptcontrib_fermion}
\frac{\partial^4 \vone_F}{\partial \Phi_i \partial \Phi_j \partial \Phi_k \partial \Phi_l} =&\ \frac{1}{2} \mathrm{Re}( y^{IJi} y^{J'Kj} y^{K'Lk} y^{L'I'l} M_{II'} M_{JJ'} M_{KK'} M_{LL'}) D_0(m_I^2,m_J^2,m_K^2,m_L^2) \nn\\
& + 2 \mathrm{Re} (y^{IJi} y^{J'Kj} y^{K'Lk} y_{LIl} M_{JJ'} M_{KK'}) F_4 (m_I^2,m_J^2,m_K^2,m_L^2) \nn\\
& + \mathrm{Re} (y^{IJi} y^{J'Kj} y_{KLk} y_{L'Il} M_{JJ'} M^{LL'}) F_4 (m_I^2,m_J^2,m_K^2,m_L^2)\nn\\
& + \frac{1}{2}\mathrm{Re} (y^{IJi} y_{JKj} y^{KLk} y_{LIl}) H_4 (m_I^2,m_J^2,m_K^2,m_L^2)+ (ijkl)\,,
\end{align}
where $F_4$ and $H_4$ are defined in terms of Euclidean momenta as
\begin{align}
F_4(m_I^2,m_J^2,m_K^2,m_L^2) \equiv&\ \lim_{\epsilon \rightarrow 0} C\int d^d k (-k^2) \left(\prod_{\mathcal{I}\in\{I,J,K,L\}} \frac{1}{k^2 + m_\mathcal{I}^2}\right) \nn\\
=&\ m_I^2D_0(m_I^2,m_J^2,m_K^2,m_L^2) - C_0 (m_J^2,m_K^2,m_L^2)\nn\\
H_4 (m_I^2,m_J^2,m_K^2,m_L^2)\equiv&\ \lim_{\epsilon \rightarrow 0} \bigg[ C\int d^d k (k^4) \left(\prod_{\mathcal{I}\in\{I,J,K,L\}} \frac{1}{k^2 + m_\mathcal{I}^2}\right) - \frac{1}{\epsilon}\bigg]\nn\\
=&- P_{SS} (m_K^2,m_L^2)+\big(m_I^2 + m_J^2\big)F_4 (m_I^2,m_J^2,m_K^2,m_L^2)\nn\\
&-m_I^2 m_J^2 D_0 (m_I^2,m_J^2,m_K^2,m_L^2)\,,\nn
\label{EQ:defH4}
\end{align}
and, as before, $(ijkl)$ denotes the 24 possible permutations of $\{i,j,k,l\}$. Note that to pass to infra-red safe expressions we replace $P_{SS}, C_0, D_0$ by $\ov{P}_{SS}, \ov{C}_0, \ov{D}_0$ in the above.

Finally the contributions from massless gauge bosons are:
\begin{align}
\frac{\partial^4 \vone_V}{\partial \Phi_i \partial \Phi_j \partial \Phi_k \partial \Phi_l}\bigg|_{\mathcal{O}(g^2)} =&-\frac{1}{4}\xi g^{aim} g^{ajn} \lt_{mnkl} P_{SS} (m_m^2, m_n^2) \nn\\
&- \frac{1}{2}\xi g^{aim} g^{ajr} a_{mnk} a_{nrl} C_0 (m_m^2,m_n^2,m_r^2) + (ijkl).
\end{align}
We omit the terms of higher order in the gauge coupling, which automatically cancel between high- and low-energy theories except when they are given in different schemes.

\section{Cancellation of infra-red divergences}
\label{APP:CANCELIR}

In this appendix we explicitly show the cancellation of infra-red divergences in the matching of quartic scalar couplings. We start by considering the case of purely scalar contributions to the different terms in the matching, before turning to the case of fermionic contributions in appendix~\ref{APP:noIRdiv_fermions}. We have already demonstrated the complete cancellation of contributions from gauge interactions in section \ref{sec:gaugedep}.

For the diagrams with only scalars in the loops, let us first summarise the infra-red divergent parts of the necessary quantities in the high-energy theory:
\begin{align}
\kappa^{-1}\delta m_{\heI\heJ}^2 =&\ \frac{1}{2} a_{\heI \lix\liy} a_{\heJ \lix\liy} P_{SS} (m_\lix^2,m_\liy^2) + \mathrm{IR\ safe}, \nn\\
\kappa^{-1}\delta a_{\heI\lij\lik} =&\ \frac{1}{2} a_{\heI\lix\liy} a_{\lij\liy\heJ} a_{\lik\heJ\lix} C_0 (m_\lix^2,m_\liy^2,m_\heJ^2) + \frac{1}{4} \tilde\lambda_{\lij\lik\lix\liy} a_{\heI\lix\liy} P_{SS} (m_\lix^2,m_\liy^2) + ( \lij \leftrightarrow \lik) + \mathrm{IR\ safe} , \\
\kappa^{-1} \delta \lt_{\lii\lij\lik\lil} =&\ \frac{1}{16} \lt_{\lii\lij\lix\liy} \lt_{\lik\lil\lix\liy} P_{SS} (m_\lix^2,m_\liy^2) + \frac{1}{4} \lt_{\lii\lij \lix\liy} a_{\lik\liy\heJ} a_{\lil\heJ\lix} C_0 (m_\lix^2,m_\liy^2,m_\heJ^2) \nn\\
&- \frac{1}{4} a_{\lii\lix\heJ} a_{\lij\heJ\liz} a_{\lik\liz\heK} a_{\lil\heK\lix} D_0 (m_\lix^2,m_\heJ^2,m_\liz^2,m_\heK^2) +  (\lii\lij\lik\lil)  + \mathrm{IR\ safe} . \nn
\end{align}
These must cancel against the calculation of $\delta \lambda_{\lii\lij\lik\lil}$ in the low-energy theory:
\begin{align}
\kappa^{-1}\delta \lambda_{\lii\lij\lik\lil} =&\ \frac{1}{16} P_{SS} (m_\lix^2,m_\liy^2)  \lambda_{\lii\lij\lix\liy} \lambda_{\lik\lil\lix\liy} +  (\lii\lij\lik\lil) \\
=&\ \frac{1}{16} P_{SS}(m_\lix^2,m_\liy^2)\bigg[ \lt_{\lii\lij\lix\liy} \lt_{\lik\lil\lix\liy} -  \frac{2}{m_\heI^2} (a_{\heI\lii\lij} a_{\heI\lix\liy} + 2 a_{\heI\lii\lix} a_{\heI\lij\liy})  \lt_{\lik\lil\lix\liy} \nn\\
& \ \ + \frac{1}{m_\heI^2 m_\heJ^2} \bigg( a_{\heI\lii\lij}  a_{\heJ\lik\lil} a_{\heI\lix\liy} a_{\heJ\lix\liy} + 4a_{\heI\lii\lij} a_{\heI\lix\liy} a_{\heJ\lik\lix} a_{\heJ\lil\liy} + 4a_{\heI\lii\lix} a_{\heI\lij\liy} a_{\heJ\lik\lix} a_{\heJ\lil\liy} \bigg)\bigg] +  (\lii\lij\lik\lil). \nn
\end{align}
Collecting these together in the matching relation for the scalar quartic coupling, equation (\ref{EQ:MatchingPerturbative}), one finds for the potentially divergent terms: 
\begin{align}
\kappa^{-1} \Delta \lambda_{\lii\lij\lik\lil} \supset &\ \frac{1}{4}\lt_{\lik\lil\lix\liy} a_{\heI\lii\lix} a_{\heI\lij\liy} \bigg[ C_0 (m_\lix^2,m_\liy^2,m_\heI^2) + \frac{1}{m_\heI^2} P_{SS} (m_\lix^2,m_\liy^2) \bigg] \nn\\
& - \frac{1}{4} a_{\heI\lii\lix} a_{\heI\lij\liy} a_{\heJ\lik\lix} a_{\heJ\lil\liy}  \frac{1}{m_\heI^{2}} \bigg[ C_0 (m_\lix^2,m_\liy^2,m_\heJ^2) + \frac{1}{m_\heJ^2} P_{SS} (m_\lix^2,m_\liy^2) \bigg] \nn\\
& - \frac{1}{4} a_{\heI\lii\lix} a_{\heI\lij\liy} a_{\heJ\lik\lix} a_{\heJ\lil\liy} \bigg[ D_0 (m_\lix^2,m_\heI^2,m_\liy^2,m_\heJ^2) + \frac{1}{m_\heI^2 m_\heJ^2} P_{SS} (m_\lix^2,m_\liy^2) \bigg]+  (\lii\lij\lik\lil)
\end{align}
The terms in square brackets are all finite as we take the limit $m_\lix,m_\liy \rightarrow 0$, and could be taken as the definitions of the functions $\ov{C}_0 (0,0,X), \ov{D}_0 (0,0,X,Y)$  which agree with our dimensional-regularisation definitions (\ref{EQ:IRSAFEDEFINITIONS}).
Note that these do not give the limiting expressions for $\ov{C}_0(0,0,0), \ov{D}_0 (0,0,0,0)$ which, as mentioned in section \ref{sec:IRsafety}, remain ambiguous but give no net contribution when we subtract the contribution of the LET from that of the HET in the matching.

\subsection{Absence of IR divergences from massless fermions}
\label{APP:noIRdiv_fermions}

We can now show that vanishing fermion masses cause no divergence in the fermion contributions to the matching of three- and four-point functions -- see eqs.~(\ref{EQ:threeptcontrib_fermion}) and (\ref{EQ:fourptcontrib_fermion}). First of all, it should be noted that terms in which all the fermion masses vanish do not pose a problem in the matching because they appear in both the high- and low-energy parts of the matching. 

Then, considering the three-point contribution in eq.~(\ref{EQ:threeptcontrib_fermion}), one can notice immediately that the first term -- of the form $M_{II'}M_{JJ'}M_{KK'}C_0(m_I^2,m_J^2,m_K^2)$ -- cannot be divergent, because by itself the function $C_0$ diverges at most as an inverse mass-squared if all its three arguments tend to 0 -- recall that $C_0(\eps,\eps,\eps)=1/2\eps$. 
For the second term, in the case where $M_{JJ'}$ vanishes, the overall term is also zero, however there remains to verify that $F_3(m_I^2,m_J^2,m_K^2)$ is regular in the limit where $m_I^2$ and $m_K^2$ go to zero (if only one of these two masses is zero, the $P_{SS}$ and $C_0$ functions are not divergent). We have then
\begin{align}
 F_3(\eps,m_J^2,\eps)=&\ \eps C_0(\eps,m_J^2,\eps)+P_{SS}(m_J^2,\eps)\underset{\mathclap{\eps\to0}}{\longrightarrow}\frac{A_0(m_J^2)}{m_J^2}\,,
\end{align}
as we know that $C_0(\eps,\eps,m_J^2)$ diverges as $\log\eps$.

Turning now to the four-point couplings, for which the fermion contribution is given in equation (\ref{EQ:fourptcontrib_fermion}), we have three types of terms to verify. First, the term $M_{II'}M_{JJ'}M_{KK'}M_{LL'}D_0(m_I^2,m_J^2,m_K^2,m_L^2)$ is not divergent even if all four mass arguments are zero because $D_0(\eps,\eps,\eps,\eps)=1/6\eps^2$. Second, we must consider the terms involving the function $F_4(m_I^2,m_J^2,m_K^2,m_L^2)$: if three (or four) of the masses are zero, the mass prefactors ensure that the contributions to the four-point coupling are not divergent. However, it is necessary to verify what happens when only two mass arguments vanish, say $m_I^2$ and $m_J^2$. We find
\begin{align}
 F_4(\eps,\eps,m_K^2,m_L^2)=&\ \eps D_0(\eps,\eps,m_K^2,m_L^2)-C_0(\eps,m_K^2,m_L^2)\nn\\
 \underset{\eps\to0}{\to}&-C_0(0,m_K^2,m_L^2)\,,
\end{align}
as $D_0(\eps,\eps,m_K^2,m_L^2)$ diverges like $\log\eps$.

There remains to show that $H_4(m_I^2,m_J^2,m_K^2,m_L^2)$ is not divergent when one or several of its arguments are zero. For only one vanishing mass, this is apparent from its definition in eq.~(\ref{EQ:defH4}). Then, we can consider the case of two vanishing masses, say $m_I^2$ and $m_J^2$:
\begin{align}
 H_4(\eps,\eps,m_K^2,m_L^2)=&- P_{SS}(m_K^2,m_L^2)+2\eps F_4(\eps,\eps,m_K^2,m_L^2) - \eps^2 D_0(\eps,\eps,m_K^2,m_L^2)\nn\\
 \underset{\mathclap{\eps\to0}}{\longrightarrow}&-P_{SS}(m_K^2,m_L^2)
\end{align}

The case with three zero masses is also simple to verify, requiring only the intermediate results
\begin{align}
 D_0(\eps,\eps,\eps,m_L^2)&\underset{\eps\to0}{\to}\frac{1}{2 m_L^2\eps}+\frac{\llog\eps}{m_L^4}-\frac{A_0(m_L^2)}{m_L^6}+\frac{1}{2m_L^4}\\
 F_4(\eps,\eps,\eps,m_L^2)&\underset{\eps\to0}{\to}\frac{\llog\eps}{m_L^2}-\frac{A_0(m_L^2)}{m_L^4}+\frac{1}{2m_L^2}
\end{align}
and we find 
\begin{align}
 H_4(\eps,\eps,\eps,m_L^2)=&-P_{SS}(\eps,m_L^2)+2\eps F_4(\eps,\eps,\eps,m_L^2)-\eps^2D_0(\eps,\eps,\eps,m_L^2)\underset{\mathclap{\eps\to0}}{\longrightarrow}-\frac{A_0(m_L^2)}{m_L^2}
\end{align}

Finally, if all four of its mass arguments are zero, the function $H_4(\eps,\eps,\eps,\eps)$ does diverge, but this does not cause a problem for the matching because a term with all fermion masses vanishing would appear both in the EFT and the UV-complete sides of the matching condition. 

\section{Threshold corrections to the electroweak gauge couplings}
\label{app:gaugethresholds}
We present in this appendix details about the derivation of the threshold corrections to $g_Y$ and $g_2$ -- the gauge couplings of $U(1)_Y$ and $SU(2)_L$\,, respectively -- given in equation (\ref{EQ:gaugethresholds}). The
radiative corrections to gauge couplings are obtained as the corrections to the gauge-boson kinetic term, i.e.
\begin{equation}
 \frac{1}{4g^2}F_{\mu\nu}F^{\mu\nu}\longrightarrow\frac{1}{4g^2}\left(1+\Delta Z_A\right)F_{\mu\nu}F^{\mu\nu},
\end{equation}
where $A_\mu$ is the gauge boson of some gauge group, $F_{\mu\nu}$ the associated field-strength tensor, and $g$ the gauge coupling. In the above relation we have $\Delta Z_A=-\Pi_{AA}^\prime (0)$, the latter being the derivative with respect to external momentum of the transverse part of the gauge-boson self-energy. 
Applying this to $U(1)_Y$, the threshold correction to $g_Y$ is found with the relation
\begin{align}
 \frac{1}{4g_Y^2}\left(1-\Pi_{BB}^{HET\,\prime}(0)\right)B_{\mu\nu}B^{\mu\nu}=\frac{1}{4(g_Y^2+\Delta_{g_Y^2})}\left(1-\Pi_{BB}^{SM\,\prime}(0)\right)B_{\mu\nu}B^{\mu\nu},
\end{align}
where $B_\mu$ is the gauge boson of $U(1)_Y$ and $B_{\mu\nu}$ the corresponding field-strength, and the self-energies on the left and on the right are computed in the high-energy theory and in the SM, respectively. We thus obtain
\begin{align}
 \Delta_{g_Y^2}=g_Y^2\left[\Pi_{BB}^{HET\,\prime}(0)-\Pi_{BB}^{SM\,\prime}(0)\right].
\end{align}
Similarly, for $SU(2)_L$, we obtain
\begin{align}
 \Delta_{g_2^2}=g_2^2\left[\Pi_{W_3W_3}^{HET\,\prime}(0)-\Pi_{W_3W_3}^{SM\,\prime}(0)\right],
\end{align}
where $W_3$ is the third component of the $SU(2)_L$ gauge boson. Now, expressing the $Z$-boson mass eigenstate in terms of $W_3$ and $B$, we have 
\begin{equation}
 (g_Y^2+g_2^2)\Pi_{ZZ}=g_2^2\Pi_{W_3W_3}-2g_Yg_2\Pi_{W_3B}+g_Y^2\Pi_{BB}.
\end{equation}
Finally, as we need in section \ref{SEC:POLEMATCHING} the gauge threshold corrections in the limit $v\to0$, we know from gauge invariance that in this limit the $\Pi_{W_3B}$ term should vanish. Taking all these intermediate results together, we obtain
\begin{align}
 \Delta_{g_Y^2}+\Delta_{g_2^2}\Big|_{v=0}=(g_Y^2+g_2^2)\bigg[\Pi_{ZZ}^{HET\prime}(0)-\Pi_{ZZ}^{EFT\prime}(0)\bigg]_{v=0},
\end{align}
which is equation (\ref{EQ:gaugethresholds}). 

\section{Dirac gaugino contributions}
\label{APP:DIRAC}

In this appendix we collect the various contributions necessary for matching the Higgs quartic coupling from the MDGSSM assuming that the low-energy theory is the SM plus higgsinos, in the limit that $\mu$ is small compared to $\Mm$ and the other masses. We split each term up according to the fields contributing: the Higgs and $S/T$ scalars $\delta_S$; fermions $\delta_F$; and sfermions (squarks and sleptons) $\delta_{\tilde{f}}$. Hence the corrections to the Higgs quartic in the MDGSSM become 
\begin{align}
\ov\delta \lt_{HH}^{HH} =&\ \ov\delta_S \lt_{HH}^{HH}+ \ov\delta_F \lt_{HH}^{HH}+ \ov\delta_{\tilde{f}}\lt_{HH}^{HH}\,,
\end{align}
and so on for the other corrections appearing in eq.~(\ref{EQ:DiracMatching}); while the derivatives of the self-energies with respect to external momentum are expanded as
\begin{align}
(\Pi^\prime)_H^H (0) =&\ (\Pi^\prime_S)_H^H (0) + (\Pi^\prime_F)_H^H (0) + (\Pi^\prime_{\tilde{f}})_H^H (0).
\end{align}

\subsection{Corrections to the singlet tadpole}

The singlet tadpole term obtains a contribution from the heavy Higgs and the squarks and sleptons
\begin{align}
\kappa^{-1}\delta t_S =&\ m_{DY} g_Y c_{2\beta}  A_0 (m_{\mathcal{H}}^2)+ m_{DY} g_Y \sum_{i=1}^3\bigg[  A_0 ( m_{Q_i}^2) - 2 A_0 ( m_{U_i}^2) + A_0 ( m_{D_i}^2) - A_0 ( m_{L_i}^2) + A_0 ( m_{E_i}^2) \bigg]
\end{align}
where $\mathcal{H}$ is the heavy Higgs doublet, the sum on the last line is over all generations $i$, and $Q, U, D, L, E$ represent the sfermion partners of the left-handed quarks, right-handed up-type quarks, right-handed down-type quarks, left-handed leptons and right-handed leptons. Note that we explicitly set the singlet expectation value $v_S$ to zero at tree level along the lines of option (3).

\subsection{Contributions from Higgs and $S/T$ scalars}

\subsubsection{Corrections to the Higgs quartic coupling}

\allowdisplaybreaks
The scalar contributions to the one-loop Higgs quartic coupling in the MDGSSM read

\begin{align}
  \kappa^{-1}\ov\delta_S \lt^{HH}_{HH}=&-4g_Y^4m_{DY}^4\big[D_0(m_{\mathcal{H}}^2,m_{\mathcal{H}}^2,m_{SR}^2,m_{SR}^2)s_{2\beta}^4+2D_0(0,m_{\mathcal{H}}^2,m_{SR}^2,m_{SR}^2)c_{2\beta}^2s_{2\beta}^2\nn\\
  &\hspace{7cm}+\ov{D}_0(0,0,m_{SR}^2,m_{SR}^2)c_{2\beta}^4\big]\nn\\
  &-8g_2^2g_Y^2m_{D2}^2m_{DY}^2\big[D_0(m_{\mathcal{H}}^2,m_{\mathcal{H}}^2,m_{SR}^2,m_{TP}^2)s_{2\beta}^4+2D_0(0,m_{\mathcal{H}}^2,m_{SR}^2,m_{TP}^2)c_{2\beta}^2s_{2\beta}^2\nn\\
  &\hspace{7cm}+\ov{D}_0(0,0,m_{SR}^2,m_{TP}^2)c_{2\beta}^4\big]\nn\\
  &-12g_2^4m_{D2}^4\big[D_0(m_{\mathcal{H}}^2,m_{\mathcal{H}}^2,m_{TP}^2,m_{TP}^2)s_{2\beta}^4+2D_0(0,m_{\mathcal{H}}^2,m_{TP}^2,m_{TP}^2)c_{2\beta}^2s_{2\beta}^2\nn\\
  &\hspace{7cm}+\ov{D}_0(0,0,m_{TP}^2,m_{TP}^2)c_{2\beta}^4\big]\nn\\
  &+4g_Y^2m_{DY}^2\lambda_S^2\big[s_{2\beta}^2C_0(m_{\mathcal{H}}^2,m_{SR}^2,m_{SR}^2)+c_{2\beta}^2C_0(0,m_{SR}^2,m_{SR}^2)\big]\nn\\
 &-8g_2g_Ym_{D2}m_{DY}\lambda_S\lambda_T\big[s_{2\beta}^2C_0(m_{\mathcal{H}}^2,m_{SR}^2,m_{TP}^2)+c_{2\beta}^2C_0(0,m_{SR}^2,m_{TP}^2)\big]\nn\\
 &+12g_2^2m_{D2}^2\lambda_T^2\big[s_{2\beta}^2C_0(m_{\mathcal{H}}^2,m_{TP}^2,m_{TP}^2)+c_{2\beta}^2C_0(0,m_{TP}^2,m_{TP}^2)\big]\nn\\
 &+\frac12g_Y^2 m_{DY}^2 \big[g_2^2 + g_Y^2 + 2 (\lambda_S^2 + \lambda_T^2) - 3 \big(g_2^2 + g_Y^2 - 2 (\lambda_S^2 + \lambda_T^2)\big)c_{4\beta}\big] s_{2\beta}^2C_0(m_{\mathcal{H}}^2, m_{\mathcal{H}}^2, m_{SR}^2)\nn\\
 &+\frac12g_2^2 m_{D2}^2\big[7g_2^2 - g_Y^2 - 2 (\lambda_S^2 - 15\lambda_T^2) - 5 \big(g_2^2 + g_Y^2 - 2 (\lambda_S^2 + \lambda_T^2)\big)c_{4\beta}\big] s_{2\beta}^2C_0(m_{\mathcal{H}}^2, m_{\mathcal{H}}^2, m_{TP}^2)\nn\\
 &+2 (g_2^2+g_Y^2-2(\lambda_S^2+\lambda_T^2))c_{2\beta}^2s_{2\beta}^2\big[3g_Y^2m_{DY}^2C_0(0,m_{\mathcal{H}}^2,m_{SR}^2)+5g_2^2m_{D2}^2C_0(0,m_{\mathcal{H}}^2,m_{TP}^2)\big]\nn\\
 &+\big((g_2^2+g_Y^2)c_{2\beta}^2+2(\lambda_S^2+\lambda_T^2)s_{2\beta}^2\big)c_{2\beta}^2\big[3g_Y^2m_{DY}^2\ov{C}_0(0,0,m_{SR}^2)+5g_2^2m_{D2}^2\ov{C}_0(0,0,m_{TP}^2)\big]\nn\\
 &+\frac{1}{32}\bigg[(-3g_2^2+g_Y^2+2\lambda_S^2-14\lambda_T^2+(g_2^2+g_Y^2-2(\lambda_S^2+\lambda_T^2))c_{4\beta})^2\nn\\
 &\qquad\qquad+4(-2(\lambda_S^2+\lambda_T^2)+(g_2^2+g_Y^2-2(\lambda_S^2+\lambda_T^2))c_{4\beta})^2\nn\\
 &\qquad\qquad+4(g_2^2+g_Y^2-2(\lambda_S^2+\lambda_T^2))^2s_{2\beta}^4\bigg]P_{SS}(m_{\mathcal{H}}^2, m_{\mathcal{H}}^2)\nn\\
 &+\frac32(g_2^2+g_Y^2-2(\lambda_S^2+\lambda_T^2))^2c_{2\beta}^2s_{2\beta}^2P_{SS}(0,m_{\mathcal{H}}^2)\nn\\
 &+\lambda_S^4\big(P_{SS}(m_{SR}^2,m_{SR}^2)+P_{SS}(m_{SI}^2,m_{SI}^2)\big)+3\lambda_T^4\big(P_{SS}(m_{TP}^2,m_{TP}^2)+P_{SS}(m_{TM}^2,m_{TM}^2)\big)\nn\\
 &+(g_2^2-2\lambda_T^2)^2c_{2\beta}^2P_{SS}(m_{TP}^2,m_{TM}^2)+2\lambda_S^2\lambda_T^2\big(P_{SS}(m_{SR}^2,m_{TP}^2)+P_{SS}(m_{SI}^2,m_{TM}^2)\big)
\end{align}

\subsubsection{Corrections to cubics}

The relevant non-zero cubic couplings are $S_R|H|^2$ and $T_P^0 |H|^2$; these are at tree level
\begin{align}
a^{S_R H}_H =& - g_Y m_{DY} c_{2\beta} + \sqrt{2} \lambda_S \mu , \nn\\
a^{T_P^0 H}_H =&\ g_2 m_{D2} c_{2\beta} + \sqrt{2} \lambda_T \mu .
\end{align}
In the following we shall set $\mu =0$.

The shifts are then 
\begin{align}
\kappa^{-1}\ov\delta_S a^{S_R H}_H =& -2g_Ym_{DY}\lambda_S^2c_{2\beta}P_{SS}(0,m_{SR}^2)-g_Y^3m_{DY}^3c_{2\beta}^3\ov{C}_0(0,0,m_{SR}^2)\nn\\
&+6g_2m_{D2}\lambda_S\lambda_Tc_{2\beta}P_{SS}(0,m_{TP}^2)-3g_2^2g_Ym_{D2}^2m_{DY}c_{2\beta}^3\ov{C}_0(0,0,m_{TP}^2)\nn\\
&+g_Y^3m_{DY}^3c_{2\beta}s^2_{2\beta}\bigg(C_0(m_{\mathcal{H}}^2,m_{\mathcal{H}}^2,m_{SR}^2)-2C_0(0,m_{\mathcal{H}}^2,m_{SR}^2)\bigg)\nn\\
&+3g_Yg_2^2m_{DY}m_{D2}^2c_{2\beta}s^2_{2\beta}\bigg(C_0(m_{\mathcal{H}}^2,m_{\mathcal{H}}^2,m_{TP}^2)-2C_0(0,m_{\mathcal{H}}^2,m_{TP}^2)\bigg)\nn\\
& + \frac{1}{8} g_Y m_{DY} c_{2\beta} \bigg( 3 g_2^2 - g_Y^2 +2 \lambda_S^2 + 18 \lambda_T^2 - 3 (g_2^2 + g_Y^2 - 2 (\lambda_S^2 + \lambda_T^2))c_{4\beta}\bigg) P_{SS}(m_{\mathcal{H}}^2,m_{\mathcal{H}}^2)\nn\\
& - \frac{3}{2} g_Y m_{DY} c_{2\beta}s_{2\beta}^2 \big(g_2^2 + g_Y^2 - 2(\lambda_S^2 + \lambda_T^2)\big) P_{SS}(0,m_{\mathcal{H}}^2) \label{EQ:DGcorrectionsaSRHH}\\
\kappa^{-1}\ov\delta_S a^{T_P^0 H}_H =&-2\lambda_S\lambda_Tg_Ym_{DY}c_{2\beta}P_{SS} (0,m_{SR}^2)+g_2 g_Y^2 m_{D2} m_{DY}^2 c_{2\beta}^3\ov{C}_0 (0,0,m_{SR}^2) \nn\\
&+2\lambda_T^2 g_2m_{D2}c_{2\beta}P_{SS} (0,m_{TP}^2)- g_2^3 m_{D2}^3 c_{2\beta}^3\ov{C}_0 (0,0,m_{TP}^2) \nn\\
& - g_2 g_Y^2 m_{D2} m_{DY}^2 c_{2\beta} s_{2\beta}^2 \bigg( C_0 (m_{\mathcal{H}}^2, m_{\mathcal{H}}^2, m_{SR}^2) - 2 C_0 (0, m_{\mathcal{H}}^2, m_{SR}^2) \bigg) \nn\\
& + g_2^3 m_{D2}^3 c_{2\beta} s_{2\beta}^2 \bigg(C_0 (m_{\mathcal{H}}^2, m_{\mathcal{H}}^2, m_{TP}^2) - 2 C_0 (0, m_{\mathcal{H}}^2, m_{TP}^2) \bigg) \nn\\
& + \frac{1}{8} g_2 m_{D2} c_{2\beta}  \bigg(3 g_2^2 -  g_Y^2 - 6 \lambda_S^2 + 10 \lambda_T^2 + \big(g_2^2 + g_Y^2 - 2 (\lambda_S^2 + \lambda_T^2)\big)c_{4\beta}\bigg) P_{SS}(m_{\mathcal{H}}^2,m_{\mathcal{H}}^2)\nn\\
& + \frac{1}{2} g_2 m_{D2} c_{2\beta}s_{2\beta}^2  \big(g_2^2 + g_Y^2 - 2(\lambda_S^2 + \lambda_T^2)\big) P_{SS}(0,m_{\mathcal{H}}^2)
\end{align}
Note that due to our choice of working around the tree-level value of the singlet VEV $v_S=0$ -- following the option 3 described in section \ref{SEC:GENERAL} -- there is an additional shift to the trilinear coupling $a^{S_RH}_H$ not included in eq.~(\ref{EQ:DGcorrectionsaSRHH}), as shown in eq.~(\ref{EQ:DiracTrilinearShift}). However, as can be see in the last line of equation~(\ref{EQ:DiracMatching}), we have already included this shift separately in the threshold correction to the Higgs quartic coupling. 

\subsubsection{Self-energy correction}

The derivative of the Higgs self-energy with respect to momentum is 
\begin{align}
\kappa^{-1}(\Pi^\prime_S)^H_H (0) =& - g_Y^2 m_{DY}^2 c_{2\beta}^2  B^\prime (0;0,m_{SR}^2) - 3 g_2^2 m_{D2}^2 c_{2\beta}^2 B^\prime (0;0,m_{TP}^2) \nn\\
& - g_Y^2 m_{DY}^2 s_{2\beta}^2 B^\prime (0;m_{\mathcal{H}}^2,m_{SR}^2) - 3 g_2^2 m_{D2}^2s_{2\beta}^2 B^\prime (0;m_{\mathcal{H}}^2,m_{TP}^2)
\end{align}

\subsubsection{Corrections to masses}

The corrections to the masses of $S_R$ and $T_P^0$ are
\begin{align}
\kappa^{-1}\ov\delta_S m_{SR}^2 =&\ 2 \lambda_S^2 A_0 (m_{\mathcal{H}}^2) + 2g_Y^2 m_{DY}^2 \big( c_{2\beta}^2P_{SS}(m_{\mathcal{H}}^2,m_{\mathcal{H}}^2) + 2 s_{2\beta}^2 P_{SS}(0,m_{\mathcal{H}}^2)\big) \nn\\
\kappa^{-1}\ov\delta_S m_{T_P^0}^2 =&\ 2 \lambda_T^2 A_0 (m_{\mathcal{H}}^2) + 2 g_2^2 A_0 (m_{T_M}^2) + 2g_2^2 m_{D2}^2 \big( c_{2\beta}^2P_{SS}(m_{\mathcal{H}}^2,m_{\mathcal{H}}^2) + 2 s_{2\beta}^2 P_{SS}(0,m_{\mathcal{H}}^2)\big)
\end{align}

\subsection{Contributions from fermions}

\subsubsection{Corrections to the Higgs quartic coupling}

In the limit of $\mu \rightarrow 0$, we have
\begin{align}
 \kappa^{-1}\ov\delta_F \lt_{HH}^{HH} =&\ \big( g_Y^4 + 4 \lambda_S^4 \big)H_4(0,0,m_{DY}^2,m_{DY}^2)+2 (g_2^2 g_Y^2 + 4 \lambda_S^2 \lambda_T^2) H_4(0,0,m_{D2}^2,m_{DY}^2)\nn\\
 &+\bigg[ 4(g_2^4 + g_2^2 \lambda_T^2 + 4 \lambda_T^4 ) + (g_2^2 - 2 \lambda_T^2)^2 c_{4\beta}\bigg] H_4(0,0,m_{D2}^2,m_{D2}^2) \nn\\ 
 &+4g_Y^2m_{DY}^2\lambda_S^2F_4(0,0,m_{DY}^2,m_{DY}^2)-8 g_Y g_2 \lambda_S \lambda_T m_{DY} m_{D2}F_4(0,0,m_{D2}^2,m_{DY}^2)\nn\\
 &+4g_2^2m_{D2}^2\lambda_T^2(3+2c_{4\beta})F_4(0,0,m_{D2}^2,m_{D2}^2)\,.
\end{align}

\subsubsection{Self-energies and cubic terms}

The fermionic contribution to the derivative of the Higgs self-energy is given (still in the limit $\mu\to0$) by
\begin{align}
\kappa^{-1}(\Pi_F^\prime)_H^H(0)= &-\frac12(g_Y^2+2\lambda_S^2)G'(0;0,m_{DY}^2)-\frac32(g_2^2+2\lambda_T^2)G'(0;0,m_{D2}^2)\nn\\
=&-\frac{1}{4} \bigg[ g_Y^2 + 2\lambda_S^2 + 3 g_2^2 + 6 \lambda_T^2 - 2 (g_Y^2 + 2 \lambda_S^2) \blog m_{DY}^2 - 6 ( g_2^2 + 2 \lambda_T^2) \blog m_{D2}^2 \bigg]
\end{align}
and for the cubic couplings we have
\begin{align}
\kappa^{-1}\ov\delta_F a^{S_R H}_H =&\ \sqrt{2}\lambda_S\bigg[\big(g_Y^2+2\lambda_S^2+\sqrt{2}g_Y\lambda_Sc_{2\beta}\big)F_3(0,0,m_{DY}^2)\nn\\
&\qquad\qquad-\big(g_2^2+2\lambda_T^2+3\sqrt{2}g_2\lambda_Tc_{2\beta}\big)F_3(0,0,m_{D2}^2)\bigg]\nn\\
\kappa^{-1}\ov\delta_F a^{T_P^0 H}_H =&\ \sqrt{2}\lambda_T\bigg[\big(g_Y^2+2\lambda_S^2+\sqrt{2}g_Y\lambda_Sc_{2\beta}\big)F_3(0,0,m_{DY}^2)\nn\\
&\qquad\qquad+\big(3g_2^2+6\lambda_T^2+\sqrt{2}g_2\lambda_Tc_{2\beta}\big)F_3(0,0,m_{D2}^2)\bigg]\nn\\
&-4g_2\big(\sqrt{2}g_2\lambda_T+(g_2^2+2\lambda_T^2)c_{2\beta}\big)F_3(0,m_{D2}^2,m_{D2}^2)
\end{align}

\subsection{Contributions from sfermions}
Here we give the contributions to the different terms in the matching of the Higgs quartic coupling arising from sfermions, in the approximation that $y_t$ -- the top Yukawa coupling in the MDGSSM -- is the only non-vanishing Yukawa coupling. 

\subsubsection{Corrections to the Higgs quartic coupling}

The contribution to $\ov\delta_{\tilde{f}} \lt_{HH}^{HH}$ is unchanged from the MSSM, see for example the result in \cite{Bagnaschi:2014rsa} -- but note that the tree-level expressions therein are given in terms of the SM electroweak couplings, and not in terms of the couplings of the high-energy theory as in this appendix.

\subsubsection{Corrections to cubics}
\begin{align}
 \kappa^{-1}\ov\delta_{\tilde{f}} a_H^{S_RH}\supset&\ y_t^2s_\beta\bigg[-3\sqrt{2}\lambda_SA_tc_\beta P_{SS}(m_{Q_3}^2,m_{U_3}^2)\nn\\
 &\hspace{1cm}+g_Ym_{DY}s_\beta\Big(A_t^2\big[C_0(m_{Q_3}^2,m_{Q_3}^2,m_{U_3}^2)-4C_0(m_{Q_3}^2,m_{U_3}^2,m_{U_3}^2)\big]\nn\\
 &\hspace{2.5cm}+P_{SS}(m_{Q_3}^2,m_{Q_3}^2)-4P_{SS}(m_{U_3}^2,m_{U_3}^2)\Big)\bigg]\nn\\
 &-\frac16g_Y^3m_{DY}c_{2\beta}\sum_{i=1}^3\bigg[P_{SS}(m_{Q_i}^2,m_{Q_i}^2)+8P_{SS}(m_{U_i}^2,m_{U_i}^2)+2P_{SS}(m_{D_i}^2,m_{D_i}^2)\nn\\
 &\hspace{5cm}+3P_{SS}(m_{L_i}^2,m_{L_i}^2)+6P_{SS}(m_{E_i}^2,m_{E_i}^2)\bigg]\nn\\
 \kappa^{-1}\ov\delta_{\tilde{f}} a_H^{T_P^0H}\supset&\ 3y_t^2s_\beta\bigg[-\sqrt{2}\lambda_TA_tc_\beta P_{SS}(m_{Q_3}^2,m_{U_3}^2)\nn\\
 &\hspace{3cm}+g_2m_{D2}s_\beta\Big(A_t^2C_0(m_{Q_3}^2,m_{Q_3}^2,m_{U_3}^2)+P_{SS}(m_{Q_3}^2,m_{Q_3}^2)\Big)\bigg]\nn\\
 &+\frac12g_2^3m_{D2}c_{2\beta}\sum_{i=1}^3\bigg[3P_{SS}(m_{Q_i}^2,m_{Q_i}^2)+P_{SS}(m_{L_i}^2,m_{L_i}^2)\bigg]
\end{align}

\subsubsection{Higgs self-energy corrections}
The sfermion contribution to the derivative of the Higgs self-energy (with respect to momentum) is
\begin{align}
 \kappa^{-1}(\Pi_{\tilde{f}}^\prime)_H^H(0)=-3y_t^2A_t^2s_\beta^2B'(0;m_{Q_3}^2,m_{U_3}^2).
\end{align}

\subsubsection{Corrections to masses}
\begin{align}
 \kappa^{-1}\delta_{\tilde{f}} m_{SR}^2\supset&\ \frac23g_Y^2m_{DY}^2\sum_{i=1}^3\bigg[P_{SS}(m_{Q_i}^2,m_{Q_i}^2)+8P_{SS}(m_{U_i}^2,m_{U_i}^2)+2P_{SS}(m_{D_i}^2,m_{D_i}^2)\nn\\
 &\hspace{3cm}+3P_{SS}(m_{L_i}^2,m_{L_i}^2)+6P_{SS}(m_{E_i}^2,m_{E_i}^2)\bigg]\\
 \kappa^{-1}\delta_{\tilde{f}} m_{T_P^0}^2\supset&\ 2g_2^2m_{D2}^2\sum_{i=1}^3\bigg[3P_{SS}(m_{Q_i}^2,m_{Q_i}^2)+P_{SS}(m_{L_i}^2,m_{L_i}^2)\bigg].
\end{align}

\newpage
\bibliographystyle{utphys}
\bibliography{BGS}

\end{document}